\documentclass[aps,prr,showpacs,preprintnumbers,twocolumn,superscriptaddress,noamssymb,noamsmath]{revtex4-2}

\usepackage{amsmath,amssymb}
\usepackage{graphicx}
\usepackage{color}
\usepackage{natbib}
\usepackage{hyperref}
\hypersetup{
	colorlinks,
	citecolor=blue,
	linkcolor=blue,
	urlcolor=blue}
\usepackage{times}
\usepackage[T1]{fontenc}
\usepackage[toc,page,titletoc]{appendix}
\usepackage{multirow}
\usepackage{placeins}
\usepackage{physics}
\usepackage{bbold}
\usepackage{array}
\usepackage{mathtools}

\let\oricorresponds\corresponds
\let\oridiv\div
\let\corresponds\relax 
\let\div\relax 
\usepackage{mathabx}
\let\corresponds\oricorresponds
\let\div\oridiv
\usepackage[mathscr]{euscript}

\newcolumntype{C}[1]{>{\centering\arraybackslash}m{#1}}
\newcolumntype{L}[1]{>{\raggedright\arraybackslash}m{#1}}
\newcolumntype{R}[1]{>{\raggedleft\arraybackslash}m{#1}}

\newcommand{\cfig}[1]{Fig.~\ref{fig:#1}}
\newcommand{\csubfig}[2]{\cfig{#1}\,(#2)}
\newcommand{\csubfigrange}[3]{\cfig{#1}\,(#2)--(#3)}
\newcommand{\csec}[1]{Sec.~\ref{sec:#1}}

\newcommand{\ceqn}[1]{Eq.~(\ref{eq:#1})}
\newcommand{\capp}[1]{App.~\ref{app:#1}}
\newcommand{\ctab}[1]{Tab.~\ref{tab:#1}}
\newcommand{\cref}[1]{Ref.~\onlinecite{#1}}

\newcommand{\conj}{^{\ast}}
\newcommand{\pconj}{^{\phantom{\ast}}}
\newcommand{\pdag}{^{\phantom{\dagger}}}
\renewcommand{\dag}{^{\dagger}}
\newcommand{\pr}{^{\prime}}
\renewcommand{\vec}[1]{\boldsymbol{#1}}

\begin{document}
\title{A Novel Perspective on Ideal Chern Bands with Strong Short-Range Repulsion: \\ Applications to Correlated Metals, Superconductivity, and Topological Order}

\author{Patrick H. Wilhelm}
\email{patrick.wilhelm@uibk.ac.at}
\affiliation{Institut f\"ur Theoretische Physik, Universit\"at Innsbruck, A-6020 Innsbruck, Austria}
\author{Andreas M. L\"{a}uchli}
\affiliation{Laboratory for Theoretical and Computational Physics, Paul Scherrer Institute, 5232 Villigen, Switzerland}
\affiliation{Institute of Physics, \'{E}cole Polytechnique F\'{e}d\'{e}rale de Lausanne (EPFL), 1015 Lausanne, Switzerland}
\author{Mathias S. Scheurer}
\affiliation{Institute for Theoretical Physics III, University of Stuttgart, 70550 Stuttgart, Germany }

\begin{abstract}
Motivated by recent experiments on correlated van der Waals materials, including twisted and rhombohedral graphene and twisted WSe$_2$, we perform an analytical and numerical study of the effects of strong on-site and short-range interactions in fractionally filled ideal Chern bands. We uncover an extensive non-trivial ground state manifold within the band filling range $0 < \nu < 1$ and introduce a general principle, the ``three-rule'', for combining flatband wave functions, which governs their zero-energy property on the torus geometry. Based on the structure of these wave functions, we develop a variational approach that reveals distinct phases under different perturbations: metallic behavior emerges from a finite dispersion, and superconductivity is induced by attractive Cooper channel interactions. Our approach, not reliant on the commonly applied mean-field approximations, provides an analytical expression for the macroscopic wave function of the off-diagonal long-range order correlator, attributing pairing susceptibility to the set of non-trivial zero-energy ground state wave functions. Extending to finite screening lengths and beyond the ideal limit using exact diagonalization simulations, we demonstrate the peculiar structure in the many-body wave function's coefficients to be imprinted in the low-energy spectrum of the topologically ordered Halperin spin-singlet state. Our findings also make connections to frustration-free models of non-commuting projector Hamiltonians, potentially aiding the future construction of exact ground states for various fractional fillings.
\end{abstract}

\date{\today}

\maketitle

\section{Introduction} 

Topological flatbands with finite Berry curvature have emerged as a fascinating playground for strongly correlated quantum many-body physics~\cite{BergholtzReview,SidReview,BalentSCReview}. The interplay of non-trivial quantum geometry and strong interactions facilitated by the quenched kinetic energy of the electrons paves the way for a plethora of interesting many-body phenomena. However, the inherently correlated nature of the problem complicates the analysis of ground state wave functions, such that our understanding of their general structure remains incomplete. The development of model Hamiltonians entirely made of sums of projector operators historically provided important insights into the structure of otherwise largely inaccessible many-body wave functions. Of particular interest are so called frustration-free models, where the ground state is a simultaneous ground state of each and every individual projector, making it an eigenstate of eigenvalue zero of the Hamiltonian. This class of Hamiltonians is comprised of comparatively simple models such as the paradigmatic toric code and Levin-Wen models~\cite{Kitaev2003,Levin2005}, where all projectors commute, but also non-commuting ones such as the Rokhsar-Kivelson and AKLT models~\cite{Affleck1987,Rokhsar1988,Castelnovo2005}. In this work, we will explore how the ideal quantum geometry of certain flat Chern bands subject to on-site interactions leads to a projector structure of the latter type. This will inherently constrain the ground state selection in systems with spinful flatbands such as twisted bilayer graphene and related moir\'e superlattices~\cite{andreiGrapheneBilayersTwist2020, balentsSuperconductivityStrongCorrelations2020,kennes2020moir,MoireTopology}.

A $\mathcal{C}=1$ ideal flatband is an exactly flat Bloch band with Chern number unity whose quantum geometric tensor exhibits a peculiar structure in its real and imaginary parts. The Berry curvature is generally not homogeneous, but necessarily positive definite and, moreover, it is bound to fluctuate in sync with the Fubini-Study metric.
\begin{figure}[tb]
	\centering
	\includegraphics[width=\columnwidth]{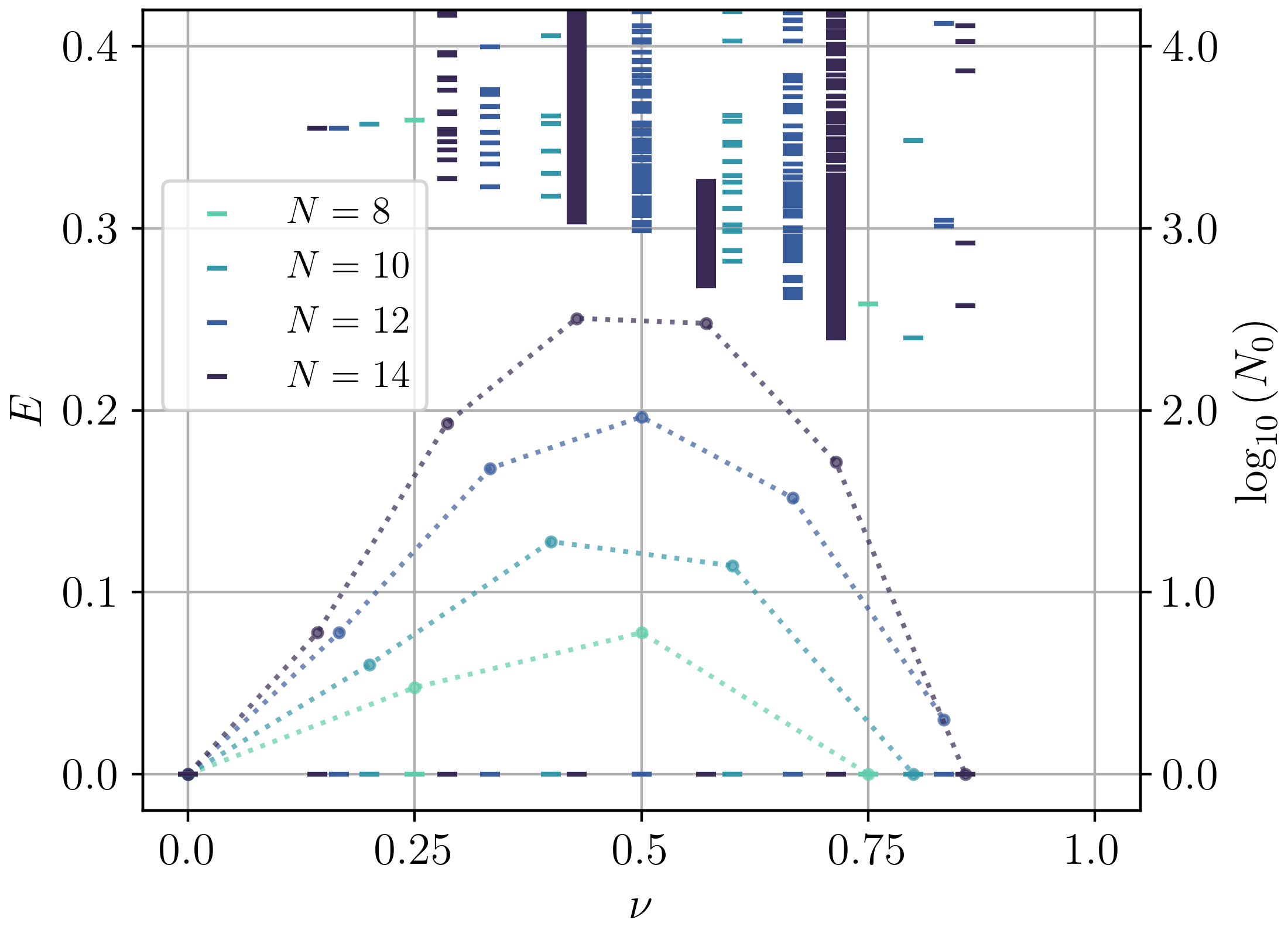}
	\caption{\textbf{Interacting many-body spectrum and zero-degeneracy.} Eigenvalues (individual symbols) of $H$ and zero-energy degeneracy $N_0$ (dotted lines) in the singlet sector with vanishing center of mass momentum of the CTBG model with on-site interaction as a function of band filling $\nu$ for multiple system sizes $N$. $N_0$ is finite for all $0 \leq \nu < 1$, with a maximum likely in the vicinity of $\nu=1/2$ and appears to grow exponentially with system size. Note that the dimension of the Hilbert space itself grows until $\nu=1$. States in the zero-energy manifold are separated by a pronounced gap to excited states at $E>0$. There is no singlet with zero energy at $\nu=1$.} 
	\label{fig:spectrum_filling}
\end{figure}
These properties were found to be related to the notion of vortexability~\cite{Ledwith2023,Okuma2024} and to inherently constrain the band wave functions to have the form~\cite{Tarnopolsky2019,Wang2021a,Wang2021}
\begin{equation}
    \label{eq:psi_ideal_flatband}
    \psi_{\vec{k},\alpha}(\vec{r}) = \mathcal{N}_{\vec{k}} \mathcal{B}_{\alpha}(\vec{r})\Phi_{\vec{k}}(\vec{r}),
\end{equation}
where $\mathcal{B}_{\alpha}(\vec{r})$ is a model- and purely $\vec{r}$-dependent function that is periodic on the lattice up to a phase, $\alpha$ denotes internal indices of the wave function such as sublattice or layer, $\mathcal{N}_{\vec{k}}$ is a normalization factor and $\Phi_{\vec{k}}(\vec{r})$ the lowest Landau level (LLL) wave function. In the symmetric gauge, the LLL wave function on the torus is given by 
\begin{equation}
    \label{eq:formOfPhi}
    \Phi_{\vec{k}}(\vec{r}) = \sigma(r + i k) e^{i k\conj r} e^{-\frac{1}{2}(\abs*{r}^2 + \abs*{k}^2)}, 
\end{equation}
with $\sigma(z)$ denoting the modified Weierstrass sigma function~\cite{Haldane2018} and the unbold symbols indicate the (isotropically) complexified real- and momentum-space coordinates (see \capp{derivation_three_rule} for details).

\subsection{Relation to twisted van der Waals materials}
The chiral model for twisted bilayer graphene (CTBG) constitutes a crucial limit of the TBG continuum model (cf. \capp{continuum_model}) that exhibits ideal flatbands with Chern number $\mathcal{C}=1$ and emerged as a paradigmatic model in the vast family of moir\'e materials with topologically non-trivial bands. At the magic angle, a complete suppression of the intra-sublattice tunneling amplitude $w_{AA}$ (while keeping the sublattice-off-diagonal coupling $w_{AB}$ finite) results in two exactly dispersionless bands per spin and valley at charge neutrality that can be shown to have ideal quantum geometry. 
Furthermore, the CTBG flatband wave functions can be grouped according to the sign of their Chern number $\mathcal{C}= \pm 1$, labeled by a combination of valley and sublattice indices~\cite{Bultinck2020}. A sublattice staggered potential, like it may be caused by the alignment with a hexagonal boron nitride substrate, energetically separates the bands in each valley -- resulting in effective two-component layer spinor wave functions.
Consequently, the band wave function $\psi_{\alpha,\vec{k}}(\vec{r})$ must be of the form \ceqn{psi_ideal_flatband}, with $\alpha$ denoting the moir\'e layer $\alpha=\mathrm{top}/\mathrm{bottom}$. 
Indeed, due to the additional intravalley inversion symmetry in this ideal model, the spinor parts are not independent but $\mathcal{B}_{\mathrm{top}}(\vec{r})=i \mathcal{G}(\vec{r})$ and
$\mathcal{B}_{\mathrm{bottom}}(\vec{r})= \mathcal{G}(-\vec{r})$ at the first magic angle~\cite{Wang2021a}. 
Given the absence of a dispersive term, these flatband systems provide a natural setting to study the effect of comparatively strong density-density interactions in topologically non-trivial bands. Furthermore, the apparently universal connection to Landau levels in \ceqn{psi_ideal_flatband} suggests we can draw from previous insights about quantum Hall wave functions. Crucially, however, in contrast to the LLL case \emph{no magnetic field} is required for quenching the kinetic energy of the charge carriers. As a result, spin is no longer necessarily polarized such that interactions may favor states with vanishing total spin.

Indeed, a plethora of exotic correlated phases have been reported in fractionally filled moir\'e flatbands. Notably, spontaneous symmetry breaking tendencies at half-integer filling observed in twisted mono-bilayer graphene~\cite{Polshyn2022} have been identified to originate from an emergent manifold of spin-charge-modulated topological density waves~\cite{Wilhelm2023}, which was subsequently related to the decomposition into ideal flatbands~\cite{Dong2023,Wang2023}. Similar charge ordered states at half-integer fillings have since been found in a triple-bilayer graphene system~\cite{Wang2024}. 
Furthermore, the ideal quantum geometry was found to play a key role in the construction of fractional Chern insulator (FCI) wave functions~\cite{Ledwith2020,Wang2021,Ledwith2023} -- lattice analogues of fractional quantum Hall states. These topologically ordered states were promoted to be ground state contenders via numerical exact diagonalization (ED) simulations~\cite{Abouelkomsan2020,Repellin2020,Wilhelm2021,Zhao2021,Wang2021,Wang2022,Sharma2024,Liu2024,Shen2024,Reddy2023,Dong2023a} and have been reported to be realized in TBG at a finite magnetic field~\cite{Xie2021} and even at zero field in rhombohedral multi-layer graphene~\cite{Lu2024, Xie2024} as well as MoTe\textsubscript{2}~\cite{Cai2023,Zeng2023,Park2023,Xu2023} moir\'e superlattices. The connection to Landau levels in twisted MoTe\textsubscript{2} was highlighted in \cref{Li2024} by using a variational approach, though the exact parameters of the single-particle model and their relation to the emergence of FCIs is still an open question~\cite{Yu2024} and intensively studied in density functional theory calculations~\cite{Jia2024,Wang2024a}. The importance of lattice effects was also assessed in a composite fermion picture in \cref{Lu2024a}.
Finally, the observation of superconductivity in graphene-based superlattices~\cite{Cao2018, Lu2019, Yankowitz2019, Arora2020, Park2021, Hao2021, Oh2021, JIAsTrilayerScreening, Kim2022, SCDiodesMoireExp} and, very recently, twisted WSe\textsubscript{2}~\cite{Xia2024,WSe2Experiment2} and in rhombohedral graphene~\cite{rhombohedralgrapheneSC,Han2024a} also begs the question how the formation of the superconducting condensate is affected by the interplay of strong correlations and close to ideal quantum geometry. Progress in this direction was made through the concept of ``Quantum Geometric Nesting'' \cite{Han2024}, which relates geometric aspects of the band structure to the stabilization of ordered phases such as superconductors. We emphasize that our study of superconductivity in a single valley is not as artificial as it might seem: first, the observation of a superconducting diode effect at zero field \cite{SCDiodesMoireExp} in certain samples of multi-layer graphene has been shown \cite{Scammell_2022} to be most naturally consistent with valley polarization in the normal state and, more recently, signs of single-valley superconductivity in rhombohedral graphene have been reported \cite{Han2024a}. 

However, while providing important insights on the ground state selection in (quasi) ideal flatbands, a detailed understanding of their microscopic structure remains lacking. In this work, we take a step back by initially considering only the effect of ultra-local on-site interactions, which typically constitute the dominant part of the screened Coulomb interaction. As noted in previous works~\cite{Ledwith2023,Dong2023}, the position- and momentum-separated product representation of \ceqn{psi_ideal_flatband} renders the spatially antisymmetric ferromagnets to be \emph{trivial} zero-energy states under on-site interactions. This poses the natural problem, whether there is a general mechanism for an analogous annihilation of wave functions in multiplets with lower total spin and how it would constrain the structure of the many-body wave function.

\subsection{Summary of main results}
We address this question by focusing on the potentially most \emph{non-trivial} symmetry sector of vanishing total spin, making several significant findings. Firstly, we identify a vast array of singlet zero-energy ground states within the band filling range $0 < \nu < 1$ through ED simulations of on-site interactions in the valley-polarized flatband of the CTBG model (see~\cfig{spectrum_filling}). 
We uncover a novel principle, dubbed the ``three-rule,'' which governs the relations among coefficients in these many-body wave functions and is thus essential for the presence of non-trivial zero-energy states. As opposed to previous works, this mechanism is based on the functional properties of ideal flatband wave functions, rather than purely their product form.
By developing a projector reformulation of the ground state condition, we provide a new perspective on ideal flatbands with short-range interactions as constrained Hilbert (sub-)spaces. This approach links our findings to other frustration-free models of non-commuting projector Hamiltonians like the AKLT and Rokhsar-Kivelson models~\cite{Affleck1987,Rokhsar1988,Castelnovo2005} and the problem of quantum satisfiability~\cite{Bravyi2006,Bravyi2010,Laumann2010,Sattath2016}.
Analytically studying the interplay with perturbations in a simple subspace incorporating the three-rule principle, a finite dispersion is found to introduce correlated metallic behavior, with correlations tied to the three-rule structure. Furthermore, attractive Cooper channel interactions can lead to superconductivity, with the non-trivial zero-energy condition on the ground state wave functions enabling susceptibility to pairing. We derive an analytical expression for the macroscopic pair wave function of the off-diagonal long-range order correlator.
 Extending our analysis to finite screening lengths and beyond the ideal chiral limit using large-scale ED simulations, we confirm that the peculiar structure in the many-body wave function’s coefficients remains evident in the low-energy spectrum of the topologically ordered Halperin spin-singlet state.

\section{Alternative form of on-site interactions in ideal flatbands}\label{sec:alternative}

We consider ideal flatband fermions with spin $S=1/2$ subject to a density-density interaction. The canonical $\mathrm{SU}(2)$-symmetric band-projected Hamiltonian for such a system is 
\begin{align}
    H = \frac{1}{2 \Omega}\sum_{\substack{\vec{k}_1,\vec{k}_2 \\ \vec{k}_3,\vec{k}_4 \in \mathrm{BZ}}} V_{\vec{k}_1,\vec{k}_2,\vec{k}_3,\vec{k}_4} \sum_{\sigma,\sigma\pr=\updownarrows} c\dag_{\vec{k}_1,\sigma}  c\dag_{\vec{k}_2,\sigma\pr} c\pdag_{\vec{k}_3,\sigma\pr} c\pdag_{\vec{k}_4,\sigma}
\end{align}
with $c^{(\dagger)}_{\vec{k},\sigma}$ denoting the fermionic annihilation (creation) operator of momentum $\vec{k}$ in the Brillouin zone (BZ) and spin $\sigma$. 
Restricting to on-site interactions, $U(\vec{r} - \vec{r}\pr) = \delta(\vec{r} - \vec{r}\pr)$, leads to particularly simple matrix elements in ideal flatbands
\begin{align}
    \label{eq:matrix_element_ideal_flatband}
    V_{\vec{k}_1,\vec{k}_2,\vec{k}_3,\vec{k}_4} =& \delta_{\vec{k}_1+\vec{k}_2 -\vec{k}_3 -\vec{k}_4, \delta \vec{b}}\frac{\mathcal{N}_{\vec{k}_1}\mathcal{N}_{\vec{k}_2}\mathcal{N}_{\vec{k}_3}\mathcal{N}_{\vec{k}_4}}{\Omega} \nonumber\\
    & \int_{\Omega}d^2\vec{r} \, \mathcal{F}^2(\vec{r}) \Phi_{\vec{k}_1}\conj(\vec{r}) \Phi_{\vec{k}_2}\conj(\vec{r}) \Phi_{\vec{k}_3}\pconj(\vec{r}) \Phi_{\vec{k}_4}\pconj(\vec{r}),
\end{align}
with $\Phi_{\vec{k}}(\vec{r})$ as given in \ceqn{formOfPhi}, the Kronecker delta implements momentum conservation modulo a reciprocal lattice vector $\delta \vec{b}$, $\Omega$ is the total area of the system and $\mathcal{F}(\vec{r}) = \sum_{\alpha} \abs*{\mathcal{B}_{\alpha}(\vec{r})}^2 \geq 0$. Since $\mathcal{B}_{\alpha}(\vec{r})$ is quasiperiodic on the lattice, $\mathcal{F}(\vec{r})$ is representable as a Fourier series $\mathcal{F}(\vec{r}) = \sum_{\vec{b}} w_{\vec{b}} \exp(i \vec{b} \cdot \vec{r})$, with $w_{\vec{b}}$ also determining the normalization $\mathcal{N}_{\vec{k}}$ (cf. \capp{continuum_model}). This provides a natural connection to the pure LLL case by setting $w_{\vec{0}}=1$ and $w_{\vec{b} \neq \vec{0}}=0$, such that $\psi_{\vec{k}}(\vec{r})=\Phi_{\vec{k}}(\vec{r})$. For the CTBG model specifically, the decomposition of $\mathcal{F}(\vec{r}) = \abs*{\mathcal{G}(\vec{r})}^2 + \abs*{\mathcal{G}(-\vec{r})}^2$ with $w_{\vec{b}} \in \mathbb{R}$ was obtained in \cref{Wang2021}.

Clearly, \ceqn{matrix_element_ideal_flatband} is invariant under $\vec{k}_1 \leftrightarrow \vec{k}_2$ and $\vec{k}_3 \leftrightarrow \vec{k}_4$ separately, resulting in identical Hartree and Fock contributions. This naturally leads to a vanishing of the spin-diagonal parts of $H$ and consequently has trivial zero modes in the form of ferromagnetic product states $\ket*{\Psi_{\mathrm{FM}, \sigma}} = \prod_{\vec{k}} c\dag_{\vec{k},\sigma} \ket*{0}$ (as well as all other states with maximal total spin). 

\begin{figure}[t]
	\centering
	\includegraphics[width=1\columnwidth]{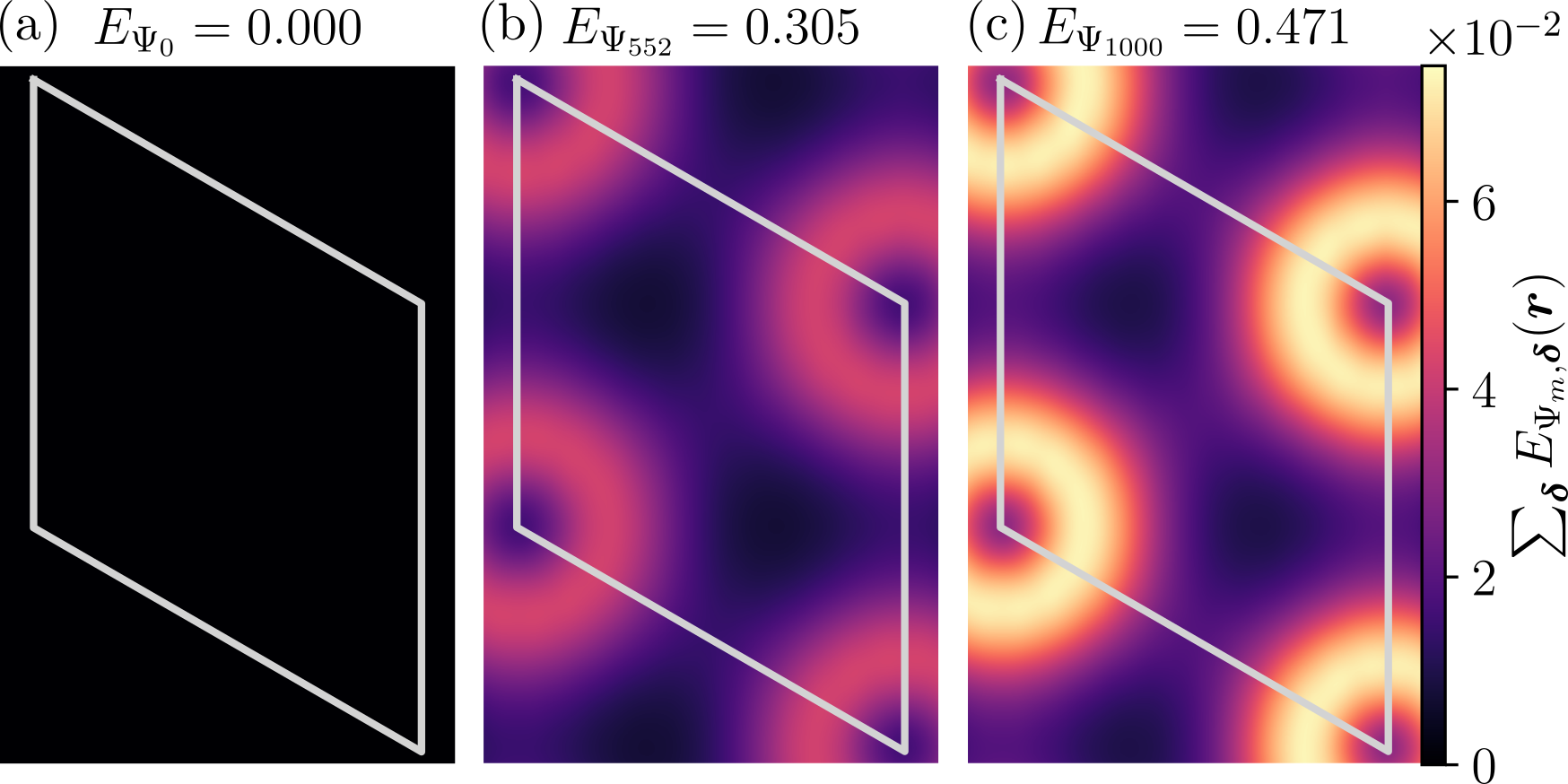}
	\caption{\textbf{Energy density of the IPO Hamiltonian.}
 Locally measured $\sum_{\vec{\delta}}E_{\Psi_m,\vec{\delta}}(\vec{r})$ on representative states in the spectrum. At $\nu=2/5$ for the $N=15$ CTBG model, there are 552 zero-energy states with vanishing total spin and center of mass momentum for which (a) $E_{\Psi_m,\vec{\delta}}(\vec{r})$ has to vanish everywhere. (b) For $m\geq552$, rings of finite signal appear to develop around the vertices of the underlying triangular lattice, resulting in $E_{\Psi_m}>0$ . (c) The intensity successively increases with $m$, leading to higher energies. The qualitative textures in the energy density of excited states persist for other filling fractions. The gray parallelogram marks the unit cell.} 
	\label{fig:local_energy_Q}
\end{figure}

The non-zero part of the Hamiltonian can be expressed via the singlet pair operators $\chi_{\vec{k}, \vec{k}\pr}\pdag = c\pdag_{\vec{k},\uparrow} c\pdag_{\vec{k}\pr,\downarrow}+c\pdag_{\vec{k}\pr,\uparrow} c\pdag_{\vec{k},\downarrow}$ as
\begin{equation}
    H = \frac{1}{4 \Omega}\sum_{\substack{\vec{k}_1,\vec{k}_2 \\ \vec{k}_3,\vec{k}_4}} V_{\vec{k}_1,\vec{k}_2,\vec{k}_3,\vec{k}_4} \chi\dag_{\vec{k}_1,\vec{k}_2} \chi\pdag_{\vec{k}_3,\vec{k}_4}.
\end{equation}
Given the locality of the matrix elements in \ceqn{matrix_element_ideal_flatband}, it is beneficial to switch to a hybrid real-space picture; to this end, we introduce the ``ideal pair operator'' (IPO)
\begin{align}
    \label{eq:Q_definition}
    Q\pdag_{\vec{\delta}}(\vec{r}) =& \sum_{\vec{k} \in \mathrm{BZ}}  \tilde{\Phi}_{\vec{k}+\vec{\delta}}\pconj(\vec{r}) \tilde{\Phi}_{-\vec{k}+\vec{\delta}}\pconj(\vec{r})\chi\pdag_{\vec{k}+\vec{\delta},-\vec{k}+\vec{\delta}}, \nonumber \\
    \text{with} \quad \tilde{\Phi}_{\vec{k}}(\vec{r}) =& \mathcal{N}_{\vec{k}}  e^{i \theta(\vec{k})}\, \Phi_{\vec{k}}(\vec{r})
\end{align}
and $e^{i \theta(\vec{k})} = \Phi_{\vec{k}+\vec{b}}\pconj(\vec{r})/\Phi_{\vec{k}}\pconj(\vec{r})$, where $\vec{b}$ is a reciprocal lattice vector chosen such that $\vec{k} + \vec{b}$ is located in the first BZ.
Using \ceqn{Q_definition}, the Hamiltonian may be written in an alternative form
\begin{equation}
    \label{eq:H_Q_definition}
    H = \sum_{\vec{\delta} \in \mathrm{BZ}} \int_{\Omega}d^2\vec{r} \, \frac{\mathcal{F}^2(\vec{r})}{4 \Omega^2} Q\dag_{\vec{\delta}}(\vec{r}) Q\pdag_{\vec{\delta}}(\vec{r}) = \sum_{\vec{\delta}\in \mathrm{BZ} } H_{\vec{\delta}}.
\end{equation}
It follows that the energy of a state $\ket*{\Psi}$ is given by
\begin{subequations}\begin{equation}
    E_{\Psi} = \bra*{\Psi} H \ket*{\Psi} = \sum_{\vec{\delta} \in \mathrm{BZ}} \int_{\Omega}d^2\vec{r} \, E_{\Psi,\vec{\delta}}(\vec{r}),
\end{equation}
with the energy density in each momentum channel $\vec{\delta}$ amounting to
\begin{equation}
    E_{\Psi,\vec{\delta}}(\vec{r}) = \frac{\mathcal{F}^2(\vec{r})}{4\Omega^2} \norm{Q\pdag_{\vec{\delta}}(\vec{r}) \ket*{\Psi}}^2 \geq 0.
\end{equation}\label{eq:energywithQ}\end{subequations}
Finally, we arrive at the central ground state condition for $\ket*{\Psi}$, which can only be located at zero energy if and only if
\begin{align}
    \label{eq:Q_zero_condition}
    Q\pdag_{\vec{\delta}}(\vec{r}) \ket*{\Psi} = 0, \quad \forall\, \vec{r} \in \Omega, \vec{\delta} \in \mathrm{BZ}.
\end{align}
Remarkably, \ceqn{Q_zero_condition} is not only completely independent of $\mathcal{F}(\vec{r})$, but the normalization factors $\mathcal{N}_{\vec{k}}$ do not alter the dimension of the kernel of $Q\pdag_{\vec{\delta}}(\vec{r})$ (they effectively just alter the normalization of the basis states), such that \ceqn{Q_zero_condition} is \emph{universal for all ideal flatband} systems on a given lattice. 
In stark contrast to ferromagnetic states such as $\ket*{\Psi_{\mathrm{FM}, \sigma}}$, which are annihilated by $\chi\pdag_{\vec{k}+\vec{\delta},-\vec{k}+\vec{\delta}}$ and thus do not require any sort of interference of contributions from different orbitals, there is no obvious reason why states from multiplets with lower total spin should be annihilated by $Q\pdag_{\vec{\delta}}(\vec{r})$. Most notably, states from the spin-singlet sector should, if anything, couple strongly to the singlet pair annihilation operators. 

Counter intuitively though, as evident from the ED spectra in \cfig{spectrum_filling}, there is a substantial amount of many-body singlets with zero energy for a range of fillings $0 < \nu < 1$, separated by a clear gap to higher excited states~\footnote{Although the excitation gap above $E=0$ in \cfig{spectrum_filling} decreases until $\nu \simeq 0.8$, it does not appear to vanish for $\nu \rightarrow 1$. Still, our finite-size data is insufficient for definite statements about the regime just below $\nu=1$.}. 
What is more, the degeneracy appears to grow rapidly with systems size, while being largely insensitive to the geometry of the simulation cell. This is reminiscent of models implementing a form of supersymmetry on the lattice via hardcore fermions~\cite{Fendley2003, Huijse2008,Huijse2012,Galanakis2012,Wilhelm2023a}, where the constrained nature of the Hilbert space can lead to an extensive entropy of the ground state manifold.
In \cfig{local_energy_Q}, we plot the local energy density of the IPO operator measured on representative eigenvectors in the spectrum. As reflected by \csubfig{local_energy_Q}{a}, in order to have $E=0$, $E_{\Psi,\vec{\delta}}(\vec{r})$ is required to vanish identically everywhere. In contrast, $E>0$ implies a finite signal to be present somewhere in the unit cell, which, for the CTBG model studied here, according to \csubfigrange{local_energy_Q}{b}{c} develops rings of increasing intensity centered on vertices of the triangular moir\'e lattice.

\section{Origin of singlet zero-energy modes} \label{sec:origin}

The unexpected presence of robust singlet zero modes begs the question, whether there is a general mechanism for \ceqn{Q_zero_condition} to be satisfied while coupling to $\chi\pdag_{\vec{k}+\vec{\delta},-\vec{k}+\vec{\delta}}$. 

\subsection{The three-rule principle for combining LLL wave functions}
To this end, we start off by considering the commutator of the IPO in \ceqn{Q_definition} with the singlet pair creation operator
\begin{align}
    \label{eq:Q_chi_commutator}
    &[Q\pdag_{\vec{\delta} + \vec{\Delta}}(\vec{r}), \chi\dag_{\vec{k}+\vec{\delta},-\vec{k}+\vec{\delta}}] = \delta_{\vec{\Delta},\vec{0}} \, 4 \, \tilde{\Phi}_{\vec{k}+\vec{\delta}}\pconj(\vec{r}) \tilde{\Phi}_{-\vec{k}+\vec{\delta}}\pconj(\vec{r})
    \nonumber\\
    &-2\sum_{\pm} 
     \tilde{\Phi}_{\pm\vec{k}+\vec{\delta}}(\vec{r})\tilde{\Phi}_{\mp\vec{k}+\vec{\delta} + 2 \vec{\Delta}}(\vec{r}) \sum_{\sigma=\updownarrows} c\dag_{\mp \vec{k} + \vec{\delta},\sigma} c\pdag_{\mp \vec{k} + \vec{\delta} + 2 \vec{\Delta},\sigma}.
\end{align}
For a single pair created on top of the vacuum state, only the first term contributes, resulting in
\begin{align}
    \label{eq:Q_naive_single_pair}
    Q\pdag_{\vec{\delta} + \vec{\Delta}}(\vec{r}) \chi\dag_{\vec{k}+\vec{\delta},-\vec{k}+\vec{\delta}} \ket*{0} = &\delta_{\vec{\Delta},\vec{0}} \, 4 \, \textstyle\prod_{\pm} \mathcal{N}_{\pm \vec{k} + \vec{\delta}} e^{i \theta(\pm \vec{k} + \vec{\delta})} \nonumber \\
    &\Phi_{\vec{k}+\vec{\delta}}(\vec{r}) \Phi_{-\vec{k}+\vec{\delta}}(\vec{r}) \ket*{0},
\end{align}
where we reinserted the definition of $\tilde{\Phi}_{\vec{k}}(\vec{r})$.
For $\vec{\Delta} = \vec{0}$ the IPO maps the singlet back to the vacuum state, changing its center of mass momentum by $-2 \vec{\delta}$. Although $\Phi_{\vec{k}+\vec{\delta}}(\vec{r})$ necessarily becomes zero at specific positions shifted by the momentum argument $\vec{k}+\vec{\delta}$~\cite{Wang2021a}, the expression in \ceqn{Q_naive_single_pair} would need to \emph{vanish at all $\vec{r}$} to result in zero energy, see \ceqn{energywithQ}. It is customary to thus seek a minimal superposition over some set of momentum orbitals $\{\vec{k}\}$, which satisfies the IPO annihilation condition \ceqn{Q_zero_condition} and thus serves as a basic building block for the zero-energy manifold.
Remarkably, resolving \ceqn{Q_zero_condition} boils down to a single, quintessential principle for combining LLL wave functions on the torus:

\begin{figure}[t]
	\centering
	\includegraphics[width=1\columnwidth]{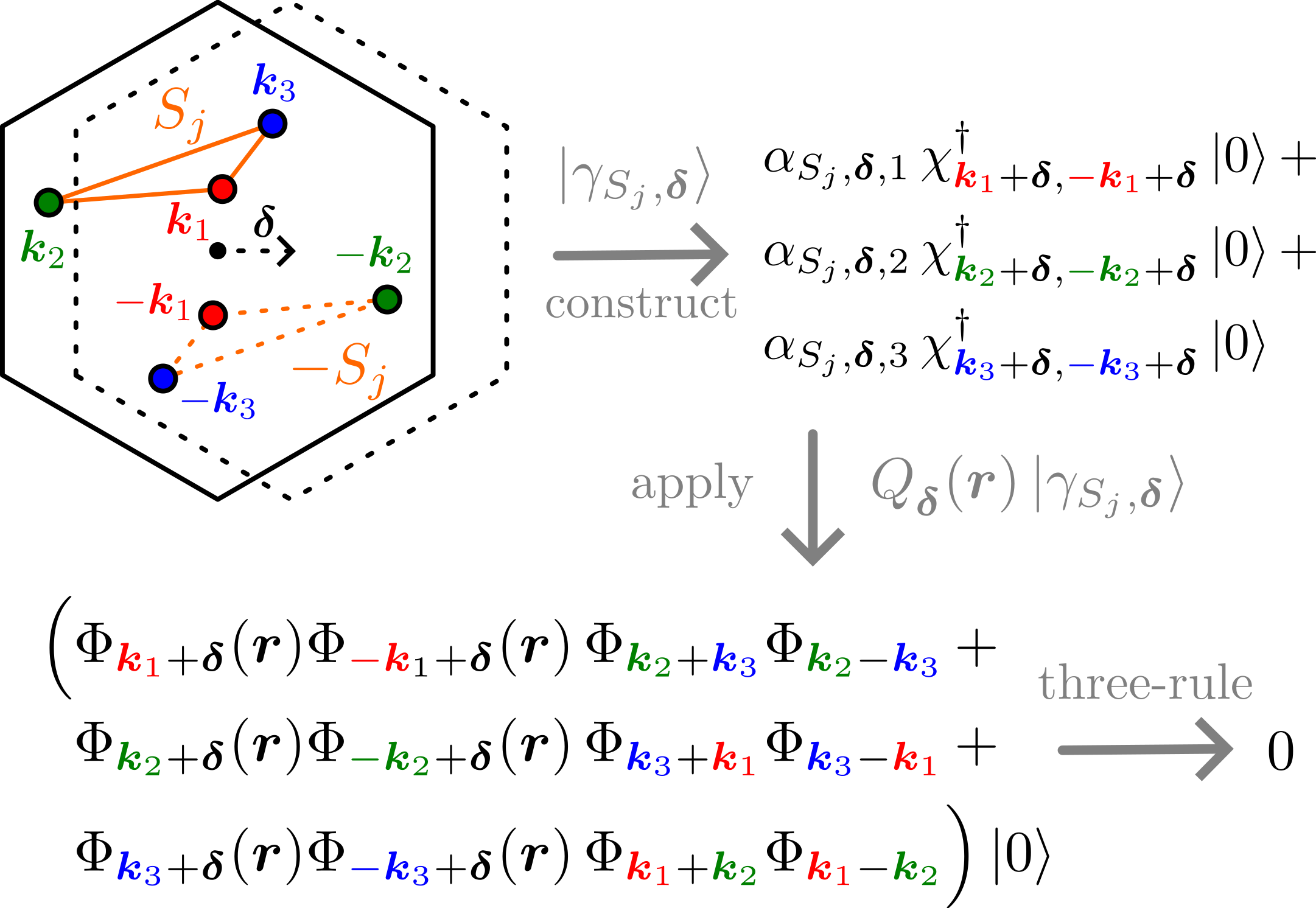}
	\caption{\textbf{Three-rule principle for a single pair.} Illustration of the construction of a two-particle singlet state based on the three-rule principle. States corresponding to three distinct momentum orbitals $\vec{k}_{1,2,3}$ (red,green,blue) from the BZ (solid black hexagon) are combined in a way to exactly reproduce the identity in \ceqn{three_rule_Phi}. A finite momentum transfer $\vec{\delta}$ effectively just shifts the original Brillouin zone (dashed black hexagon). The joint set of three orbitals (and its inverse) $S_j$ is indicated in orange.} 
	\label{fig:three_rule_illustration}
\end{figure}

\emph{Let $\vec{k}_1, \vec{k}_2, \vec{k}_3$ be three arbitrary momentum orbitals, then $\forall\, \vec{r} \in \Omega, \vec{\delta} \in \mathrm{BZ}$ it holds that}
\begin{align}
    \label{eq:three_rule_Phi}
    &\Phi_{\vec{k}_1 + \vec{\delta}}(\vec{r}) \Phi_{-\vec{k}_1 + \vec{\delta}}(\vec{r}) \,\Phi_{\vec{k}_2+\vec{k}_3} \Phi_{\vec{k}_2 -\vec{k}_3} \, +\nonumber \\
    &\Phi_{\vec{k}_2 + \vec{\delta}}(\vec{r}) \Phi_{-\vec{k}_2 + \vec{\delta}}(\vec{r}) \,\Phi_{\vec{k}_3+\vec{k}_1} \Phi_{\vec{k}_3 -\vec{k}_1} \, +\nonumber\\
    &\Phi_{\vec{k}_3 + \vec{\delta}}(\vec{r}) \Phi_{-\vec{k}_3 + \vec{\delta}}(\vec{r}) \,\Phi_{\vec{k}_1+\vec{k}_2} \Phi_{\vec{k}_1 -\vec{k}_2} = 0,
\end{align}
\emph{with $\Phi_{\vec{k}} = \Phi_{\vec{k}}(\vec{0})$.} 

The derivation of this curious property can be found in \capp{derivation_three_rule}, but, in essence, it is inherited from a related identity involving Weierstrass sigma functions. 

We can make use of \ceqn{three_rule_Phi} by choosing a set of three \emph{distinct} momenta $S_j = \{\vec{k}_1^j, \vec{k}_2^j, \vec{k}_3^j\}$ from the Brillouin zone and defining
\begin{align}
     \alpha_{S_j, \vec{\delta}, l} = \mathcal{N}_{S_j,\vec{\delta}} \sum_{m,n=1}^3 \epsilon_{l m n} \frac{\Phi_{\vec{k}_m^j+\vec{k}_n^j} \Phi_{\vec{k}_m^j -\vec{k}_n^j}} {\textstyle\prod_{\pm} \mathcal{N}_{\pm \vec{k}_l^j + \vec{\delta}} e^{i \theta( \pm \vec{k}_l^j + \vec{\delta})}},
\end{align}
where $\epsilon_{l m n}$ is the Levi-Civita symbol compensating for the additional sign acquired by $\Phi_{\vec{k}_m^j-\vec{k}_n^j}$ when exchanging $\vec{k}_m^j \leftrightarrow \vec{k}_n^j$ and $\mathcal{N}_{S_j,\vec{\delta}}$ is a normalization such that $\sum_{l=1}^3 \abs*{ \alpha_{S_j, \vec{\delta}, l}}^2=1$.
As can be straightforwardly verified, $ \alpha_{S_j, \vec{\delta}, l}$ is invariant under the separate inversion of any individual momentum $\vec{k}_l^j \rightarrow -\vec{k}_l^j$. Furthermore, as long as $\vec{k}_m^j \neq \pm \vec{k}_n^j$ for $m \neq n$, $ \alpha_{S_j, \vec{\delta}, l}$ is non-zero and the entangled singlet pair operator
\begin{align}
    \label{eq:definition_gamma}
     \gamma\dag_{S_j,\vec{\delta}} =& \frac{1}{\sqrt{2}}\sum_{l=1}^3 \,  \alpha_{S_j,\vec{\delta},l}  \, \chi\dag_{\vec{k}_l^j + \vec{\delta}, -\vec{k}_l^j + \vec{\delta}}
\end{align}
is well defined \footnote{For simplicity of notation, we assume $S_j$ in $ \gamma\dag_{S_j,\vec{\delta}}$ does not contain inversion-invariant momenta, as this would change the prefactor in the norm $1/\sqrt{2} \rightarrow 1/2$.}. 

Applying $Q\pdag_{\vec{\delta}}(\vec{r}) $ to $\ket*{ \gamma_{S_j,\vec{\delta}}}=  \gamma\dag_{S_j,\vec{\delta}} \ket*{0}$ naturally gives
\begin{align}
    \label{eq:Q_single_pair}
    Q\pdag_{\vec{\delta}\pr}(\vec{r}) \ket*{ \gamma_{S_j,\vec{\delta}}} =
    &\delta_{\vec{\delta}\pr, \vec{\delta}}\, 2 \sqrt{2} \,\mathcal{N}_{S_j,\vec{\delta}}\nonumber\\
    \Bigg[\sum_{l,m,n=1}^3 \epsilon_{l m n}  & \Phi_{\vec{k}_l^j + \vec{\delta}}(\vec{r}) \Phi_{-\vec{k}_l^j + \vec{\delta}}(\vec{r}) \Phi_{\vec{k}_m^j+\vec{k}_n^j} \Phi_{\vec{k}_m^j -\vec{k}_n^j}\Bigg] \ket*{0} \nonumber\\
    =& \, 0,
\end{align}
where the expression in square brackets is exactly of the type contained in \ceqn{three_rule_Phi} and thus vanishes identically by the three-rule. The steps of the procedure are illustrated graphically in \cfig{three_rule_illustration}. 
By virtue of the zero energy condition of \ceqn{Q_zero_condition}, $\ket*{ \gamma_{S_j,\vec{\delta}}}$ is a singlet ground state of $H$. Furthermore, for this two-particle case we can explicitly quantify the amount of entanglement stemming purely from the three-rule as
\begin{align}
    \Delta\mathscr{S}_{ \gamma_{S_j,\vec{\delta}}} = -\sum_{l=1}^3 \abs*{ \alpha_{S_j, \vec{\delta}, l}}^2 \log \abs*{ \alpha_{S_j, \vec{\delta}, l}}^2 > 0,
\end{align}
with $\abs*{ \alpha_{S_j, \vec{\delta}, l}}^2<1$ (cf. \capp{entanglement_single_pair}). Naturally, $\Delta\mathscr{S}_{ \gamma_{S_j,\vec{\delta}}}$ becomes maximal if all $\abs*{ \alpha_{S_j, \vec{\delta}, l}}^2$ are equal, which will be the case if, e.g., all $\vec{k}_l^j \in S_j$ are related by point-group symmetry operations.

\subsection{Projector form of the ground state condition}

Even though the preceding arguments are rigorous, the full theory eludes an exact analytical treatment. The predicament arises from the second part of the commutator in \ceqn{Q_chi_commutator}, which vanishes identically for a single pair, but remains active for more particles. This prohibits a straightforward generalization to more than one pair. The central role of the three-rule, however, can be still be verified numerically on finite systems.

To this end, we note that $Q\pdag_{\vec{\delta}}(\vec{r})$ maps a state $\ket*{\Psi}$ from the Hilbert-space $\mathcal{H}_{\bar{\vec{k}}}^{N_C}$ with $N_C$ pairs and center of mass momentum $\bar{\vec{k}}$ to $\ket*{\Xi_{\vec{\delta}}(\vec{r})}= Q\pdag_{\vec{\delta}}(\vec{r}) \ket*{\Psi}$ located in the Hilbert-space $\mathcal{H}_{\bar{\vec{k}} - 2 \vec{\delta}}^{N_C-1}$ with $N_C-1$ pairs and $\bar{\vec{k}}$ shifted by $- 2 \vec{\delta}$ without changing its spin polarization (which we assume to be zero as we are exclusively concerned with the singlet sector). The ground state condition \ceqn{Q_zero_condition} implies that all basis coefficients $\Xi_{n,\vec{\delta}}(\vec{r}) \in \mathbb{C}$ in $\ket*{\Xi_{\vec{\delta}}(\vec{r})} = \sum_{n}\Xi_{n,\vec{\delta}}(\vec{r})\ket*{n,\vec{\delta}}$ must vanish identically, where $n$ labels the Fock-space configurations of the $Q\pdag_{\vec{\delta}}(\vec{r})$-scattered Hilbert-space.
Given the structure of the IPO in \ceqn{Q_definition}, each $\Xi_{n,\vec{\delta}}(\vec{r})$ individually is affected only by comparatively few components of $\ket*{\Psi}$, whose subspace we label by $\mathcal{W}_{n,\vec{\delta}} \subset \mathcal{H}_{\bar{\vec{k}}}^{N_C}$.
In this sense, the orientation of $\ket*{\Psi}$ within $\mathcal{W}_{n,\vec{\delta}}$ uniquely determines the value of $\Xi_{n,\vec{\delta}}(\vec{r})$, with only select alignments resulting in $\Xi_{n,\vec{\delta}}(\vec{r})=0$ for all $\vec{r}$.

In order to test the degree to which the three-rule governs the zero-energy manifold, we henceforth make the ansatz that the zero-subspaces $\mathcal{V}_{n,\vec{\delta}} \subset \mathcal{W}_{n,\vec{\delta}}$ leading to $\Xi_{n,\vec{\delta}}(\vec{r})=0$ are spanned in accordance with the three-rule, analogous to the single-pair case in \ceqn{Q_single_pair}. Constructing the projectors $\mathbb{1}_{n,\vec{\delta}}$ on $\mathcal{W}_{n,\vec{\delta}}$ as well as $\mathcal{P}_{n,\vec{\delta}}$ on $\mathcal{V}_{n,\vec{\delta}}$ (for details we refer to \capp{projector_construction}), the condition of $\mathbb{1}_{n,\vec{\delta}} \ket*{\Psi}$ being fully contained in $\mathcal{V}_{n,\vec{\delta}}$ can be formulated as an eigenvalue problem
\begin{align}
    \label{eq:projector_condition_subspace}
    \mathcal{P}_{n,\vec{\delta}}\ket*{\Psi}=\mathbb{1}_{n,\vec{\delta}}\ket*{\Psi},
\end{align}
where we used $\mathcal{P}_{n,\vec{\delta}} \mathbb{1}_{n,\vec{\delta}} = \mathcal{P}_{n,\vec{\delta}}$. 
Intuitively, if \ceqn{projector_condition_subspace} is satisfied, $\ket*{\Psi}$ is guaranteed to lead to a structure similar to \ceqn{three_rule_Phi} under the application of the IPO, resulting in a vanishing of the associated basis coefficient $\Xi_{n,\vec{\delta}}(\vec{r})=0$. 

Using zero-energy states $\ket*{\Psi}$ obtained from performing ED on the original Hamiltonian $H$ for various fillings and system sizes in \cfig{spectrum_filling}, we can numerically test \ceqn{projector_condition_subspace} by measuring
\begin{align}
    \label{eq:weight_three_rule}
    w_{\Psi,n,\vec{\delta}} = \frac{\norm{\mathcal{P}_{n,\vec{\delta}} \ket*{\Psi}}^2}{\norm{\mathbb{1}_{n,\vec{\delta}} \ket*{\Psi}}^2},
\end{align}
which turns out to be \emph{exactly unity} for all $\vec{\delta}$ and $n$ if $\norm{\mathbb{1}_{n,\vec{\delta}} \ket*{\Psi}}^2>0$, else zero. What is more, $w_{\Psi,n,\vec{\delta}}$ may serve as a proxy for the pertinence of the three-rule in the analysis of general many-body wave functions. We make use of this metric below by applying it to the ground state and its lowest excitations at filling $\nu=2/5$ of the TBG flatband with finite screening length and away from the chiral limit. 
On this note, while  $w_{\Psi,n,\vec{\delta}}=1$ is well defined in the thermodynamic limit, in perturbed models with $w_{\Psi,n,\vec{\delta}}<1$, we expect the relative alignment to scale to zero according to the arguments of the orthogonality catastrophe~\cite{Anderson1967}. For a given system size,  $w_{\Psi,n,\vec{\delta}}$ then serves as a lower bound to the corresponding size-invariant quantity.

Building on \ceqn{projector_condition_subspace}, with $\bar{\mathcal{P}}_{n,\vec{\delta}} = \mathbb{1}_{n,\vec{\delta}} - \mathcal{P}_{n,\vec{\delta}}$ denoting the zero-subspace orthogonal projector, $\bar{\mathcal{P}}_{n,\vec{\delta}}\ket*{\Psi} = 0$ must equally hold. Mandating this requirement for all $\ket*{n,\vec{\delta}}$ simultaneously, we arrive at a projector reformulation of the full zero-energy condition \ceqn{Q_zero_condition}:
\begin{align}
    \label{eq:projector_condition}
    \bar{\mathcal{P}} \ket*{\Psi} = \sum_{\vec{\delta} \in \mathrm{BZ}} \sum_{n} \bar{\mathcal{P}}_{n,\vec{\delta}}\ket*{\Psi} = 0.
\end{align}
Notably, the orthogonal subspace projectors $\bar{\mathcal{P}}_{n,\vec{\delta}}$ generally do not commute, rendering \ceqn{projector_condition} a genuine quantum many-body problem.
By numerically diagonalizing the full orthogonal projector $\bar{\mathcal{P}}$ we can examine the completeness of the conditions imposed by the three-rule. Assuming the three-rule is at the root of \emph{all} non-trivial zero-energy states, the kernel of $\bar{\mathcal{P}}$ must be identical to the one of $Q\pdag_{\vec{\delta}}(\vec{r})$ and thus $H$. 
Comparing with the low-energy spectra obtained via ED \footnote{Restricted to $N\leq12$ due to the rapid increase in complexity of constructing $\bar{\mathcal{P}}$.}, the degeneracy of the zero eigenvalue of $\bar{\mathcal{P}}$ \emph{exactly matches} the dimension of the zero-energy manifold of $H$ \emph{for every multiplet sector} -- reassuring the three-rule \ceqn{three_rule_Phi} to be the \emph{unique} principle governing the appearance of non-trivial zero modes. 
States with maximal total spin are fully antisymmetric under the exchange of any pair of momentum orbitals and hence trivially contained in the kernel of $\bar{\mathcal{P}}$. A more in depth characterization of the mathematical structure in \ceqn{projector_condition} is reserved for future works, though it would be interesting to analyze it in the context of the quantum satisfiability problem~\cite{Bravyi2006,Bravyi2010,Laumann2010} and potentially relate it to classical interaction graphs as has been conducted for local projector Hamiltonians in \cref{Sattath2016}. 

\section{Analytic study of perturbations to the degenerate manifold} \label{sec:analytic}

Given the substantial degeneracy of the zero-energy states, it is crucial to analyze the qualitative effect of three-rule correlations, induced by the dominant density-density interaction, on phases stabilized by perturbations from the degenerate manifold. In order to keep the problem analytically tractable while retaining the essence of the zero-energy manifold, in this section we exclusively work in the subspace spanned by a set of non-trivial zero modes of $H_{\vec{0}}$, i.e.~we exclusively consider states that inherently rely on the three-rule to satisfy \ceqn{Q_zero_condition} for $\vec{\delta}=\vec{0}$. 
Analogous arguments can be made for $\vec{\delta} \neq \vec{0}$, though $\vec{\delta}=\vec{0}$ is the most relevant for the perturbations considered in the following.

To begin with, we partition the Brillouin zone, or any subset of momenta, $K$ into $N_{K}$ disjoint sets $K_j = S_j \cup (-S_j)$, such that $K = \bigcup_{j=1}^{N_K} K_j$ (each $S_j$ only contains three momentum orbitals not related by inversion).
Subselecting a combination of $N_C$ distinct sets of $S_j$, we use $\mathcal{S} = \{S_{j_1},...,S_{j_{N_C}}\}$ to form a zero-energy subspace basis from simple states that naturally incorporate the three-rule:
\begin{align}
    \label{eq:definition_S}
    \ket*{\mathcal{S}} = \prod_{S_j \in \mathcal{S}} \gamma\dag_{S_j,\vec{0}} \ket*{0}, \quad \mathrm{with} \quad \braket*{\mathcal{S}}{\mathcal{S}\pr}=\delta_{\mathcal{S},\mathcal{S}\pr}.
\end{align}
The orthogonality property is a direct consequence of $S_j$ and $S_{j\pr}$ not overlapping for $j \neq j\pr$. The total number of configurations $\mathcal{S}$, and hence the dimension of this subspace, is $D_{\mathcal{S}} = {N_K \choose N_C}$. Applying $ \gamma\dag_{S_j,\vec{0}}$ to the vacuum state populates it with a pair of electrons across six distinct momentum orbitals (neglecting inversion invariant momenta), corresponding to $N_K = N/6$ and a maximum (electron) filling fraction of $\nu=1/3$ when considering the entire Brillouin zone. It should be emphasized that this maximum filling restriction is an artifact of the way the subspace is set up as non-overlapping combinations of three-rule orbitals.
Higher fillings can, for example, be reached by introducing double-occupation of orbitals by exploiting a three-rule-like relation of coefficients in the wave function [cf.~\ceqn{three_rule_pure_momentum_space}]. Furthermore, our numerical results in \cfig{spectrum_filling} indicate that such constructions should, at least in principle, be feasible until just below $\nu = 1$.

\subsection{Dispersion-induced correlated metal} \label{sec:disperion}

We now consider the effect of introducing a small but finite dispersive term
\begin{align}
    H_{\mathrm{kin}} = \sum_{\vec{k} \in \mathrm{BZ}} \sum_{\sigma=\updownarrows} \epsilon_{\vec{k}} \,c\dag_{\vec{k},\sigma}c\pdag_{\vec{k},\sigma},
\end{align}
while keeping the band geometry ideal. We calculate the commutator with $ \gamma\dag_{S_j,\vec{0}}$ to be (details in \capp{derivation_dispersion})
\begin{align}
    \left[H_{\mathrm{kin}},  \gamma\dag_{S_j,\vec{0}}  \right] =  \frac{1}{\sqrt{2}} \sum_{l=1}^3 (\epsilon_{\vec{k}_l^j} + \epsilon_{-\vec{k}_l^j}) \,  \alpha_{S_j,\vec{0},l} \,\chi\dag_{\vec{k}_l^j, -\vec{k}_l^j},
\end{align}
which implies for the action on a many-body state
\begin{align}
    H_{\mathrm{kin}} \ket*{\mathcal{S}} =& H_{\mathrm{kin}} \prod_{S_j \in S} \gamma\dag_{S_j,\vec{0}} \ket*{0} \nonumber \\
    =& \frac{1}{\sqrt{2}} \sum_{j=1}^{N_C} \sum_{l=1}^3 (\epsilon_{\vec{k}_l^j} + \epsilon_{-\vec{k}_l^j}) \,  \alpha_{S_j,\vec{0},l} \,\chi\dag_{\vec{k}_l^j, -\vec{k}_l^j} \ket*{\mathcal{S} \setminus S_j},
\end{align}
with $\ket*{\mathcal{S} \setminus S_j} = \prod_{S_i \in \mathcal{S}, S_i\neq S_j } \gamma\dag_{S_i,\vec{0}} \ket*{0}$. As a result of the orthonormality relation in \ceqn{definition_S}, the matrix elements turn out to be
\begin{align}
    \label{eq:matrixelement_Hkin}
    \bra*{\mathcal{S}} H_{\mathrm{kin}} \ket*{\mathcal{S}\pr} = \delta_{\mathcal{S},\mathcal{S}\pr} \sum_{j=1}^{N_C} \sum_{l=1}^3 (\epsilon_{\vec{k}_l^j} + \epsilon_{-\vec{k}_l^j}) \, \abs*{ \alpha_{S_j,\vec{0},l}}^2.
\end{align}

As evident from \ceqn{matrixelement_Hkin}, a spin independent dispersive term is diagonal in the subspace spanned by all $\ket*{\mathcal{S}}$. Consequently, the splitting of the degenerate manifold of states will be governed by the values of $\epsilon_{\vec{k}_l^j} + \epsilon_{-\vec{k}_l^j}$ weighted by the squared amplitude of the corresponding three-rule-coefficients $\abs*{ \alpha_{S_j,\vec{0},l}}^2$. The absolute ground state $\ket*{\Psi_0^{\mathrm{kin}}}$ in the subspace of non-trivial zero modes of $H_{\vec{0}}$ is found by choosing an optimal configuration of $S_j$ in $\mathcal{S}$ such that this weighted average is minimized. Consequently, the ground state $\ket*{\Psi_0^{\mathrm{kin}}} = \ket*{\mathcal{S}}$ is entangled beyond the degree of an ordinary Fermi-sea by respecting the three-rule on the level of each $ \gamma\dag_{S_j,\vec{0}}$, but is ultimately composed as a simple product of these operators; we will see below that this is very different when additional attractive interactions are taken into account. With the ground state energy of the form contained in \ceqn{matrixelement_Hkin}, there exist gapless excitations in the thermodynamic limit as long as the three-rule condition of \ceqn{three_rule_Phi} is respected.

\begin{figure}[t]
	\centering
	\includegraphics[width=\columnwidth]{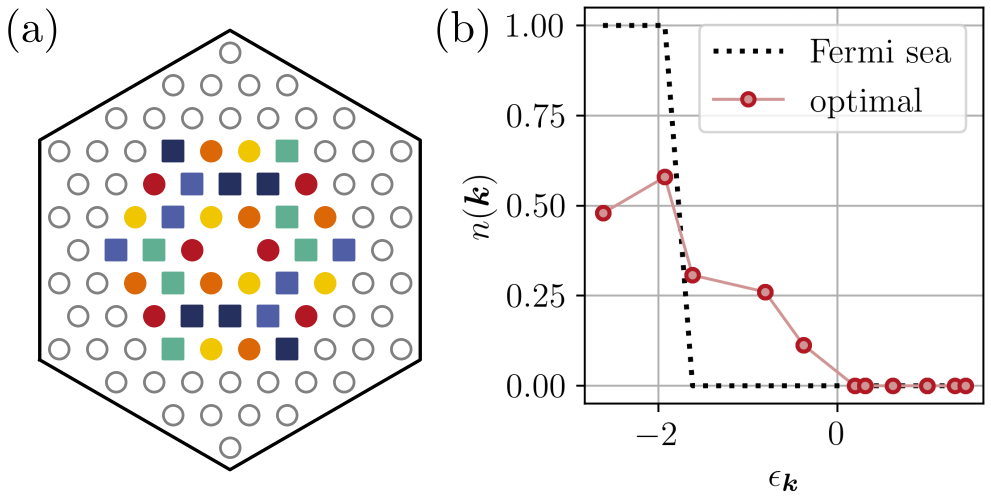}
	\caption{\textbf{Optimal three-rule configuration and occupation number for a finite dispersion.} (a) An exemplary configuration of three-rule orbitals minimizing \ceqn{matrixelement_Hkin} for the cosine dispersion in \ceqn{cosine_dispersion}. (b) The corresponding electron occupation number exhibits a softened step as a function of $\epsilon_{\vec{k}}$. For reference, the dashed line indicates the occupation of the same momentum grid for a Fermi sea without double-occupation. The displayed system has 12 electrons on effectively 90 momentum orbitals. In (a), individual three-rule sets $S_j$ (involving six momenta including inversion) are indicated by different colors. Sets related by rotational symmetry operations share the same symbol.} 
\label{fig:dispersion_configuration_occupation}
\end{figure}

Before addressing additional interactions, we first validate these considerations for the CTBG model by choosing a simple form of $\epsilon_{\vec{k}}$:
\begin{align}
    \label{eq:cosine_dispersion}
    \epsilon_{\vec{k}} = - \sum_{j=0}^2 \cos( \vec{k}  \cdot \vec{a}_1^{(j)}), \quad \mathrm{with} \quad \vec{a}_1^{(j)} = R_{2 \pi j /3} \vec{a}_1.
\end{align}
Similar to the dispersion of the TBG valence band away from the chiral limit, $\epsilon_{\vec{k}}$ has a minimum at the origin of the Brillouin zone and increases towards the boundary. Figure~\ref{fig:dispersion_configuration_occupation}~(a) shows the optimal configuration of three-rule orbitals for an exemplary system (for details on the implementation, we refer to \capp{pairing_optimization}, where the depth-first search procedure is discussed in the context of a pairing Hamiltonian). While all orbitals close to the origin are occupied, apparently preserving the rotational symmetries of the lattice, the ones at larger $\abs*{\vec{k}}$ are empty. This translates to an electron occupation number $n(\vec{k}_l^j) = \abs*{ \alpha_{S_j,\vec{0},l}}^2$ [cf.~\csubfig{dispersion_configuration_occupation}{b}] that resembles metallic behavior as a function of $\epsilon_{\vec{k}}$, with correlations rooted in the three-rule construction significantly softening the step. While our finite-size results in \csubfig{dispersion_configuration_occupation}{b} illustrate the general occupation tendency of the system, the non-overlapping constraint imposed on the orbital selection will be less severe for larger systems. We can thus expect $n(\vec{k})$ to be smoothened in the thermodynamic limit, while the overall trend remains the same.

\subsection{Singlet pairing induced superconductivity} 
\label{sec:pairing}

We now turn to a different scenario, where we assume an attractive pair interaction term on top of the dominant repulsive interaction in $H_{\vec{0}}$. Specifically, we consider a pairing Hamiltonian of the form
\begin{align}
    \label{eq:Hpair}
    H_{\mathrm{pair}} =& -g \, \eta\dag \eta\pdag, \quad  g>0, \nonumber \\
    \eta\pdag =& \sum_{\vec{k} \in \mathrm{BZ}} f_{\vec{k}} \braket*{\psi_{-\vec{k}}\conj} {\tau \, \psi_{\vec{k}}} \, \chi_{-\vec{k},\vec{k}}, \quad
\end{align}
where $f_{\vec{k}}$ is an even, complex-valued function encoding the form factors and in the ideal flatband-projected setting 
\begin{align}
    \braket*{\psi_{-\vec{k}}\conj} {\tau \, \psi_{\vec{k}}} = &\frac{\mathcal{N}_{-\vec{k}} \mathcal{N}_{\vec{k}} }{\Omega} \nonumber \\
    &\int_{\Omega} d^2\vec{r} \Big[\sum_{\alpha,\beta} \tau_{\alpha,\beta} \mathcal{B}_{\alpha}(\vec{r}) \mathcal{B}_{\beta}(\vec{r})\Big] \Phi_{-\vec{k}}(\vec{r}) \Phi_{\vec{k}}(\vec{r}) 
\end{align}
with $\tau$ acting on internal indices such as layer for TBG (see \capp{eta_pairing} for details). The simplest $s$-wave channel is obtained for $f_{\vec{k}}=1$, with $H_{\mathrm{pair}}$ corresponding to an on-site pairing interaction in real-space.  
Again, we evaluate the commutator with the entangled pair operator to obtain
\begin{align}
    \label{eq:commutator_xi_Gamma}
    \left[\eta,  \gamma\dag_{S_j,\vec{0}}  \right] =  \sqrt{2} \sum_{l=1}^3 f_{\vec{k}_l^j} \braket*{\psi_{-\vec{k}_l^j}\conj} {\tau \, \psi_{\vec{k}_l^j}} \,  \alpha_{S_j,\vec{0},l} \, \Big(2 - \sum_{\pm} n_{\pm\vec{k}_l^j} \Big),
\end{align}
where $n_{\vec{k}}=\sum_{\sigma=\updownarrows}c\dag_{\vec{k},\sigma}c\pdag_{\vec{k},\sigma}$, while the commutation with $\eta\dag$ is trivial \footnote{Due to the anticommutation of pairs of fermionic creation operators.}. Acting with $\eta$ on $\ket*{\mathcal{S}}$ now gives
\begin{align}
    \label{eq:betaSj}
    \eta \ket*{\mathcal{S}} =& \sum_{j=1}^{N_C} \beta_{S_j} \ket*{\mathcal{S} \setminus S_j}, \nonumber \\
    \beta_{S_j} =& \sqrt{2} \sum_{l=1}^3 f_{\vec{k}_l^j} \braket*{\psi_{-\vec{k}_l^j}\conj} {\tau \, \psi_{\vec{k}_l^j}} \,  \alpha_{S_j,\vec{0},l} \in \mathbb{C}. 
\end{align}
Due to the self-adjoint structure of \ceqn{Hpair}, we can immediately conclude that
\begin{align}
    \label{eq:matrixelement_Hpair}
    \bra*{\mathcal{S}} H_{\mathrm{pair}} \ket*{\mathcal{S}\pr} = -g \sum_{i,j=1}^{N_C} \beta_{S_i}\conj \beta_{S\pr_{j}} \delta_{\mathcal{S} \setminus S_i,\mathcal{S}\pr \setminus S\pr_j},
\end{align}
where $\delta_{\mathcal{S} \setminus S_i,\mathcal{S}\pr \setminus S\pr_j} = 1$ if $\mathcal{S}$ without $S_i$ is the same set as $\mathcal{S}\pr$ without $S\pr_j$, otherwise it is zero.
Note that as long as $\abs*{\beta_{S_j}}>0$, which, for a given $f_{\vec{k}}$ and $\tau$, is exclusively determined by the partitioning of orbitals into sets $S_j$, the three-rule states spanning this subspace are bound to couple to $\eta$. This ultimately stems from the fact that the pairing Hamiltonian in \ceqn{Hpair} has a similar operator structure to $H_{\vec{0}}$ and all $\ket*{\mathcal{S}}$ are non-trivial zero modes. Importantly, however, considering the isotropic $f_{\vec{k}}=1$ case, we can straightforwardly see that $\beta_{S_j}$ will realize a three-rule structure as in \ceqn{three_rule_Phi}, resulting in a vanishing of the matrix elements in \ceqn{matrixelement_Hpair}. This is intuitively understood in the current framework, as the correlation facilitating the avoidance of on-site repulsion drives electrons of opposite spin apart, such that the local attractive interaction corresponding to this pairing channel cannot contribute. 
For $f_{\vec{k}}\neq \mathrm{const.}$, enabled by more complex forms of pair interactions, the identity no longer holds and states from this subspace generically couple non-trivially to $H_{\mathrm{pair}}$.

Our goal is to solve for the smallest eigenvalue $E$ in
\begin{align}
    \sum_{\mathcal{S}\pr} \bra*{\mathcal{S}} H_{\mathrm{pair}} \ket*{\mathcal{S}\pr} \, \psi_{\mathcal{S}\pr} = E \, \psi_{\mathcal{S}}.
\end{align}
To explore the separable nature of $H_{\mathrm{pair}}$ in this subspace, we set
\begin{align}
    \label{eq:bar_trafo}
    \bar{\psi}_{\mathcal{S}} = \Big(\prod_{j=1}^{N_C} e^{i \phi_{S_j}} \Big) \psi_{\mathcal{S}}, \quad e^{i \phi_{S_j}} = \frac{\beta_{S_j}}{\abs*{\beta_{S_j}}},
\end{align}
which is well defined as long as $\abs*{\beta_{S_j}} >0$ for all sets of orbitals $S_j$, such that
\begin{align}
    \label{eq:matrixelement_Hpairbar}
    \sum_{\mathcal{S}\pr} &\bra*{\mathcal{S}} \bar{H}_{\mathrm{pair}} \ket*{\mathcal{S}\pr} \, \bar{\psi}_{\mathcal{S}\pr} = E \, \bar{\psi}_{\mathcal{S}}, \nonumber \\
    \bra*{\mathcal{S}} &\bar{H}_{\mathrm{pair}} \ket*{\mathcal{S}\pr} = -g \sum_{i,j=1}^{N_C} \big|\beta_{S_i}\big| \big|\beta_{S\pr_{j}}\big| \delta_{\mathcal{S}\setminus S_i,\mathcal{S}\pr \setminus S\pr_j} \leq 0.
\end{align}

Since $\bra*{\mathcal{S}} \bar{H}_{\mathrm{pair}} \ket*{\mathcal{S}\pr} \leq 0$ for all $\mathcal{S},\mathcal{S}\pr$ \emph{and} $\bar{H}_{\mathrm{pair}}$ is irreducible in this subspace (cf. \capp{pairing_irreducibility}), the Perron-Frobenius theorem implies that the \emph{unique} ground state has to have the form
\begin{align}
    \label{eq:ground_state_pairing}
    \ket*{\Psi_0^{\mathrm{pair}}} = \sum_{\mathcal{S}} w_{\mathcal{S}} \, e^{-i \phi_{\mathcal{S}} } \, \ket*{\mathcal{S}},
\end{align}
for some set of weights $w_{\mathcal{S}} > 0$, but the angle $\phi_{\mathcal{S}} = \sum_{j=1}^{N_C} \phi_{S_j}$ is fixed via \ceqn{bar_trafo}.
The fact that the Perron-Frobenius theorem mandates $w_S > 0$ is in sharp contrast to the case of adding a dispersive term. While the latter stabilizes a ground state composed of a single, optimal product of entangled pair operators, the ground state wave function in the present scenario is necessarily spread across all possible configurations $\ket*{\mathcal{S}}$ spanning the subspace basis.

\begin{figure*}[t]
	\centering
	\includegraphics[width=1\linewidth]{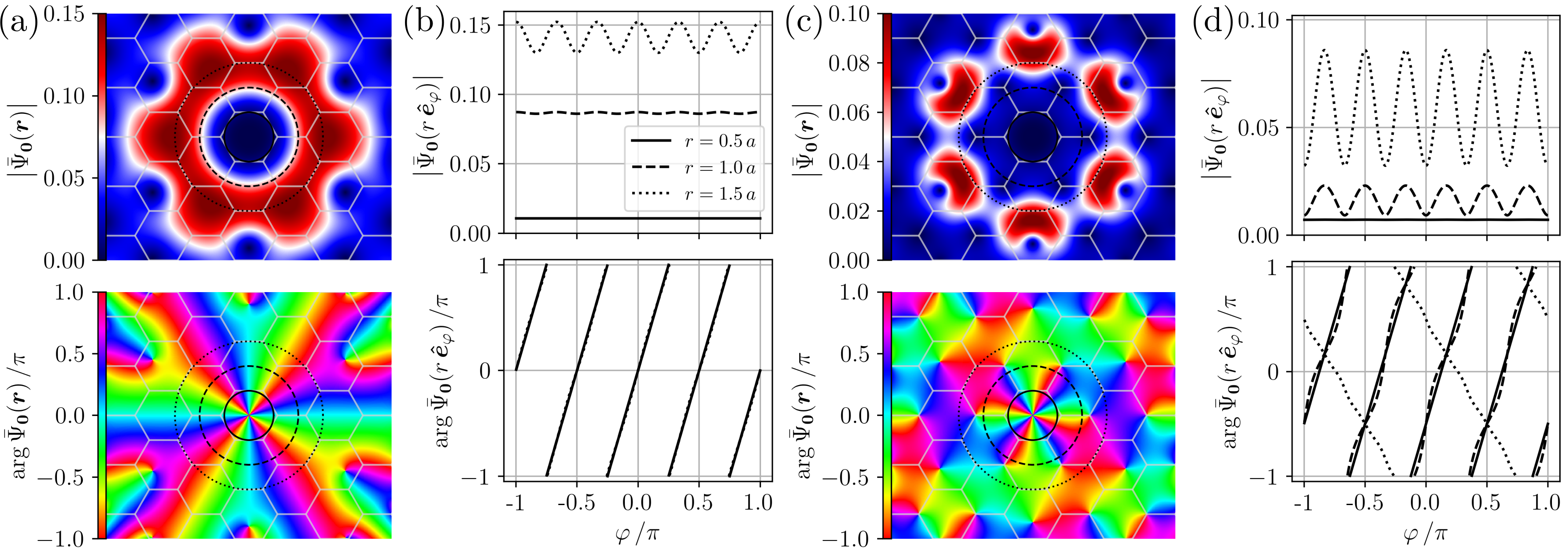}
	\caption{\textbf{Pair wave functions in the LLL and CTBG model}. Real-space plots of the macroscopic pair wave function $\bar{\Psi}_{\vec{0}}(\vec{r})$ from \ceqn{pair_wave_function} with $\vec{r}_0=\vec{0}$ in (a)--(b) the pure LLL setting $\psi_{\vec{k}}(\vec{r}) = \Phi_{\vec{k}}(\vec{r})$ and (c)--(d) the CTBG ideal flatband from \ceqn{psi_ideal_flatband} [with $\bar{\Psi}_{\vec{0}}(\vec{r}) = \bar{\Psi}_{\vec{0}}^{\mathrm{top},\mathrm{top}}(\vec{r})$]. The analysis of its magnitude and phase along circular contours of different radii in (b) and (d) indicates an approximate circular symmetry $\sim (x + i y)^4$ close to the origin. At larger $r$, $\abs*{\bar{\Psi}_{\vec{0}}(\vec{r})}$ is modulated and the contour encloses additional zeros of the wave function, resulting in the different winding number of $\arg{\bar{\Psi}_{\vec{0}}(r \, \vec{\hat{e}}_{\varphi})}$ at $r=1.5\, a$ (dotted line) in (d). The gray hexagons indicate the Wigner-Seitz cell of the triangular lattice.
 }
	\label{fig:pair_wave_function}
\end{figure*}

In order to further investigate the energetics analytically, we assume the weights $w_{\mathcal{S}}$ do not vary strongly with $\mathcal{S}$, such that the ground state is well approximated by $w_{\mathcal{S}}  \simeq w = 1/\sqrt{D_{\mathcal{S}}}$.
The pairing energy $E_{\mathrm{pair}} = \bra*{\Psi_0^{\mathrm{pair}}} H_{\mathrm{pair}}\ket*{\Psi_0^{\mathrm{pair}}}$ is then given by
\begin{align}
    \label{eq:pairing_energy}
     E_{\mathrm{pair}}=& - g w^2 \sum_{\mathcal{S},\mathcal{S}\pr} \sum_{i,j=1}^{N_C} \big|\beta_{S_i}\big| \big|\beta_{S\pr_{j}}\big| \delta_{\mathcal{S} \setminus S_i,\mathcal{S}\pr \setminus S\pr_j} \nonumber \\
    \simeq& \begin{aligned}[t]
        - g \Bigg(&\nu_{K}^2 \sum_{j=1}^{N_K} \big|\beta_{S_j}\big|^2 + \nu_{K} (1 - \nu_{K}) \Big(\sum_{j=1}^{N_K} \big|\beta_{S_j}\big| \Big)^2 \Bigg)
    \end{aligned}
\end{align}
where $\nu_{K} = N_C/N_K$ ($\nu_K=3 \,\nu$ if $N_K = N/6$) and the last equality becomes exact in the thermodynamic limit ($N_K \rightarrow \infty$, $\nu_K = \mathrm{const.}$, see \capp{pairing_energy} for details). Interestingly, the sum over pairs of electrons is replaced by a sum over all sets of three-rule orbitals, where the filling only enters as a prefactor. While the first (`diagonal') term strictly grows with filling $\nu_K$, the second (`off-diagonal') one has a maximum at $\nu_K=1/2$, but vanishes towards $\nu_K=0,1$. 
Provided that  $\nu_K \neq 0,1$, clearly the second term dominates for large enough systems as it scales with $N_K^2$. As a result, in order to minimize $E_{\mathrm{pair}}$ while vanishing (non-trivially) under $H_{\vec{0}}$, the choice of three-rule orbitals $S_j$ is optimized such that $\sum_j |\beta_{S_j}|$ is maximal. 

To demonstrate the superconducting nature of the ground state in \ceqn{ground_state_pairing}, we can investigate the correlator indicating off-diagonal long-range order (ODLRO)~\cite{penroseBoseEinsteinCondensationLiquid1956, penroseCXXXVIQuantumMechanics1951,yangConceptOffDiagonalLongRange1962} (cf. \capp{pair_wave_function})
\begin{align}
    \label{eq:odlro_correlator}\rho_{\vec{r}_0,\vec{r}_0\pr}^{\alpha,\beta,\alpha\pr,\beta\pr}(\vec{r},\vec{r}\pr) =& \bra*{\Psi_0^{\mathrm{pair}}} \hat{\rho}_{\vec{r}_0,\vec{r}_0\pr}^{\alpha,\beta,\alpha\pr,\beta\pr}(\vec{r},\vec{r}\pr) \ket*{\Psi_0^{\mathrm{pair}}}\nonumber \\
    =&\begin{aligned}[t]
        \frac{1}{\Omega^2} &\sum_{\vec{k},\vec{k}\pr \in \mathrm{BZ}} \zeta_{\vec{k},\alpha,\beta}\conj(\vec{r}_0,\vec{r}) \zeta_{\vec{k}\pr,\alpha\pr,\beta\pr}\pdag(\vec{r}_0\pr,\vec{r}\pr) \\
    &\bra*{\Psi_0^{\mathrm{pair}}} c\dag_{\vec{k},\downarrow} c\dag_{-\vec{k},\uparrow} c\pdag_{-\vec{k}\pr,\uparrow} c\pdag_{\vec{k}\pr,\downarrow} \ket*{\Psi_0^{\mathrm{pair}}},
    \end{aligned}    
\end{align}
where
\begin{align}
    \zeta_{\vec{k},\alpha,\beta}\pdag(\vec{r}_0,\vec{r}) =& \psi_{\vec{k},\alpha}\pdag(\vec{r}_0 + \vec{r}/2) \, \psi_{-\vec{k},\beta}\pdag(\vec{r}_0 - \vec{r}/2),
\end{align}
$\vec{r}_0^{(\prime)}$ is a reference position in the unit cell, $\vec{r}^{(\prime)}$ the relative spatial separation of the paired electrons, and $\psi_{\vec{k},\alpha}(\vec{r})$ the ideal flatband wave function from \ceqn{psi_ideal_flatband}. Evaluating the matrix elements
\begin{align}
    \label{eq:correlator_matrixelement}
    &\bra*{\mathcal{S}} \hat{\rho}_{\vec{r}_0,\vec{r}_0\pr}^{\alpha,\beta,\alpha\pr,\beta\pr}(\vec{r},\vec{r}\pr) \ket*{\mathcal{S}\pr} = \begin{aligned}[t]
    \frac{1}{\Omega^2}  &\sum_{i,j=1}^{N_C}\omega_{S_i;\alpha,\beta}\conj(\vec{r}_0,\vec{r}) \\        &\omega_{S\pr_j;\alpha\pr,\beta\pr}\pdag(\vec{r}_0\pr,\vec{r}\pr)
    \delta_{\mathcal{S} \setminus S_i, \mathcal{S}\pr \setminus S\pr_j},
    \end{aligned} \nonumber \\
    &\mathrm{with} \quad \omega_{S_j;\alpha,\beta}\pdag(\vec{r}_0,\vec{r}) = \frac{1}{\sqrt{2}}\sum_{l=1}^3 \sum_{\pm} \zeta_{\pm \vec{k}_l^j,\alpha,\beta}\pdag(\vec{r}_0,\vec{r})\,  \alpha_{S_j,\vec{0},l},
\end{align}
we find an identical structure to \ceqn{matrixelement_Hpairbar} (with $w_{\mathcal{S}}=w$), such that we can immediately rewrite the correlator in the thermodynamic limit as
\begin{align}
    \rho_{\vec{r}_0,\vec{r}_0\pr}^{\alpha,\beta,\alpha\pr,\beta\pr}(\vec{r},\vec{r}\pr) \simeq& \nu_K (1 - \nu_K) \bar{\Psi}_{\vec{r}_0}^{\alpha,\beta}(\vec{r})\conj \, \bar{\Psi}_{\vec{r}_0\pr}^{\alpha\pr,\beta\pr}(\vec{r}\pr) \nonumber \\
    &+ \mathcal{O}(1/{\Omega}).
\end{align}
Evidently, the correlation function separates into a pair wave function form with
\begin{widetext}
\begin{align}
    \label{eq:pair_wave_function}
    \bar{\Psi}_{\vec{r}_0}^{\alpha,\beta}(\vec{r}) 
    =& \frac{\mathcal{B}_{\alpha}(\vec{r}_0 + \vec{r}/2) \mathcal{B}_{\beta}(\vec{r}_0 - \vec{r}/2)}{\sqrt{2}\Omega} \sum_{j=1}^{N_K} e^{-i \phi_{S_j}} \mathcal{N}_{S_j,\vec{0}} \sum_{l,m,n=1}^3 \epsilon_{l m n} \Bigg[\sum_{\pm} \Phi_{\pm\vec{k}_l^j}(\vec{r}_0 + \vec{r}/2) \Phi_{\mp\vec{k}_l^j}(\vec{r}_0 - \vec{r}/2) \Bigg] \Phi_{\vec{k}_m^j + \vec{k}_n^j} \Phi_{\vec{k}_m^j - \vec{k}_n^j},
\end{align}
\end{widetext}
where the LLL wave functions $\Phi_{\vec{k}}(\vec{r})$ are $\mathcal{O}(1)$ quantities and both the total system area $\Omega$ as well as $N_K$ are proportional to $N$ such that for $N \rightarrow \infty$, $\bar{\Psi}_{\vec{r}_0}^{\alpha,\beta}(\vec{r})$ generically remains finite. 
Crucially, the finiteness of $\rho_{\vec{r}_0,\vec{r}_0\pr}^{\alpha,\beta,\alpha\pr,\beta\pr}(\vec{r},\vec{r}\pr)$ in the thermodynamic limit demonstrates the presence of ODLRO, which in turn implies the defining superconducting phenomenology---the Meissner effect~\cite{sewellOffdiagonalLongrangeOrder1990}, flux quantization~\cite{niehOffdiagonalLongrangeOrder1995}, as well as the Josephson effect and the existence of persistent currents~\cite{sewellOffdiagonalLongRange1997}; furthermore,
$\bar{\Psi}_{\vec{r}_0}^{\alpha,\beta}(\vec{r})$ is the macroscopic pair wave function in the superconducting phase.
This is in stark contrast to the variational state stabilized by a finite dispersion discussed in the previous section~\ref{sec:disperion}: in that case, only a single $\mathcal{S}$ contributes, which can be straightforwardly shown to only include $\mathcal{O}(1/{\Omega})$ terms in $\rho_{\vec{r}_0,\vec{r}_0\pr}^{\alpha,\beta,\alpha\pr,\beta\pr}(\vec{r},\vec{r}\pr)$ and hence cannot exhibit ODLRO.

We perform the optimization of three-rule orbitals $\{S_j\}_j$ numerically for the CTBG model via a depth-first-search on finite systems (see \capp{pairing_optimization} for details). Crucially, with $\tau_j$ denoting Pauli matrices in layer space, the symmetries of the system mandate $\braket*{\psi_{-\vec{k}}\conj} {\tau_{0,y} \, \psi_{\vec{k}}} = 0$, prohibiting pairing in these channels. Instead, we focus on symmetric inter-layer pairing $\tau=\tau_x$, which is non-zero in the CTBG model except for $\vec{k}=\vec{0}$. This suggests a minimal coupling form factor $f_{\vec{k}}=(k/\abs*{k})^2$, having $\abs*{f_{\vec{k}}}=1$ (similar to the vanishing $s$-wave case) for finite $\abs*{\vec{k}}$ \footnote{Note that we work in a single $\mathcal{C}=1$ band, such that time-reversal symmetry is already explicitly broken.}. With this choice, we find the system to favor a configuration of three-rule orbitals predominantly related by the effective sixfold rotational symmetry. 
Using this optimal configuration to construct the pairing Hamiltonian from \ceqn{matrixelement_Hpairbar}, we find that its finite-size ground state \ceqn{ground_state_pairing} is almost evenly spread across all configurations, with $w_S$ deviating from $w=1/\sqrt{D_{\mathcal{S}}}$ by only a few percent. This provides a level of corroboration to the approximation $w_{\mathcal{S}}  \simeq w$.

In \csubfig{pair_wave_function}{c}, we show $\bar{\Psi}_{\vec{0}}^{\mathrm{top},\mathrm{top}}(\vec{r})$ for the CTBG flatband wave function alongside the pure LLL form (essentially setting $\mathcal{B}_{\alpha}(\vec{r})=1,\mathcal{N}_{\vec{k}}=1$) in \csubfig{pair_wave_function}{a} for a pairing-optimized set of orbitals $\{S_j\}_{j=1}^{N_K}$. 
In both cases, the pair wave function goes to zero for $\vec{r} \rightarrow \vec{r}_0=\vec{0}$, which is a manifestation of the on-site avoidance due to the three-rule inherent to the construction. It also explicitly demonstrates that our variational procedure captures the expected energetics of strong on-site repulsion in the resulting superconducting ground state. The phases of the pair wave function [lower panels in \csubfig{pair_wave_function}{a,c}] further allow us to determine the symmetry of the underlying superconducting order parameter. For CTBG, the combination of the physical three-fold rotational symmetry $C_{3}$ and the emergent, intra-valley inversion symmetry $I$ in the chiral limit~\cite{Wang2021a}, with representation $I=\tau_y$, lead to the point group $S_6$. Upon noting that $\bar{\Psi}_{\vec{0}}^{\alpha,\alpha}(\vec{r})$ picks up a non-trivial phase under $C_3$ and taking into account the additional sign change under $\alpha = \mathrm{top} \rightarrow \mathrm{bottom}$ coming from the prefactors $\mathcal{B}_\alpha(\vec{r})$ in \ceqn{pair_wave_function}, we conclude that the superconductor corresponds to a chiral state associated with the (complex) irreducible representation $E_g$ of $S_6$. Such a superconductor is commonly referred to as a ``chiral $d$-wave state''. For the LLL case, there is only a single component (no layer index) and the chiral $d$-wave nature can be directly inferred from the behavior at small $\abs*{\vec{r}}$ visible in \csubfig{pair_wave_function}{a}.

In fact, both for the LLL and the CTBG case, the behavior close to the origin is similarly circularly symmetric, 
with the wave function well described by $\bar{\Psi}_{\vec{0}}(\vec{r})\sim (x + i y)^4$ (up to a global phase shift). 
However, for larger relative separations the ideal flatband case is qualitatively different. The connected region of high weight in \csubfig{pair_wave_function}{a} is broken down into separate, relatively localized islands in \csubfig{pair_wave_function}{c}.
Furthermore, the presence of the $\mathcal{B}_{\alpha}(\vec{r}/2) \mathcal{B}_{\alpha}(-\vec{r}/2) = \pm \mathcal{G}(\vec{r}/2) \mathcal{G}(-\vec{r}/2)$ factors in \ceqn{pair_wave_function} introduces additional vortices in the form of unmovable zeros~\cite{Wang2021a}, such that $\bar{\Psi}_{\vec{0}}^{\mathrm{top},\mathrm{top}}(r \, \vec{\hat{e}}_{\varphi})$ in \csubfig{pair_wave_function}{d} winds by $-4 \pi$ along a circular contour with $r = 1.5\,a$ ($r=\abs*{\vec{r}}$ and $\vec{\hat{e}}_{\varphi}$ is a unit vector here) around the origin while the LLL pair wave function in \csubfig{pair_wave_function}{b} is still well captured by the small $\abs*{\vec{r}}$ behavior, winding by only $8 \pi$. For even larger $\abs*{\vec{r}}$, $\bar{\Psi}_{\vec{0}}(\vec{r})$ appears to decay (until half the linear extent of the torus).

We finally point out that the precise form of the resulting pairing state crucially depends on the choice of $f_{\vec{k}}$ and that also other pairing symmetries, such as extended $s$-wave, can likely be realized. While the expression for the resulting pair wave function in \ceqn{pair_wave_function} holds for arbitrary $f_{\vec{k}}$, which enter through the phases $\phi_{S_j}$, we leave the explicit study of other form factors to future works.

\section{Beyond the ideal limit} 
\label{sec:non_ideal}
In this section, we analyze the impact of interactions with finite range, discuss the relation of the three-rule principle to fractional Chern insulator states, and study the impact of going beyond ideal Chern bands.

\subsection{Finite screening length and fractional Chern insulators}
Considering the scenario of a finite-range interaction, for short screening lengths $\lambda$ compared to the lattice constant (i.e. the moir\'e scale for TBG), the Fourier transform of a Yukawa-type screened Coulomb interaction potential $V(\vec{q}) = (1 + \abs*{\vec{q}}^2 \lambda^2)^{-1/2}$ is well represented by the expansion
\begin{align}
    \label{eq:yukawa_series}
    V(\vec{q}) = 1 - \frac{\lambda^2}{2} \abs*{\vec{q}}^2 + \mathcal{O}(\lambda^4 \abs*{\vec{q}}^4) \simeq 1 + V_1(\vec{q}),
\end{align}
such that the interaction Hamiltonian may be approximated as
\begin{align}
    H_{\lambda} \simeq H + H_1^{\lambda},
\end{align}
with $H$ being the pure on-site interaction in \ceqn{H_Q_definition} and $H_1^{\lambda}$ incorporating the effect of $V_1(\vec{q}) =  - \lambda^2  \abs*{\vec{q}}^2 / 2$.
A flat electronic band with Chern number unity and ideal band geometry is a natural setting for the emergence of FCIs in the presence of a short-ranged density-density interaction potential. In the context of twisted bilayer graphene, these lattice generalizations of fractional quantum Hall states were found to be contenders for the spin- and valley-polarized ground state at certain filling fractions~\cite{Abouelkomsan2020,Repellin2020,Ledwith2020,Wilhelm2021,Xie2021}. 
For spin-polarized FCIs, $H$ is inactive due to the antisymmetry of the fermionic wave function. However, for the spinful system studied in this work, this is not the case and the ground state is required to minimize both contributions, $H$ as well as $H_1^{\lambda}$, simultaneously.
In fact, for small enough $\lambda$, $H$ will be dominant and $H_1^{\lambda}$ acts as a perturbation on the degenerate zero-energy manifold of $H$. In this sense, any singlet ground state should be expected to abide by the three-rule \ceqn{three_rule_Phi}. 

\begin{figure}[t]
	\centering
	\includegraphics[width=\columnwidth]{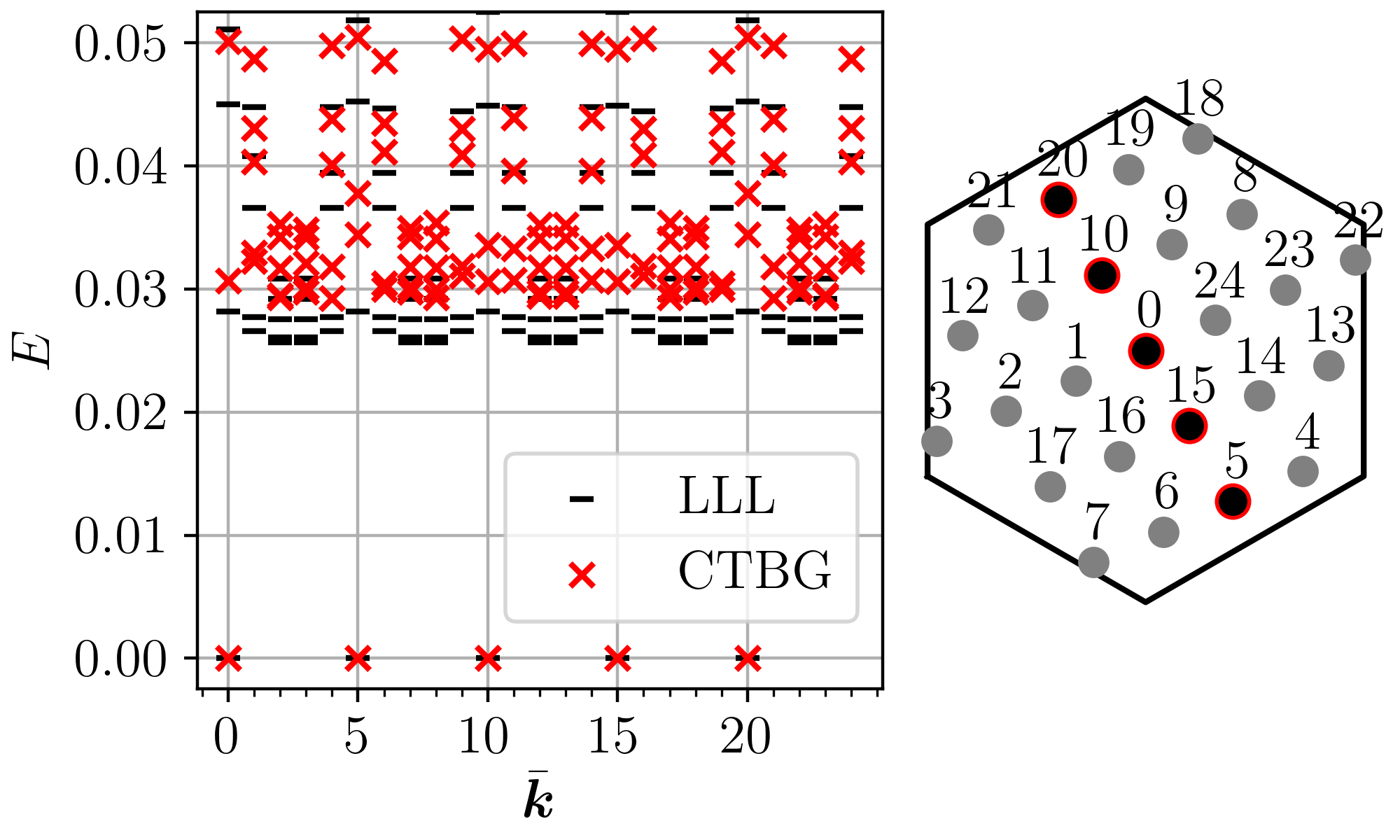}
	\caption{\textbf{Comparison of many-body spectra at $\boldsymbol{\nu=2/5}$.} Exact diagonalization spectrum of $H + H_1^{\lambda}$ at $\nu=2/5$ across all center of mass momenta $\bar{\vec{k}}$ in the singlet sector of a $N=25$ system. Both the LLL and CTBG models exhibit an exact five-fold zero-energy degeneracy, consistent with a Halperin $(3,3,2)$ spin-singlet FCI. We choose $\lambda = a/5$ such that the constant term remains the dominant energy scale in the expansion in \ceqn{yukawa_series}. The location of ground state momentum sectors in the Brillouin zone for this system is indicated on the right.} 
\label{fig:spectrum_singlet_nu_2_over_5}
\end{figure}

In the related setting of a LLL, there exists a well-established candidate for a singlet ground state wave function known as the Halperin $(m+1,m+1,m)$-state~\cite{Halperin1983}. It is located at a band filling of $\nu=2/(2\, m + 1)$ and for $m\geq 2$ it is an exact zero-mode of the pseudo-potential interactions $U_0(\vec{r}) = \delta(\vec{r})$ as well as $U_1(\vec{r}) = \frac{\lambda^2}{2} \delta^{\prime \prime}(\vec{r})$, whose Fourier transforms are just the lowest-order terms of \ceqn{yukawa_series}. 
Analogous to the correspondence of the spin polarized FCI at $\nu=1/3$ in the LLL and the CTBG model presented in \cref{Wang2021}, we perform ED simulations of $H + H_1^{\lambda}$ in the spin-singlet sector and compare it to the case of a LLL on the torus. Analyzing \cfig{spectrum_singlet_nu_2_over_5}, while the excitation energies are similar but different for the LLL and CTBG models, the total number and center of mass momentum sectors of zero-energy states are identical. As expected for the topological degeneracy of a Halperin (3,3,2)-state, we find five states with exactly zero energy in the spectrum of $H + H_1^{\lambda}$. The topological nature of the ground state degeneracy is further underlined by its insensitivity to the geometry of the finite-size cluster. 

\begin{figure}[b]
	\centering
	\includegraphics[width=\columnwidth]{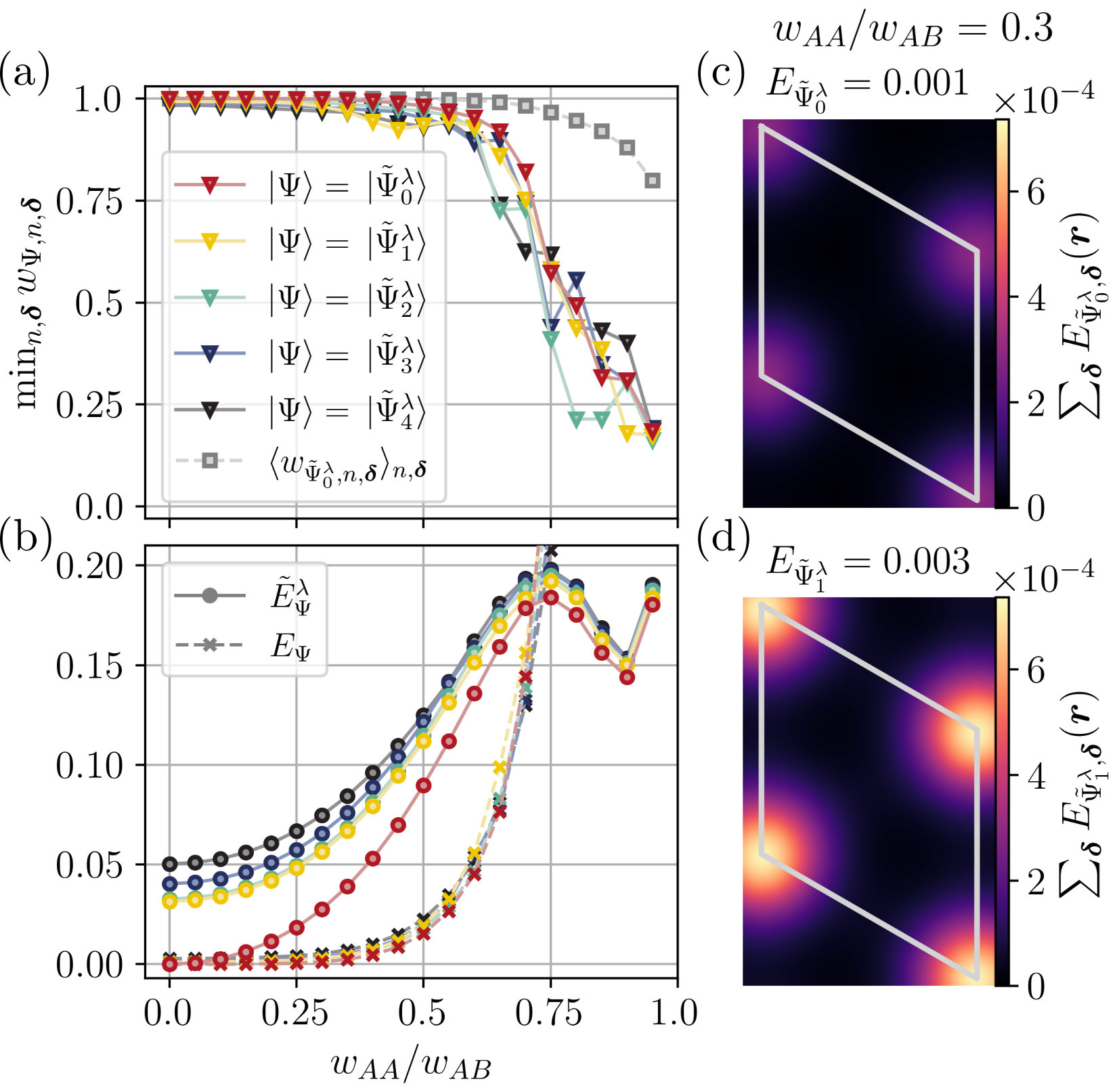}
	\caption{\textbf{Subspace weights and IPO-energy of the FCI spectrum away from the chiral limit.} (a) Minimum (triangles, solid lines) and average (squares, dashed line) relative weights in the three-rule subspaces $w_{\Psi,n,\vec{\delta}}$ from \ceqn{weight_three_rule} measured on the ground state and lowest excitations in momentum sector $\bar{\vec{k}}=\vec{0}$ of the TBG Hamiltonian $\tilde{H} + \tilde{H}_1^{\lambda}$ as a function of the interlayer tunneling ratio $w_{AA}/w_{AB}$. (b) Corresponding actual energies $\tilde{E}_{\Psi}^{\lambda}= \bra*{\Psi} \tilde{H} + \tilde{H}_1^{\lambda} \ket*{\Psi}$ (circles, solid lines) versus their energy under the IPO Hamiltonian $E_{\Psi} = \bra*{\Psi} H \ket*{\Psi}$ (crosses, dashed lines). Locally measured IPO energy of the (c) ground state and (d) first excited state at $w_{AA}/w_{AB}=0.3$. All measurements were performed in the singlet symmetry sector of a $N=20$ system. We again choose $\lambda = a/5$. The measurements of the average weight $\langle w_{\Psi,n,\vec{\delta}} \rangle_{n, \vec{\delta}}$ are only plotted for the ground state as they are very similar for all considered states. In (c) and (d) the unit cell is indicated in gray.} 
\label{fig:subspace_weight_energy_Q}
\end{figure}

Being a singlet with zero energy in an ideal flatband with dominantly on-site interactions, the ground state wave functions of $H + H_1^{\lambda}$ should be required to conform to the three-rule from \ceqn{three_rule_Phi}. We explicitly test this by measuring the relative three-rule subspace weights $w_{\Psi,n,\vec{\delta}}$ from \ceqn{weight_three_rule} on the lowest energy states obtained from ED. Taking a closer look at \csubfig{subspace_weight_energy_Q}{a} at $w_{AA}/w_{AB}=0$, we find that $w_{\Psi,n,\vec{\delta}}=1$ for all $n$ and $\vec{\delta}$ when measured on the FCI ground state $\ket*{\Psi_0^{\lambda}}$. For states with $E>0$, the relative weight may still be unity for some $n,\vec{\delta}$, but generically $\min_{n,\vec{\delta}} w_{\Psi,n,\vec{\delta}}<1$, e.g.~for the first excited state we find $\min_{n, \vec{\delta}} w_{\Psi,n,\vec{\delta}} \simeq 0.99$. Nevertheless, they may still be considered `quite aligned' with the three-rule such that, as visible in \csubfig{subspace_weight_energy_Q}{b} at $w_{AA}/w_{AB}=0$, they have a minimal energy penalty from $H$ alone.  

\subsection{Away from chiral limit}
In the context of the ideal flatband of TBG, there is the natural question of how relevant the physics in the chiral limit is to what happens in the more realistic model, where no analytic relation to the LLL is known. For the TBG continuum model considered here, we can allow for a finite intra-sublattice tunneling amplitude $w_{AA}$ in order to assess the robustness of our results. We denote by $\tilde{H}$ the band-projected on-site part of the interaction while $\tilde{H}_1^{\lambda}$ captures the effect of $V_1(\vec{q})$ for more general $w_{AA}$ (we denote the eigenvectors as $\ket*{\tilde{\Psi}_{m}^{\lambda}}$). Additionally, in order to study purely the effect of the altered band wave functions, we suppress the potentially subordinate dispersive term caused by a detuning from the chiral limit. 

Judging from \csubfig{subspace_weight_energy_Q}{a}, the three-rule subspace weights $w_{\Psi,n,\vec{\delta}}$ at $w_{AA}/w_{AB}>0$ of the low-energy eigenvectors appear to stay remarkably close to one until $w_{AA}/w_{AB} \simeq 0.5$ (note that we show the minimum across all subspaces), with their average 
\begin{align}
    \langle w_{\Psi,n,\vec{\delta}} \rangle_{n, \vec{\delta}} = \frac{\sum_{n, \vec{\delta}} \norm{\mathcal{P}_{n,\vec{\delta}} \ket*{\Psi}}^2}{\sum_{n, \vec{\delta}} \norm{\mathbb{1}_{n, \vec{\delta}} \ket*{\Psi}}^2}
\end{align}
falling below 0.9 only at $w_{AA}/w_{AB} > 0.8$ for all considered eigenstates. 
Correspondingly, their on-site energies $E_{\Psi}$ under the IPO-operator in \csubfig{subspace_weight_energy_Q}{b} remain minimal compared to their actual energies $\tilde{E}_{\Psi}^{\lambda}$ for the majority of this parameter range. This is intuitively understood in the current framework, as, according to \csubfig{subspace_weight_energy_Q}{a}, these states almost perfectly fulfill the three-rule condition from \ceqn{three_rule_Phi}. This provides a strong hint that not only the FCI ground state, but its lowest excitations likewise are built from the degenerate manifold in the limit $\lambda \rightarrow 0, w_{AA} \rightarrow 0$, i.e. the IPO Hamiltonian of \ceqn{H_Q_definition}. 
The deviation from the perfect satisfaction of the three-rule at finite $w_{AA}/w_{AB}$ also provides intuition on the ferromagnetic tendency at higher $w_{AA}/w_{AB}$ observed in e.g. the study of the TBG model at filling $\nu=1/3$ in \cref{Repellin2020}. Since fully spin-polarized states do not require a fine-tuning of coefficients to vanish under on-site interactions, they can be expected to be less sensitive to a detuning from the chiral limit compared to singlets.

In \csubfigrange{subspace_weight_energy_Q}{c}{d} we plot the local energy density $E_{\Psi}(\vec{r}) = \sum_{\vec{\delta} \in \mathrm{BZ}} E_{\Psi,\vec{\delta}}(\vec{r})$ of $H$ measured on the many-body ground and first excited states at $w_{AA}/w_{AB} = 0.3$. In both cases, the system appears to develop small, but finite signals in proximity to the lattice vertices. Not only is the energy density more pronounced in the excited state, but, in stark contrast to the measurement on the ground state, it stays finite in the chiral limit $w_{AA} \rightarrow 0$.
It is interesting to contrast the texture of $E_{\Psi}(\vec{r})$ for the first excitation above the Halperin FCI in \csubfig{subspace_weight_energy_Q}{d} with the local energy density of the first state above the highly degenerate zero-energy manifold of $H$ itself in \csubfig{local_energy_Q}{b}. In both cases the signal is centered on vertices of the underlying triangular lattice, however, the distinct rings in \csubfig{local_energy_Q}{b} are absent in \csubfig{subspace_weight_energy_Q}{d}. Combined with the orders-of-magnitude difference in amplitude, we conclude that they develop for different reasons. 
In \csubfig{local_energy_Q}{b} the manifold of zero-energy states of $H$ with vanishing total spin and center of mass momentum is fully exhausted by the first 552 states in the spectrum and the very definition of $\ket*{\Psi_{552}}$ having $E>0$ requires it to have a significant IPO-energy density. In contrast, $\ket*{\tilde{\Psi}_{1}^{\lambda}}$ in \csubfig{subspace_weight_energy_Q}{d} is not necessarily orthogonal, but, as suggested by \csubfig{subspace_weight_energy_Q}{a}, actually highly aligned to the kernel of $H$. In this sense, its finite signal in $E_{\Psi}(\vec{r})$ is caused by the wave function simultaneously minimizing for $H_{1}^{\lambda}$, inducing distinct, minor deviations from the perfect three-rule relation.

\section{Discussion and outlook} 
To conclude, we performed a thorough study of repulsive on-site interactions and its interplay with multiple perturbations in flat ideal Chern bands on the torus.
We reformulated the problem in terms of ideal pair operators, see \ceqn{H_Q_definition}, whose associated energy density uniquely dictates the energetics for on-site repulsion. This exposes an alternative ground state condition, which intuitively requires the energy density to vanish identically in every momentum channel. This condition is trivially satisfied by spin-polarized states, as they do not couple to the ideal pair operators in the first place. However, for multiplets with non-maximal total spin, in particular the singlet sector, this imposes a non-trivial structure on the many-body wave function. Combining analytical considerations with numerical techniques, we deduced a ``three-rule'' principle, see \cfig{three_rule_illustration}, for combining ideal flatband wave functions which is at the origin of these non-trivial zero modes. We devised the associated projector formulation [see \ceqn{projector_condition}] of the zero-energy condition, which we find to be composed as a sum of generally non-commuting projectors orthogonal to the subspaces spanned according to the three-rule. The relevance of these subspaces is substantiated by explicitly verifying the complete alignment of many-body wave functions from the zero-energy manifold obtained via exact diagonalization of the on-site interacting ideal flatband Hamiltonian of magic-angle twisted bilayer graphene. Furthermore, we numerically computed the dimension of the projector kernel and find that it exactly coincides with the degeneracy in the exact diagonalization spectrum. This suggests a connection of strongly interacting ideal flatbands to other Hamiltonians composed of non-commuting projectors like the AKLT and Rokhsar-Kivelson models~\cite{Affleck1987,Rokhsar1988,Castelnovo2005}.

We assess the effect of perturbations to the degenerate manifold analytically by working in a subspace that is exclusively composed of states that non-trivially satisfy the zero-energy condition for vanishing momentum transfer. Within this variational approach, a finite dispersion is found to stabilize a correlated metal, where the structure imposed by the three-rule causes a softening of the step in the occupation number; we further find the zero-energy manifold to be susceptible to an attractive singlet interaction -- ultimately causing the condensation of paired electrons. We explicitly computed the off-diagonal long-range order correlator comprising the macroscopic pair wave function of the superconductor. Crucially, we conclude the non-trivial zero-energy property of the investigated states to imply the avoidance of on-site pair interactions, while coupling to a wider class of singlet pairing Hamiltonians. In this sense, superconductivity is demonstrated to emerge as a cooperative phenomenon, where the dominant screened Coulomb interaction sets the stage in the form of a (almost) degenerate manifold that is susceptible to a subordinate attraction.

Finally, we investigated the universality of the developed principle to the situation of a finite-range Coulomb potential as well as flatband wave functions tuned away from ideality. Specifically, we examined the low-energy singlet spectrum of the short-range interacting TBG Hamiltonian at conduction band filling $\nu=2/5$. Most notably, we establish the perfect compliance of the topologically ordered spin-singlet ground state wave function with the three-rule principle and the lowest excitations to respect it to a remarkable degree -- a feature that is completely absent in the spin polarized case. Correspondingly, their energy densities under the ideal pair operators are minimal, suggesting the low-energy spectrum to be built from states directly evolving from the originally degenerate zero-energy manifold of the on-site interacting Hamiltonian. What is more, we find this principle to remain surprisingly well satisfied upon tuning the TBG model away from the chiral limit, where no direct analogy to the LLL is known.

In the future, we hope that the developed perspective on singlet ground states in fractionally filled ideal flatbands can also be used for the construction of ground state wave functions of other types of short-range interaction Hamiltonians. 
Notably, similar to TBG, chirally stacked twisted multi-layer graphene systems like twisted mono-bilayer or double-bilayer graphene have been found to possess flat electronic bands near charge neutrality, with a plethora of correlated phases emerging at certain filling fractions~\cite{Rubio2022,Polshyn2022,Wilhelm2023}. Although, these bands have Chern number $\mathcal{C}=2$ as opposed to $\mathcal{C}=1$ for TBG, an ideal flatband with Chern number $\mathcal{C}$ can essentially be decomposed into $\mathcal{C}$ (potentially non-orthogonal) ideal flatbands with $\mathcal{C}=1$~\cite{Wang2023,Dong2023}. 
Consequentially, the arguments presented in this work should be readily applicable to a more general class of flatband systems -- possibly enriched by an additional emergent flavor degree of freedom. 
In the context of TBG, a naturally interesting future research direction would be to examine the role of the three-rule principle upon including both flatbands in both valleys. In this more general setting, there are two ideal flatbands per spin with Chern number $\mathcal{C}=\pm 1$, providing one additional degree of freedom in each Chern sector for the fulfilment of the three-rule -- ultimately extending the possibilities to reach a vanishing on-site energy contribution in the Coulomb interaction. This time-reversal symmetric setting could be especially important for superconductivity, where the more general non-trivial zero-modes can similarly be expected to be consistent with pairing instabilities. 

\section*{Acknowledgements}
This research was funded in whole or in part by the Austrian Science Fund (FWF), Grant DOI 10.55776/W1259. For open access purposes, the author has applied a CC BY public copyright license to any author accepted manuscript version arising from this submission. The computational results presented have been achieved in part using the Vienna Scientific Cluster (VSC). M.S.S.~acknowledges funding by the European Union (ERC-2021-STG, Project 101040651---SuperCorr). Views and opinions expressed are however those of the authors only and do not necessarily reflect those of the European Union or the European Research Council Executive Agency. Neither the European Union nor the granting authority can be held responsible for them. P. H. W. is grateful for valuable discussions with E. Starchl, S. Banerjee as well as J. A. Sobral. 

\setcounter{equation}{0}
\setcounter{figure}{0}
\setcounter{table}{0}
\makeatletter
\renewcommand{\theequation}{A\arabic{equation}}
\renewcommand{\thefigure}{A\arabic{figure}}

\appendixpageoff
\appendixtitleoff
\renewcommand{\appendixtocname}{Appendix}
\begin{appendices}

\section*{Appendix}

\subsection{Derivation of the three-rule for LLL wave functions}
\label{app:derivation_three_rule}

In the symmetric gauge, the LLL wave function takes the form
\begin{equation}
	\Phi_{\vec{k}}(\vec{r}) = \sigma(r + i k) e^{i k\conj r} e^{-\frac{1}{2}(\abs*{r}^2 + \abs*{k}^2)},
\end{equation}
with $\sigma(z) = \tilde{\sigma}(z) e^{-\bar{G}(\mathbb{A}) z^2 / 2}$ the modified form of the original Weierstrass sigma function $\tilde{\sigma}(z)$, $\bar{G}(\mathbb{A})$ being a modular independent complex number~\cite{Haldane2018} and $r = (\vec{r}_x + i \vec{r}_y)/\sqrt{2S}$ and $k = \sqrt{S/2}\, (\vec{k}_x + i \vec{k}_y)$ where $2 \pi S = \norm{\vec{a}_1 \times \vec{a}_2}$ and $\vec{a}_{1,2}$ the lattice vectors. We use the following normalization for $\Phi_{\vec{k}}(\vec{r})$: 
\begin{align}
    \int_{\Omega} d^2\vec{r} \, \Phi_{\vec{k}}\conj(\vec{r}) \Phi_{\vec{k}\pr}\pdag(\vec{r}) = \Omega \,\delta_{\vec{k}, \vec{k}\pr}.
\end{align}

Combining multiple LLL wave functions like
\begin{align}
	&\Phi_{\vec{k}_1 + \vec{\delta}}(\vec{r}) \Phi_{-\vec{k}_1 + \vec{\delta}}(\vec{r}) \Phi_{\vec{k}_2+\vec{k}_3}\Phi_{\vec{k}_2-\vec{k}_3} = \Big[e^{2 i \delta\conj z}\nonumber\\
	&e^{\tilde{G}(\mathbb{A}) (k_1^2 + k_2^2 + k_3^2 - z^2 + (\delta - i z)^2}  e^{-(\abs*{k_1}^2+\abs*{k_2}^2 +\abs*{k_3}^2+\abs*{z}^2 + \abs*{\delta}^2)} \nonumber \Big]\\
	& \tilde{\sigma}(r + i k_1 + i \delta) \tilde{\sigma}(r - i k_1 + i \delta) \tilde{\sigma}(i k_2 + i k_3) \tilde{\sigma}(i k_2 - i k_3),
\end{align} 
we can see that the prefactor in square brackets is invariant under permutation of the three momentum orbitals, e.g. $(\vec{k}_1, \vec{k}_2, \vec{k}_3) \rightarrow ( \vec{k}_3,\vec{k}_1, \vec{k}_2) \rightarrow (  \vec{k}_2,\vec{k}_3,\vec{k}_1)$, with $\Phi_{\vec{k}} = \Phi_{\vec{k}}(\vec{0})$ as in the main text. With $\Big[ e^{...} ... \Big]$ denoting the common prefactor, a sum of such permutations naturally leads to

\begin{align}
	\label{eq:derivation_LLL}
	&\Phi_{\vec{k}_1 + \vec{\delta}}(\vec{r}) \Phi_{-\vec{k}_1 + \vec{\delta}}(\vec{r}) \Phi_{\vec{k}_2+\vec{k}_3} \Phi_{\vec{k}_2 -\vec{k}_3} \nonumber \\
	+\,&\Phi_{\vec{k}_2 + \vec{\delta}}(\vec{r}) \Phi_{-\vec{k}_2 + \vec{\delta}}(\vec{r}) \Phi_{\vec{k}_3+\vec{k}_1} \Phi_{\vec{k}_3 -\vec{k}_1} \nonumber\\
	+\,&\Phi_{\vec{k}_3 + \vec{\delta}}(\vec{r}) \Phi_{-\vec{k}_3 + \vec{\delta}}(\vec{r}) \Phi_{\vec{k}_1+\vec{k}_2} \Phi_{\vec{k}_1 -\vec{k}_2} = \Big[ e^{...} ... \Big]\nonumber \\
	\Big\{\, & \tilde{\sigma}(r + i k_1 + i \delta) \tilde{\sigma}(r - i k_1 + i \delta) \tilde{\sigma}(i k_2 + i k_3) \tilde{\sigma}(i k_2 - i k_3) \nonumber \\
	+ \, &\tilde{\sigma}(r + i k_2 + i \delta) \tilde{\sigma}(r - i k_2 + i \delta) \tilde{\sigma}(i k_3 + i k_1) \tilde{\sigma}(i k_3 - i k_1) \nonumber \\
	+ \, &\tilde{\sigma}(r + i k_3 + i \delta) \tilde{\sigma}(r - i k_3 + i \delta) \tilde{\sigma}(i k_1 + i k_2) \tilde{\sigma}(i k_1 - i k_2)	\Big\}.
\end{align}

Defining $u = r + i \delta, v= i k_1, x = i k_2, y= i k_3$ and using $\tilde{\sigma}(z)=-\tilde{\sigma}(-z)$, the second part of \ceqn{derivation_LLL} exactly takes the form of a combination of Weierstrass sigma functions, for which an important identity was derived in \cref{Lawden1989}:
\begin{align}
	&\tilde{\sigma}(u+v)\tilde{\sigma}(u-v)\tilde{\sigma}(x+y)\tilde{\sigma}(x-y) \nonumber \\
	+\,& \tilde{\sigma}(v+x) \tilde{\sigma}(v-x) \tilde{\sigma}(u+y) \tilde{\sigma}(u-y) \nonumber \\
	+\,& \tilde{\sigma}(x+u) \tilde{\sigma}(x-u) \tilde{\sigma}(v+y) \tilde{\sigma}(v-y) = 0 \quad \forall u,v,x,y \in \mathbb{C}.
\end{align}
This directly implies that the curly braces in \ceqn{derivation_LLL} are zero, thus proving the three-rule \ceqn{three_rule_Phi}.

Furthermore, by choosing $u = i k_1, v= i k_2, x = i k_3, y= i k_4$ and rewinding the arguments above, we straightforwardly arrive at an alternative identity for the coefficients of the wave functions, i.e. purely in terms of $\Phi_{\vec{k}}$:
\begin{align}
    \label{eq:three_rule_pure_momentum_space}
    &\Phi_{\vec{k}_1 + \vec{k}_2}\Phi_{\vec{k}_1 - \vec{k}_2} \Phi_{\vec{k}_3+\vec{k}_4} \Phi_{\vec{k}_3 -\vec{k}_4} \nonumber \\
    +\,&\Phi_{\vec{k}_2 + \vec{k}_3} \Phi_{\vec{k}_2 - \vec{k}_3} \Phi_{\vec{k}_1+\vec{k}_4} \Phi_{\vec{k}_1 -\vec{k}_4} \nonumber\\
    +\,&\Phi_{\vec{k}_3 + \vec{k}_1} \Phi_{\vec{k}_3 - \vec{k}_1} \Phi_{\vec{k}_2+\vec{k}_4} \Phi_{\vec{k}_2 -\vec{k}_4} = 0.
\end{align}

Analogously, there exists a pure real-space relation with $u = r_1, v= r_2, x = r_3, y= r_4$ and $\Phi(\vec{r}) = \Phi_{\vec{0}}(\vec{r})$:
\begin{align}
    &\Phi(\vec{r}_1 + \vec{r}_2)\Phi(\vec{r}_1 - \vec{r}_2) \Phi(\vec{r}_3+\vec{r}_4) \Phi(\vec{r}_3 -\vec{r}_4) \nonumber \\
    +\,&\Phi(\vec{r}_2 + \vec{r}_3) \Phi(\vec{r}_2 - \vec{r}_3) \Phi(\vec{r}_1+\vec{r}_4) \Phi(\vec{r}_1 -\vec{r}_4) \nonumber\\
    +\,&\Phi(\vec{r}_3 + \vec{r}_1) \Phi(\vec{r}_3 - \vec{r}_1) \Phi(\vec{r}_2+\vec{r}_4) \Phi(\vec{r}_2 -\vec{r}_4) = 0,
\end{align}
which might be relevant to the fulfillment of the Fock cyclic condition~\cite{Prange1990} guaranteeing the singlet nature of the Halperin $(m+1,m+1,m)$-state on the torus.

\subsection{Entanglement of single-pair three-rule states}
\label{app:entanglement_single_pair}
We can straightforwardly quantify the entanglement of single-pair states produced by the three-rule operators $\ket*{ \tilde{\gamma}_{S_j,\vec{\delta}}} =  \tilde{\gamma}\dag_{S_j,\vec{\delta}} \ket*{0}$ and $\ket*{ \gamma_{S_j,\vec{\delta}}} =  \gamma\dag_{S_j,\vec{\delta}} \ket*{0}$, with
\begin{align}
    \tilde{\gamma}\dag_{S_j,\vec{\delta}} =& \sum_{l=1}^3 \,  \alpha_{S_j,\vec{\delta},l}  \, c\dag_{\vec{k}_l^j + \vec{\delta},\uparrow} c\dag_{-\vec{k}_l^j + \vec{\delta},\downarrow}
\end{align}
and $\gamma\dag_{S_j,\vec{\delta}}$ defined as in the main text in \ceqn{definition_gamma}, by constructing their respective one-body reduced density matrices $\rho_{ \tilde{\gamma}/ \gamma_{S_j,\vec{\delta}}}$. Computing the von Neumann entropy $\mathscr{S}_{\tilde{\gamma}/ \gamma_{S_j,\vec{\delta}}}$ yields
\begin{align}
    \label{eq:entropy_single_pair}
    \mathscr{S}_{ \tilde{\gamma}/ \gamma_{S_j,\vec{\delta}}} =& - \tr(\rho_{\tilde{\gamma}/ \gamma_{S_j,\vec{\delta}}} \log \rho_{\tilde{\gamma}/ \gamma_{S_j,\vec{\delta}}}) \nonumber \\
    =& \log s_{ \tilde{\gamma}/ \gamma} + \sum_{l=1}^3 \abs*{ \alpha_{S_j, \vec{\delta}, l}}^2 \log \frac{1}{\abs*{ \alpha_{S_j, \vec{\delta}, l}}^2}.
\end{align}
The first part is just the entropy generated by the entanglement of a pure two-body Slater-determinant $\log s_{ \tilde{\gamma}}=\log 2$, or that of an ordinary singlet $\log s_{ \gamma}=\log 4$. On the other hand, the second part originates from the three-rule principle. Since $0 < \abs*{ \alpha_{S_j, \vec{\delta}, l}}^2 < 1$ for valid choices of $S_j$, the second term in \ceqn{entropy_single_pair} is always positive and hence $\mathscr{S}_{ \tilde{\gamma}/ \gamma_{S_j,\vec{\delta}}}$ necessarily exceeds $\log s_{ \tilde{\gamma}/ \gamma}$.
In fact, the entanglement is maximized by choosing $S_j$ such that all $\abs*{ \alpha_{S_j, \vec{\delta}, l}}^2$ are equal, which for $\vec{\delta}=\vec{0}$ can be reached if all $\vec{k}_l^j \in S_j$ are related by point-group symmetry operations.

\subsection{Three-rule projector construction}
\label{app:projector_construction}
We consider the $Q\pdag_{\vec{\delta}}(\vec{r})$-scattered state $\ket*{\Xi_{\vec{\delta}}(\vec{r})}= Q\pdag_{\vec{\delta}}(\vec{r}) \ket*{\Psi}$ in the ordinary Fock-space basis of $N_C$ pairs of electrons with vanishing spin polarization and center of mass momentum $\bar{\vec{k}} - 2 \vec{\delta}$. Then  $\ket*{\Xi_{\vec{\delta}}(\vec{r})} = \sum_{n}\Xi_{n,\vec{\delta}}(\vec{r})\ket*{n,\vec{\delta}}$ with $\ket*{n,\vec{\delta}} = \prod_{\vec{k}_i \in S_n^{\uparrow}}\prod_{\vec{k}_j \in S_n^{\downarrow}}  c\dag_{\vec{k}_i,\uparrow} c\dag_{\vec{k}_j,\downarrow} \ket*{0}$, $\sum_{l \in S_n^{\uparrow,\downarrow}} \vec{k}_{l}=\bar{\vec{k}} - 2 \vec{\delta}$ and $\abs*{S_n^{\uparrow}}=\abs*{S_n^{\downarrow}}=N_C-1$. The relevant subspace for $\Xi_{n,\vec{\delta}}(\vec{r})$ is defined as $\mathcal{W}_{n,\vec{\delta}} = \mathrm{span}\{\ket*{n,\vec{\delta},\vec{k}}\}_{\vec{k}}$, with $\ket*{n,\vec{\delta},\vec{k}} = c\dag_{\vec{k} + \vec{\delta},\uparrow} c\dag_{-\vec{k} + \vec{\delta},\downarrow} \ket*{n,\vec{\delta}}$, $\forall \vec{k} \in \mathrm{BZ}$.
Now, for every basis element $\ket*{n,\vec{\delta}}$ we construct all states of finite norm corresponding to the possible combinations of three momentum orbitals $S_j$
\begin{align}
    \label{eq:many_body_three_rule_state}
     \ket*{n,\vec{\delta};S_j} =  \tilde{\gamma}\dag_{S_j,\vec{\delta}} \, \pi_{S_j} \, \ket*{n,\vec{\delta}},
\end{align}
with $\pi_{S_j} = \prod_{\vec{k} \in S_j} (1 - n_{\vec{k} + \vec{\delta},\uparrow})(1 - n_{-\vec{k}+ \vec{\delta},\downarrow})$ ensuring that no orbitals appearing in $ \tilde{\gamma}\dag_{S_j,\vec{\delta}}$ are occupied in $\ket*{n,\vec{\delta}}$.
The basis $\{\ket*{n,\vec{\delta};S_j}\}_j$ naturally incorporates the three-rule structure, however, it is overcomplete as combinations of different $S_j$ and $\ket*{n,\vec{\delta}}$ may produce a finite overlap. This can be easily remedied via a Gram-Schmidt procedure, yielding an orthonormal basis $\{\ket*{n,\vec{\delta};i}\}_i$ which spans the identical subspace $\mathcal{V}_{n,\vec{\delta}}$ as $\{\ket*{n,\vec{\delta};S_j}\}_j$.
The intentional choice of $ \tilde{\gamma}\dag_{S_j,\vec{\delta}}$ instead of $ \gamma\dag_{S_j,\vec{\delta}}$ in \ceqn{many_body_three_rule_state} is grounded in the more general notion of singlets in a genuine many-body problem, where the spin structure implied by $ \gamma\dag_{S_j,\vec{\delta}}$ would be too restrictive. As a result, $\mathcal{V}_{n,\vec{\delta}}$ also contains states of multiplets with finite total spin.
By constructing the projector $\mathbb{1}_{n,\vec{\delta}}$ on $\mathcal{W}_{n,\vec{\delta}}$, which simply has ones on the diagonal at every $\ket*{n,\vec{\delta},\vec{k}}$, as well as the projector $\mathcal{P}_{n,\vec{\delta}}$ on the zero-subspace $\mathcal{V}_{n,\vec{\delta}}$ as $\mathcal{P}_{n,\vec{\delta}} = \sum_i \ket*{n,\vec{\delta};i} \bra*{n,\vec{\delta};i}$, we can directly assess the significance of the three-rule property. The individual degeneracies of multiplets split according to their total spin $S$ may be analyzed by simply adding a $S^2$ term to the total orthogonal projector $\bar{\mathcal{P}}$. 

\subsection{Supplementary information for a finite dispersion}
\label{app:derivation_dispersion}

For a spin-independent dispersion we may write
\begin{align}
    H_{\mathrm{kin}} = \sum_{\vec{k} \in BZ} \epsilon_{\vec{k}} n_{\vec{k}},
\end{align}
with $n_{\vec{k}} = \sum_{\sigma=\updownarrows} c\dag_{\vec{k},\sigma}c\pdag_{\vec{k},\sigma}$. Then 
\begin{align}
    [H_{\mathrm{kin}}, \chi\dag_{\vec{k},-\vec{k}}] = \sum_{\vec{k}\pr} \epsilon_{\vec{k}\pr} [n_{\vec{k}\pr}, \chi\dag_{\vec{k},-\vec{k}}] = (\epsilon_{\vec{k}} + \epsilon_{-\vec{k}}) \chi\dag_{\vec{k},-\vec{k}}
\end{align}
and consequently
\begin{align}
    [H_{\mathrm{kin}},  \gamma\dag_{S_j,\vec{0}}] = \frac{1}{\sqrt{2}} \sum_{l=1}^3 (\epsilon_{\vec{k}_l^j} + \epsilon_{-\vec{k}_l^j})  \alpha_{S_j, \vec{0},l} \chi\dag_{\vec{k}_l^j,-\vec{k}_l^j}.
\end{align}
With this we find for the action on a many-body basis state $\ket*{\mathcal{S}}$
\begin{align}
    H_{\mathrm{kin}} \ket*{\mathcal{S}}=& H_{\mathrm{kin}} \prod_{j=1}^{N_c}  \gamma\dag_{S_j,\vec{0}} \ket*{0} \nonumber\\
    =& \Big( \gamma\dag_{S_1,\vec{0}} H_{\mathrm{kin}} + [H_{\mathrm{kin}},  \gamma\dag_{S_1,\vec{0}} ]\Big) \prod_{j=2}^{N_c}  \gamma\dag_{S_j,\vec{0}} \ket*{0} \nonumber\\
    =& \Big(\prod_{j=1}^{N_c}  \gamma\dag_{S_j,\vec{0}} \underbrace{H_{\mathrm{kin}} \ket*{0}}_{=0} + \sum_{j=1}^{N_c} [H_{\mathrm{kin}},  \gamma\dag_{S_j,\vec{0}}] \underbrace{\prod_{i\neq j}  \gamma\dag_{S_i,\vec{0}} \ket*{0}}_{\ket*{\mathcal{S} \setminus S_j}}\Big) \nonumber\\
    =& \frac{1}{\sqrt{2}} \sum_{j=1}^{N_c} \sum_{l=1}^3 (\epsilon_{\vec{k}_l^j} + \epsilon_{-\vec{k}_l^j})  \alpha_{S_j, \vec{0},l} \chi\dag_{\vec{k}_l^j,-\vec{k}_l^j} \ket*{\mathcal{S} \setminus S_j},
\end{align}
where from the second to third line we used that $[[H_{\mathrm{kin}},  \gamma\dag_{S_j,\vec{0}}], \gamma\dag_{S_i,\vec{0}}]=0$. The respective matrix element now becomes 
\begin{align}
    \bra*{\mathcal{S}} H_{\mathrm{kin}} \ket*{\mathcal{S}\pr} =& \frac{1}{\sqrt{2}} \sum_{j=1}^{N_c} \sum_{l=1}^3 
    \begin{aligned}[t]
        &(\epsilon_{\vec{k}_l^j} + \epsilon_{-\vec{k}_l^j})  \alpha_{S_j, \vec{0},l} \\&\bra*{\mathcal{S}} \chi\dag_{\vec{k}_l^j,-\vec{k}_l^j} \ket*{\mathcal{S}\pr \setminus S\pr_j} 
    \end{aligned}\nonumber \\
    =& \begin{aligned}[t]
        \frac{1}{2} &\sum_{i,j=1}^{N_c} \delta_{\mathcal{S} \setminus S_i, \mathcal{S}\pr \setminus S\pr_j} \sum_{m,l=1}^3 (\epsilon_{\vec{k}_l^j} + \epsilon_{-\vec{k}_l^j}) \\
        & \alpha_{S_i, \vec{0},m}\conj  \alpha_{S_j\pr, \vec{0},l} \underbrace{\bra*{0} \chi\pdag_{\vec{k}_m^i,-\vec{k}_m^i} \chi\dag_{\vec{k}_l^j,-\vec{k}_l^j} \ket*{0}}_{2 \delta_{S_i,S_j\pr} \delta_{l,m}}
    \end{aligned}\nonumber \\
    =& \delta_{\mathcal{S},\mathcal{S}\pr} \sum_{j=1}^{N_c}\sum_{l=1}^3 (\epsilon_{\vec{k}_l^j} + \epsilon_{-\vec{k}_l^j}) \abs*{ \alpha_{S_j, \vec{0},l}}^2,
\end{align}
using again that $S_i$ and $S_j$ contain no common orbitals for $i \neq j$.

\subsection{Supplementary information for finite singlet pairing}
\label{app:derivation_pairing}

\subsubsection{Form of the pairing Hamiltonian}
\label{app:eta_pairing}

Similar to the $\eta$-pairing form~\cite{Yang1989} made from the electron annihilation operators $d_{\alpha,\sigma}(\vec{r})$ with spin $\sigma=\updownarrows$ on e.g. layer $\alpha$ 
\begin{align}
    \tilde{\eta} = \int_{\Omega} d^2\vec{r} \, \sum_{\alpha,\beta} (\tau)_{\alpha,\beta} \, d_{\alpha,\downarrow}(\vec{r}) d_{\beta,\uparrow}(\vec{r})
\end{align}
we study a Hamiltonian of the form
\begin{align}
    H_{\mathrm{pair}} =& -g \tilde{\eta}\dag \tilde{\eta}.
\end{align}
In the single-band-projected approximation we use
\begin{align}
    \label{eq:annihilation_operator_band_projection}
    d_{\alpha,\sigma}(\vec{r}) \simeq \frac{1}{\sqrt{\Omega}} \sum_{\vec{k} \in \mathrm{BZ}} \psi_{\vec{k},\alpha}(\vec{r}) c_{\vec{k},\sigma},
\end{align}
such that we can rewrite
\begin{align}
     \tilde{\eta} \simeq \sum_{\vec{k},\vec{k}\pr} \Big[\frac{1}{\Omega} \int_{\Omega} d^2\vec{r} \, \sum_{\alpha,\beta} (\tau)_{\alpha, \beta} \, \psi_{\vec{k},\alpha}(\vec{r}) \psi_{\vec{k}\pr,\beta}(\vec{r}) \Big] c_{\vec{k},\downarrow} c_{\vec{k}\pr,\uparrow}.
\end{align}
Now, using $\psi_{\vec{k},\alpha}(\vec{r}) = e^{i \vec{k} \cdot \vec{r}} u_{\vec{k},\alpha}(\vec{r})$ where $u_{\vec{k},\alpha}(\vec{r} + \vec{a})=u_{\vec{k},\alpha}(\vec{r})$ is the $\vec{a}$-lattice periodic part of the Bloch wave functions, the expression in square brackets becomes
\begin{align}
    &\sum_{\vec{a}}  e^{i (\vec{k} + \vec{k}\pr) \cdot \vec{a}}\frac{1}{\Omega}\int_{\mathrm{UC}} d^2\vec{r} \, \sum_{\alpha,\beta} (\tau)_{\alpha,\beta} \psi_{\vec{k},\alpha}(\vec{r}) \psi_{\vec{k}\pr,\beta}(\vec{r}) = \nonumber \\
    & = \delta_{\vec{k},-\vec{k}\pr} \braket*{\psi_{-\vec{k}}\conj}{\tau \psi_{\vec{k}}\pdag},
\end{align}
where $\mathrm{UC}$ denotes the unit cell and 
\begin{align}
    \braket*{\psi_{-\vec{k}}\conj}{\tau \psi_{\vec{k}}\pdag} =& \frac{1}{\Omega} \int_{\mathrm{\Omega}} d^2\vec{r} \,  \sum_{\alpha,\beta} (\tau)_{\alpha,\beta}\psi_{\vec{k},\alpha}(\vec{r}) \psi_{-\vec{k},\beta}(\vec{r}) \nonumber \\
    =&\frac{\mathcal{N}_{-\vec{k}} \mathcal{N}_{\vec{k}}}{\Omega} \int_{\Omega} d^2\vec{r} \, \tilde{\mathcal{F}}_{\tau}(\vec{r}) \Phi_{-\vec{k}}(\vec{r}) \Phi_{\vec{k}}(\vec{r}).
\end{align}
Here
\begin{align}
    \tilde{\mathcal{F}}_{\tau}(\vec{r}) = \sum_{\alpha,\beta} (\tau)_{\alpha,\beta} \mathcal{B}_{\alpha}(\vec{r}) \mathcal{B}_{\beta}(\vec{r})
\end{align}
is different from $\mathcal{F}(\vec{r})$ defined in the main text, but still only depends on $\vec{r}$ and not $\vec{k}$. 
Considering specific options for the structure of the layer-coupling-matrix $\tau$ in the CTBG model, for the simplest case $\tau=\tau_0=\mathbb{1}$ we get
\begin{align}
    \tilde{\mathcal{F}}_{\tau_0}(\vec{r}) = -\mathcal{G}^2(\vec{r}) + \mathcal{G}^2(-\vec{r}),
\end{align}
which is clearly odd under $\vec{r} \rightarrow -\vec{r}$ and since $\Phi_{-\vec{k}}(\vec{r}) \Phi_{\vec{k}}(\vec{r})$ is even under inversion, $\braket*{\psi_{-\vec{k}}\conj}{\tau_0 \psi_{\vec{k}}\pdag}=0$ for all $\vec{k}$. Furthermore, choosing $\tau=\tau_y$ leads to
\begin{align}
    \tilde{\mathcal{F}}_{\tau_y}(\vec{r}) = \mathcal{G}(\vec{r}) \mathcal{G}(-\vec{r}) - \mathcal{G}(-\vec{r}) \mathcal{G}(\vec{r}) = 0.
\end{align}
Hence, only the $\tau_{x,z}$ components can produce a finite coupling on this level. Writing
\begin{align}
    \eta = \frac{1}{\Omega} \int_{\Omega} dr \, \tilde{\mathcal{F}}_{\tau}(\vec{r}) \sum_{\vec{k}} f_{\vec{k}} \, \mathcal{N}_{- \vec{k}} \mathcal{N}_{\vec{k}} \Phi_{-\vec{k}}(\vec{r}) \Phi_{\vec{k}}(\vec{r}) \chi_{-\vec{k},\vec{k}}\pdag.
\end{align}
we introduced an even function $f_{\vec{k}}$ in order to absorb a factor of $1/2$ stemming from the singlet pair operators $\chi_{-\vec{k},\vec{k}}\pdag$ but, more importantly, to implement modulation on the superlattice length scale ($f_{\vec{k}}=1$ would reproduce the original form). 
As we can see, the form inside the integral in $\eta$ is identical to $Q_{\vec{0}}\pdag(\vec{r})$ if $f_{\vec{k}}=1$ and thus vanishes if applied to the IPO zero-energy states. However, $f_{\vec{k}} \neq 1$ can lead to finite pairing.

\subsubsection{Pairing matrix elements}
Regarding the singlet pairing Hamiltonian, we first evaluate the commutator
\begin{align}
    [\eta, \chi\dag_{\vec{k},-\vec{k}}] =& \sum_{\vec{k}\pr} f_{\vec{k}\pr} \braket*{\psi_{-\vec{k}\pr}\conj}{\tau \psi_{\vec{k}\pr}\pdag}\,[\chi\pdag_{\vec{k}\pr,-\vec{k}\pr}, \chi\dag_{\vec{k},-\vec{k}}] \nonumber \\
    =& 2f_{\vec{k}}\braket*{\psi_{-\vec{k}}\conj}{\tau \psi_{\vec{k}}\pdag} \Big(2 - \sum_{\pm} n_{\pm\vec{k}}\Big),
\end{align}
from which \ceqn{commutator_xi_Gamma} straightforwardly follows since, without loss of generality, $f_{\vec{k}}$ is taken to be even. Applying $\eta$ to a state $\ket*{\mathcal{S}}$ then gives
\begin{align}
    \eta \ket*{\mathcal{S}} =& \eta \prod_{j=1}^{N_c}  \gamma\dag_{S_j,\vec{0}} \ket*{0} \nonumber\\
    =& \Big( \gamma\dag_{S_1,\vec{0}} \eta + [\eta,  \gamma\dag_{S_1,\vec{0}} ]\Big) \prod_{j=2}^{N_c}  \gamma\dag_{S_j,\vec{0}} \ket*{0} \nonumber\\
    =& \Big(\prod_{j=1}^{N_c}  \gamma\dag_{S_j,\vec{0}} \underbrace{\eta \ket*{0}}_{=0} + \sum_{j=1}^{N_c} \beta_{S_j} \underbrace{\prod_{i\neq j}  \gamma\dag_{S_i,\vec{0}} \ket*{0}}_{\ket*{\mathcal{S} \setminus S_j}}\Big) \nonumber\\
    =& \sum_{j=1}^{N_c} \beta_{S_j} \ket*{\mathcal{S} \setminus S_j},
\end{align}
where from the second to third line we used that $n_{\vec{k}_l^j} \prod_{i \neq j}  \gamma\dag_{S_i,\vec{0}} \ket*{0}=0$ for $\vec{k}_l^j \in \pm S_j$ and $\beta_{S_j}$ is defined as in the main text. The matrix element in \ceqn{matrixelement_Hpair} now follows simply from the Hermitian structure of $H_{\mathrm{pair}}$ and $\bra*{\mathcal{S}} \eta\dag = (\eta \ket*{\mathcal{S}})\dag$.

\subsubsection{Pairing energy}
\label{app:pairing_energy}
We now derive the pairing energy for the ground state $\ket*{\Psi_0^{\mathrm{pair}}}$ of \ceqn{ground_state_pairing}. It is given by 
\begin{align}
    \label{eq:pairing_energy_appendix}
    E_{\mathrm{pair}}=& \bra*{\Psi_0^{\mathrm{pair}}} H_{\mathrm{pair}} \ket*{\Psi_0^{\mathrm{pair}}} \nonumber\\
    =& - g w^2 \sum_{\mathcal{S},\mathcal{S}\pr} \sum_{i,j=1}^{N_C} \big|\beta_{S_i}\big| \big|\beta_{S\pr_{j}}\big| \delta_{\mathcal{S} \setminus S_i,\mathcal{S}\pr \setminus S\pr_j},
\end{align}
where we again assume $w_{\mathcal{S}} \simeq w$.
The values of matrix elements $\big|\beta_{S_i}\big| \big|\beta_{S\pr_{j}}\big|$ do not depend on the full basis configurations $\mathcal{S}$ and $\mathcal{S}\pr$, but only the subsets $S_i, S\pr_j$, such that contributions from different $\mathcal{S},\mathcal{S}\pr$ are simply added up an integer number of times. We split \ceqn{pairing_energy_appendix} into diagonal and off-diagonal contributions
\begin{align}
    E_{\mathrm{pair}}=& \begin{aligned}[t]
        - g w^2 \Bigg(&\sum_{\mathcal{S}} \sum_{S_i \in \mathcal{S}} \big|\beta_{S_i}\big|^2 + \\
        &\sum_{\mathcal{S} \neq \mathcal{S}\pr} \sum_{S_i \neq S\pr_j} \big|\beta_{S_i}\big| \big|\beta_{S\pr_{j}}\big| \delta_{\mathcal{S} \setminus S_i,\mathcal{S}\pr \setminus S\pr_j} \Bigg)
    \end{aligned}
\end{align}
and find that each diagonal term $\big|\beta_{S_i}\big|^2$ appears exactly $M_{\mathrm{D}} = {{N_K-1} \choose {N_C -1}}$ times while the off-diagonal ones do so $M_{\mathrm{OD}} = {{N_K-2} \choose {N_C -1}}$ times. Hence, we may rewrite the pairing energy as
\begin{align}
    E_{\mathrm{pair}}=& - g w^2 \Bigg(M_{\mathrm{D}} \sum_{j=1}^{N_K} \big|\beta_{S_j}\big|^2 + M_{\mathrm{OD}} \sum_{i \neq j=1}^{N_K} \big|\beta_{S_i}\big| \big|\beta_{S_{j}}\big|\Bigg) \nonumber \\
    =&\begin{aligned}[t]
        - g w^2 \Bigg((M_{\mathrm{D}} - M_{\mathrm{OD}}) &\sum_{j=1}^{N_K} \big|\beta_{S_j}\big|^2 + \\
        &M_{\mathrm{OD}} \Big(\sum_{j=1}^{N_K} \big|\beta_{S_j}\big|\Big)^2 \Bigg).
    \end{aligned}
\end{align}
With $w^2 = 1/D_{\mathcal{S}}$ and $D_{\mathcal{S}} = {N_K \choose N_C}$, it is straightforward to show that $w^2 M_{\mathrm{D}} = N_C/N_K = \nu_K$ and 
\begin{align}
    w^2 M_{\mathrm{OD}} &= \frac{N_C}{N_K} \frac{N_K - N_C}{N_K - 1} =  \frac{\nu_K ( 1 - \nu_K)}{1 - 1/N_K} \nonumber \\
    &\underset{N_K \rightarrow \infty}{\longrightarrow} \nu_K ( 1 - \nu_K).
\end{align}
Consequently, in the thermodynamic limit ($N_K \sim N \rightarrow \infty$, $\nu_K = \mathrm{const.}$) the pairing energy becomes
\begin{align}
    E_{\mathrm{pair}}= \begin{aligned}[t]
        - g \Bigg(&\nu_{K}^2 \sum_{j=1}^{N_K} \big|\beta_{S_j}\big|^2 + \nu_{K} (1 - \nu_{K}) \Big(\sum_{j=1}^{N_K} \big|\beta_{S_j}\big| \Big)^2 \Bigg).
    \end{aligned}
\end{align}

\subsubsection{Off-diagonal long-range order and pair wave function}
\label{app:pair_wave_function}

\begin{figure}[t]
	\centering
	\includegraphics[width=\columnwidth]{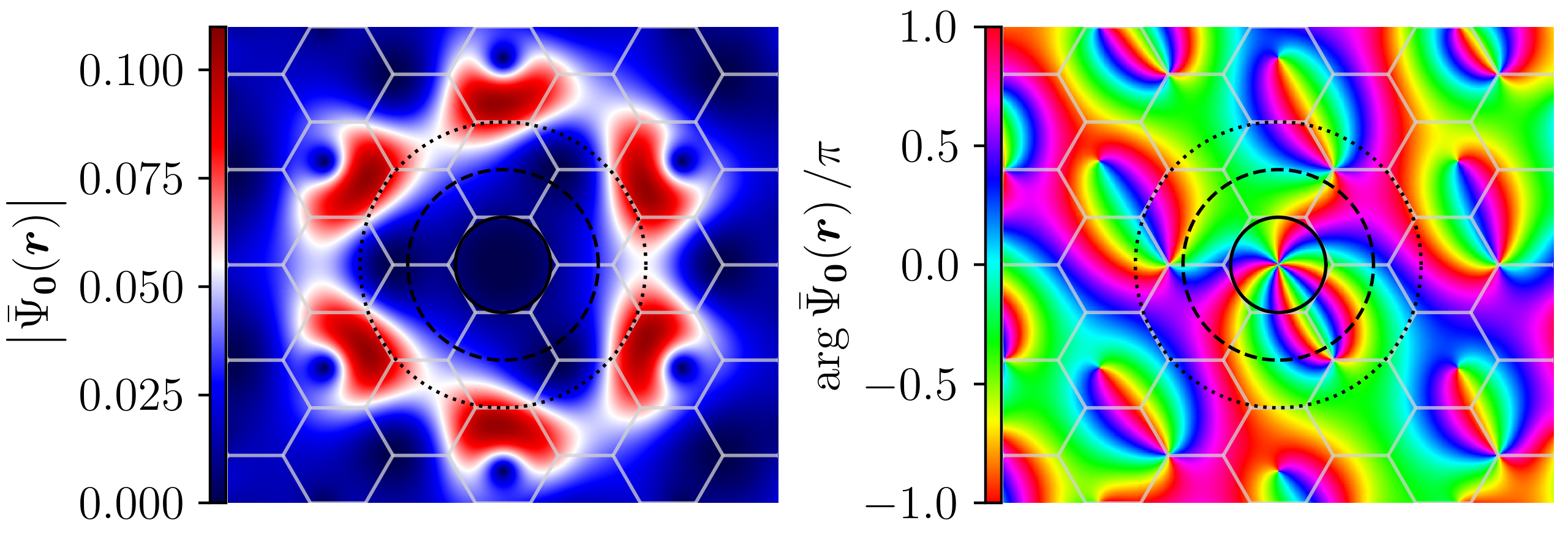}
	\caption{\textbf{Layer-off-diagonal pair wave function of CTBG.} Magnitude and phase texture of the pair wave function $\bar{\Psi}_{\vec{0}}(\vec{r})=\bar{\Psi}_{\vec{0}}^{\mathrm{top},\mathrm{bottom}}(\vec{r})$ for the identical setting to \cfig{pair_wave_function}. In contrast to the diagonal component $\bar{\Psi}_{\vec{0}}^{\mathrm{top},\mathrm{top}}(\vec{r})$, $\bar{\Psi}_{\vec{0}}^{\mathrm{top},\mathrm{bottom}}(\vec{r})$ is not an even function of $\vec{r}$.}\label{fig:pair_wave_function_CTBG_off_diagonal}
\end{figure}

To demonstrate that off-diagonal long-range order is present, we show that
\begin{align}
    \label{eq:odlro_correlator_appendix}
    \rho(\vec{r},\vec{r}') = \lim_{|\vec{R}-\vec{R}'| \rightarrow \infty} \langle &d\dag_{\downarrow}(\vec{R}+\vec{r}/2) d\dag_{\uparrow}(\vec{R}-\vec{r}/2) \nonumber\\
    &d\pdag_{\uparrow}(\vec{R}'-\vec{r}'/2) d\pdag_{\downarrow}(\vec{R}'+\vec{r}'/2) \rangle
\end{align}
is finite in the thermodynamic limit. Here $\langle \cdot \rangle = \langle\Psi_0^{\mathrm{pair}} | \cdot | \Psi_0^{\mathrm{pair}}\rangle$ and the sublattice or layer degree of freedom is suppressed for notational simplicity but put back in later. 
Inserting again the single-band-projection from \ceqn{annihilation_operator_band_projection} into \ceqn{odlro_correlator_appendix} we get
\begin{widetext}
\begin{align}
    \rho(\vec{r},\vec{r}') \simeq \frac{1}{\Omega^2} \sum_{\vec{k}_1,\vec{k}_2,\vec{k}_1\pr,\vec{k}_2\pr} &\langle c\dag_{\vec{k}_1,\downarrow} c\dag_{\vec{k}_2,\uparrow} c\pdag_{\vec{k}_2\pr,\uparrow} c\pdag_{\vec{k}_1\pr,\downarrow}\rangle e^{i (\vec{k}_1\pr - \vec{k}_2\pr) \frac{\vec{r}\pr}{2}}  e^{-i (\vec{k}_1 - \vec{k}_2) \frac{\vec{r}}{2}} \lim_{|\vec{R}-\vec{R}'| \rightarrow \infty} e^{i (\vec{k}_1\pr + \vec{k}_2\pr) \vec{R}\pr}  e^{-i (\vec{k}_1 + \vec{k}_2) \vec{R}} \nonumber \\
    & u_{\vec{k}_1}\conj(\vec{R} + \vec{r}/2) u_{\vec{k}_2}\conj(\vec{R} - \vec{r}/2) u_{\vec{k}_2\pr}\pdag(\vec{R}\pr - \vec{r}\pr/2) _{\vec{k}_1\pr}\pdag(\vec{R}\pr + \vec{r}\pr/2),
\end{align}
\end{widetext}
where, upon sending $|\vec{R}-\vec{R}'| \rightarrow \infty$, because of the $e^{i (\vec{k}_1\pr + \vec{k}_2\pr) \vec{R}\pr}  e^{-i (\vec{k}_1 + \vec{k}_2) \vec{R}}$ factors only the $\vec{k}_2 = - \vec{k}_1$ and $\vec{k}_2\pr = - \vec{k}_1\pr$ contributions survive. To properly take the limit $\vec{R} \rightarrow \infty$ (and similarly for $\vec{R}')$, we set $\vec{R} = \vec{a} + \vec{r}_0$ and send the Bravais lattice vector $\vec{a}$ to infinity. The correlator in this limit (including internal indices $\alpha,\beta,\alpha\pr,\beta\pr$ such as layer) is just \ceqn{odlro_correlator}.
By identical arguments to the derivation of the pairing energy in the previous section, the ground state expectation value for large enough systems can be recast into
\begin{widetext}
\begin{align}
    &\rho_{\vec{r}_0,\vec{r}_0\pr}^{\alpha,\beta,\alpha\pr,\beta\pr}(\vec{r},\vec{r}\pr) = \bra*{\Psi_0^{\mathrm{pair}}} \hat{\rho}_{\vec{r}_0,\vec{r}_0\pr}^{\alpha,\beta,\alpha\pr,\beta\pr}(\vec{r},\vec{r}\pr) \ket*{\Psi_0^{\mathrm{pair}}} \nonumber \\
    & \quad \simeq
    \frac{\nu_K^2}{\Omega^2}\sum_{j=1}^{N_K}\omega_{S_j;\alpha,\beta}\conj(\vec{r}_0,\vec{r})\omega_{S_j;\alpha\pr,\beta\pr}\pdag(\vec{r}_0\pr,\vec{r}\pr) +
    \nu_K (1 - \nu_K) \Bigg(\frac{1}{\Omega} \sum_{i=1}^{N_K} e^{i \phi_{S_i}} \omega_{S_i;\alpha,\beta}\conj(\vec{r}_0,\vec{r}) \Bigg)
    \Bigg(\frac{1}{\Omega} \sum_{j=1}^{N_K} e^{- i \phi_{S_j}}\omega_{S_j;\alpha\pr,\beta\pr}\pdag(\vec{r}_0\pr,\vec{r}\pr) \Bigg).
\end{align}
\end{widetext}
With each $\omega_{S_j;\alpha,\beta}\pdag(\vec{r}_0,\vec{r})$ being an $\mathcal{O}(1)$ quantity (except for zeros of the flatband wave functions), the first term vanishes as $1/\Omega$ for $\Omega \sim N \rightarrow \infty$. On the contrary, the second part remains finite and takes the form 
\begin{align}
\rho_{\vec{r}_0,\vec{r}_0\pr}^{\alpha,\beta,\alpha\pr,\beta\pr}(\vec{r},\vec{r}\pr) \simeq \nu_K (1 - \nu_K) \bar{\Psi}_{\vec{r}_0}^{\alpha,\beta}(\vec{r})\conj \, \bar{\Psi}_{\vec{r}_0\pr}^{\alpha\pr,\beta\pr}(\vec{r}\pr)
\end{align}
with the macroscopic pair wave function
\begin{align}
    \bar{\Psi}_{\vec{r}_0}^{\alpha,\beta}(\vec{r}) = \frac{1}{\Omega} \sum_{j=1}^{N_K} e^{-i \phi_{S_j}} \omega_{S_j;\alpha,\beta}(\vec{r}_0,\vec{r}),
\end{align}
written out in \ceqn{pair_wave_function}. The symmetries of the wave function become most apparent when we set $\vec{r}_0$ to high-symmetry points, e.g. $\vec{r}_0 = \vec{0}$. The layer-diagonal component in the CTBG model is shown in \cfig{pair_wave_function} while \cfig{pair_wave_function_CTBG_off_diagonal} contains a layer-off-diagonal one. For $\vec{r}_0 = \vec{0}$, layer conjugate components are related by a phase shift of $\pi$ in the diagonal case and an inversion operation in the off-diagonal one. As such, both the layer-diagonal and layer-off-diagonal component of $\bar{\Psi}_{\vec{r}_0}^{\alpha,\beta}(\vec{r})$ are even under $I=\tau_y$, which is necessary for them to transform under the same irreducible representation. This, in turn, is required for the state to be reachable by a single second order thermal phase transition from the normal state.

\subsubsection{Irreducibility of the pairing matrix}
 \label{app:pairing_irreducibility}
All elements of the pairing matrix in \ceqn{matrixelement_Hpairbar} are smaller or equal to zero, such that the Perron-Frobenius theorem for non-negative square matrices is applicable for $A = -\bar{H}_{\mathrm{pair}}$ and the Perron vector with maximal eigenvalue under $A$ corresponds to the ground state of $\bar{H}_{\mathrm{pair}}$. However, in order for the eigenvalues of $A$ to be strictly positive and the Perron vector to be unique, $A$ must be irreducible. 
Following \cref{Meyer2023}, $A$ is irreducible iff its corresponding directed graph $G(A)$ is strongly connected. The graph $G(A)$ has nodes $M_1, ..., M_{D}$ with a directed edge leading from $M_i$ to $M_j$ iff the matrix element $A_{ij}>0$. $G(A)$ is then said to be strongly connected, if for any pair of nodes $(M_i, M_j)$ there is a sequence of directed edges leading from $M_i$ to $M_j$. 

For the present case $D=D_{\mathcal{S}}$ and
\begin{align}
    A_{ij} = -\bra*{\mathcal{S}_i} \bar{H}_{\mathrm{pair}} \ket*{\mathcal{S}_j} = g \sum_{m,n=1}^{N_C} \big|\beta_{S_m}\big| \big|\beta_{S\pr_{n}}\big| \delta_{\mathcal{S} \setminus S_m,\mathcal{S}\pr \setminus S\pr_n},
\end{align}
such that $A_{ij}>0$ if $\mathcal{S}_i$ and $\mathcal{S}_j$ differ by exactly one element $S_m$ (for $i \neq j$). In this sense, every $M_i$, corresponding to the set $\mathcal{S}_i = \{S_{i_1},..., S_{i_{N_C}} \}$, has a directed edge to all nodes $M_j$ whose corresponding configuration $\mathcal{S}_j = \{S_{j_1},..., S_{j_{N_C}} \}$ can be reached from $\mathcal{S}_i$ by swapping a single $S_{i_m}$ for a $S_{j_n}$. Hence, starting from $\mathcal{S}_i$, one may always reach $\mathcal{S}_j$ from $\mathcal{S}_i$ in a maximum of $N_C$ steps by successively exchanging $S_{i_m} \in \mathcal{S}_i$ by a $S_{j_n} \in \mathcal{S}_j$ that's not already contained. Consequently, $G(A)$ is strongly connected and $A$ irreducible. 

\subsubsection{Pairing energy optimization procedure}
\label{app:pairing_optimization}

Based on the criterion of maximizing $\sum_{j=1}^{N_K} |\beta_{S_j}|$ in order to minimize the dominant contribution in \ceqn{pairing_energy}, we perform a depth-first search for an optimal configuration of non-overlapping sets of three-rule orbitals $S_j$. 
For a given finite set of available momentum orbitals $K$, e.g.~the discretized Brillouin zone, we start off by filtering out a set $K_+ = \{\vec{k}_i\}_i$, where no contained momenta are related by inversion. Next, we generate all possible, potentially overlapping sets $\tilde{S}_i= \{\vec{k}_1^i, \vec{k}_2^i, \vec{k}_3^i\}$ 
of three momenta $\vec{k}_j^i \in K_+$ and compute their corresponding $|\beta_{\tilde{S}_i}|$. Sorted by their value of $|\beta_{\tilde{S}_i}|$ in descending order, we iteratively build combinations of $N_K$ \emph{non-overlapping} sets $\{\tilde{S}_{i_1},...,\tilde{S}_{i_{N_K}}\}$. The set with maximal $\sum_{j=1}^{N_K} |\beta_{\tilde{S}_{i_j}}|$ then is the optimal pairing configuration and we set $S_j = \tilde{S}_{i_j}$. For large systems, it is useful to introduce a cutoff in the number of iterated configurations. The precautionary sorting according to $|\beta_{\tilde{S}_i}|$ ensures this introduces a minimal to no error in the optimal configuration.

\begin{figure}[t]
	\centering
	\includegraphics[width=\columnwidth]{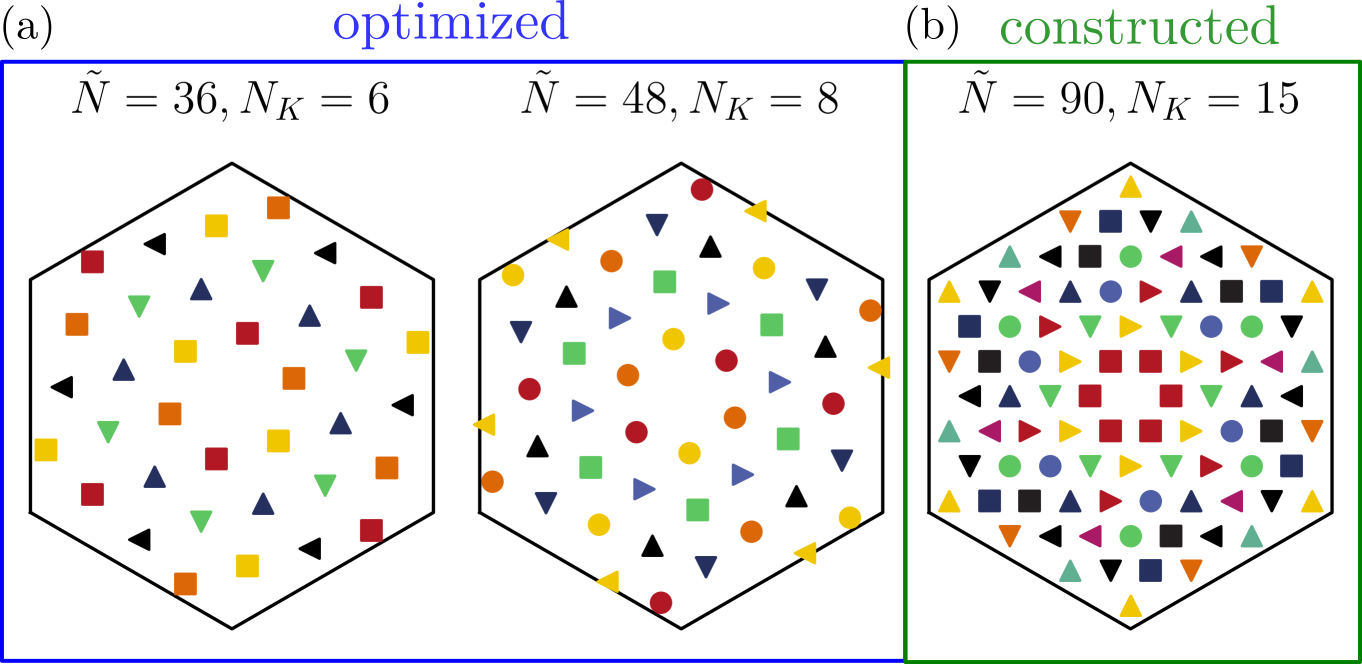}
	\caption{\textbf{Optimal pairing configuration of three-rule orbitals.} (a) Representative configurations of three-rule orbitals optimized for the dominant part in the pairing energy of \ceqn{pairing_energy}. The combination of colors and symbols indicates which set $S_j$ (or rather $K_j$, since $\beta_{S_j}$ is invariant under inversion of momenta) each momentum orbital belongs to. Except for three sets involving the orbitals closest to the origin, the system appears to favor grouping $C_6$-related momentum orbitals. (b) We make use of this tendency to construct $\{S_j\}_{j=1}^{N_K}$ on larger lattices for plotting the pair wave function in \cfig{pair_wave_function} and \cfig{pair_wave_function_CTBG_off_diagonal}. The optimization value $\sum_{j=1}^{N_K} |\beta_{S_j}|$ for such purely $C_6$-related three-rule sets is found to deviate only by about one percent for the systems in (a), with a tendency to become more exact with increasing system size. Likewise, the respective pair wave functions look almost identical.}
	\label{fig:pairing_configuration}
\end{figure}

In order to avoid the artificial breaking of rotational symmetries when combined with the non-overlapping three-rule, originating from an incommensurate number of related orbitals, we exclude orbitals at the origin and boundary of the Brillouin zone from the optimization procedure. Since the number of such excluded orbitals does not scale with the area of the system, their contribution is vanishingly small in the thermodynamic limit. The number of effective orbitals is denoted by $\tilde{N}$. 
For \csubfig{pairing_configuration}{a}, we optimize across all configurations for $\tilde{N}=36$, but introduced a cutoff of $10^8$ iterated, non-overlapping combinations starting once from every $\tilde{S}_{i}$ for $\tilde{N}=48$. In order to enable the preservation of rotational symmetries in the CTBG model, we choose $C_6$-symmetric simulation clusters spanned by $\vec{T}_1 = 2 \vec{a}_1 - 7 \vec{a}_2$ and $\vec{T}_2 = 5 \vec{a}_1 + 2 \vec{a}_2$ with $N=39$ ($\tilde{N}=36$) as well as $\vec{T}_1 = 3 \vec{a}_1 + 5 \vec{a}_2$ and $\vec{T}_2 = 8 \vec{a}_1 - 3 \vec{a}_2$ with $N=49$ ($\tilde{N}=48$) in \csubfig{pairing_configuration}{a}. Figure~\ref{fig:pairing_configuration}~(b) shows the configuration of $S_j$ underlying the pair wave function calculations displayed in \cfig{pair_wave_function} on a momentum-grid associated to the simulation cell with $\vec{T}_1 = 10 \vec{a}_1$ and $\vec{T}_2 = 10 \vec{a}_2$ with $N=100$ ($\tilde{N}=90$). The same cluster is used in \cfig{dispersion_configuration_occupation}.
The optimal configuration is not independent of the pairing Hamiltonian, but, depends on both the choice of $f_{\vec{k}}$ as well as the pairing form of the internal degrees of freedom captured by $\tau$.

\subsection{Interacting continuum and ideal flatband models of TBG}
\label{app:continuum_model}

The single particle model for the TBG flatband of a single valley is based on the continuum model formulation presented in \cref{Koshino2018}. 
We choose identical parameters to \cref{Wang2021}, with the graphene Fermi velocity $\hbar v_F/a_0 = 3.915 \, \mathrm{eV}$  ($a_0 = 1.42\,\text{\normalfont\AA}$) and $w_{AB}=110 \, \mathrm{meV}$, the first magic angle is located at $\theta \simeq 1.13^{\circ}$, where the chiral model with $w_{AA}=0$ has exactly flat bands. Additionally, we introduce a finite sublattice-staggered potential $\Delta_{AB} = 10\,\mathrm{meV}$, which can originate from an aligned hBN substrate and energetically separates the two degenerate bands at charge neutrality. 
The unit cell of the triangular moir\'e lattice is spanned by $\vec{a}_1 = a \, (0,-1)^T$ and $\vec{a}_2 = \frac{a}{2} \, (\sqrt{3},-1)^T$, where we use units of length in terms of the moir\'e lattice constant $a$, such that $\Omega = N \frac{\sqrt{3}}{2}$ for a system with $N$ unit cells.
By numerically diagonalizing the corresponding single-particle Hamiltonian, we may obtain the Bloch eigenvectors $\ket*{u_{\vec{k}}}$ of the conduction band at arbitrary $w_{AA}/w_{AB}$ (though we restrict to $w_{AA}/w_{AB}<1$ since layer corrugation is believed to enhance inter- compared to intra-sublattice tunneling processes). Defining
\begin{align}
    \label{eq:V_band_definition}
    \tilde{V}_{\vec{k}_1, \vec{k}_2, \vec{k}_3, \vec{k}_4} = &\,\delta_{\vec{k}_1+\vec{k}_2 -\vec{k}_3 -\vec{k}_4, \delta \vec{b}} \nonumber \\
    &\sum_{\vec{b} \in \mathrm{RL}} V(\vec{k}_1 - \vec{k}_4 + \vec{b}) \tilde{F}_{-\vec{b}}^{\vec{k}_1, \vec{k}_4} \tilde{F}_{\vec{b} + \delta \vec{b}}^{\vec{k}_2, \vec{k}_3}
\end{align}
with $V(\vec{q})$ the Fourier transform of the interaction potential, $\mathrm{RL}$ denoting the reciprocal lattice as well as $\tilde{F}_{\vec{b}}^{\vec{k}, \vec{k}\pr} = \braket*{u_{\vec{k}}}{u_{\vec{k}\pr+\vec{b} }}$ the overlap of the Bloch band wave functions. 
Using the series expansion from \ceqn{yukawa_series}, \ceqn{V_band_definition} is writable as a simple sum of the on-site and lowest order contributions such that $\tilde{H}_{\lambda} \simeq \tilde{H} + \tilde{H}_1^{\lambda}$. As we restrict our study to the pure interaction Hamiltonian, we choose natural units of energy where the on-site term is one. 

Exploiting the form of the ideal flatband wave function \ceqn{psi_ideal_flatband} in the chiral limit, following \cref{Wang2021} the overlaps can also be expressed as
\begin{align}
    \label{eq:overlap_chiral_limit}
    F_{\vec{b}}^{\vec{k}, \vec{k}\pr} = \mathcal{N}_{\vec{k}} \,\mathcal{N}_{\vec{k}\pr} \sum_{\vec{b}_i} w_{\vec{b}_i} f_{\vec{b}-\vec{b}_i}^{\vec{k}, \vec{k}\pr}
\end{align}
(where we dropped the tilde to indicate the chiral limit). In our numerics, we use the 19 reciprocal lattice vectors $\vec{b}$ with smallest magnitude, where the smallest $w_{\vec{b}}$ (corresponding to the largest $\abs*{\vec{b}}$) is about one percent of $w_{\vec{0}}$.
The form factors of the quantum Hall wave function are given by
\begin{align}
    f_{\vec{b}}^{\vec{k}, \vec{k}\pr} =  \xi_{\vec{b}} e^{\frac{i S}{2} (\vec{k} + \vec{k}\pr) \times \vec{b} }  e^{\frac{i S}{2} \vec{k} \times \vec{k}\pr} e^{- \frac{S^2}{4} \abs*{\vec{k} - \vec{k}\pr - \vec{b}}^2}
\end{align}
with $ \xi_{\vec{b}}=1$ if $\vec{b}/2$ is a reciprocal lattice vectors, else $-1$. With the Fourier modes $w_{\vec{b}}$ of $\abs*{\mathcal{G}(\vec{r})}^2 + \abs*{\mathcal{G}(-\vec{r})}^2$, as introduced in the main text, the $\vec{k}$-dependent normalization can be explicitly calculated as $\mathcal{N}_{\vec{k}}^{-2} = F_{\vec{0}}^{\vec{k}, \vec{k}}$.
The advantage of the form in \ceqn{overlap_chiral_limit} is that it gives a direct way to connect the ideal flatband case to the pure LLL limit by setting $w_{\vec{0}}=1$ and $w_{\vec{b}\neq\vec{0}}=0$, implying $F_{\vec{b}}^{\vec{k}, \vec{k}} = f_{\vec{b}}^{\vec{k}, \vec{k}\pr}$.

In order for the numerically obtained overlaps of Bloch wave functions in the TBG continuum model to be consistent with the ideal flatband formulation, we fix a gauge by choosing a reference momentum at the center of the Brillouin zone $\vec{k}_0 = \vec{ \Gamma}$ and setting the $\vec{k}$-dependent phase of $\ket*{u_{\vec{k}}}$ according to
\begin{align}
    \ket*{u_{\vec{k}}} \rightarrow \ket*{u_{\vec{k}}}  \frac{\abs*{\braket*{u_{\vec{k}_0}}{u_{\vec{k}}}}}{\braket*{u_{\vec{k}_0}}{u_{\vec{k}}}} \frac{F_{\vec{0}}^{\vec{k}_0, \vec{k}}}{\abs*{F_{\vec{0}}^{\vec{k}_0, \vec{k}}}}. 
\end{align}
This ensures that the measurement of e.g. $Q_{\vec{\delta}}\pdag(\vec{r})$, which naturally is in the gauge of the LLL wave functions, on ground states obtained via ED for arbitrary $w_{AA}/w_{AB}$ is well defined.

\subsection{Supplementary information on exact diagonalization simulations}

For our ED calculations, we use the identical setup to the authors previous project in \cref{Wilhelm2023}. Notably, the abundance of zero-energy states in the study of pure on-site interactions mandates the full diagonalization of the Hamiltonian, as compared to the typically performed convergence of the low-energy spectrum. This is because the vast degeneracy renders the normally used Lanczos-based algorithms impractical. As a result, the numerical results in e.g. \cfig{spectrum_filling} are limited to system sizes below those achievable with iterative algorithms. 
The introduction of a finite screening length in \csec{non_ideal} mitigates this issue by selecting a ground state that's (for most clusters) unique within its symmetry sector.
The simulation clusters corresponding to the ED data shown in the main text are compiled in \ctab{clusters}.
\begin{table}[htb]
    \centering
    \caption{Overview of simulation cells used for ED in this work. Despite varying in the number of unit cells (as well as momentum orbitals) $N$, their aspect ratio and point-group symmetry, all clusters have states with zero energy for the CTBG model with on-site interactions.  The simulation torus is spanned by $\vec{T}_1 = a \vec{a}_1 + b \vec{a}_2$ and $\vec{T}_2 = c \vec{a}_1 + d \vec{a}_2$.
    \label{tab:clusters}}
    \begin{tabular}{L{0.5cm}C{2cm}C{1cm}R{1.cm}}
        \hline
        \hline
        
        $N$\vspace{.2em} & \shortstack{Torus \\ $[[a,b],[c,d]]$} & \shortstack{Aspect \\ ratio} & \shortstack{Point \\ group} \\
        
        \hline
        8 &  $ [[1,2],[3,-2]]$ & 1.0 &  $D_2$ \\
        10 &  $[[1,2],[4,-2]]$ & 1.33 &  $C_2$ \\
        12 &  $[[2,2],[2,-4]]$ & 1.0 &  $D_6$ \\
        14 &  $[[1,3],[4,-2]]$ & 1.0 &  $C_2$\\
        15 &  $[[1,3],[4,-3]]$ & 1.0 &  $D_2$\\
        20 &  $[[2,-4],[3,4]]$ & 1.25 &  $D_2$ \\
        25 &  $[[3,2],[5,-5]]$ & 1.00 &  $C_2$ \\
        \hline
        \hline
    \end{tabular}
\end{table}

Concerning the operator form of $Q_{\vec{\delta}}\pdag(\vec{r})$, in the main text we choose to work with a symmetric form in $\pm \vec{k} + \vec{\delta}$ for our analytical arguments. However, in our numerics on finite systems, it is more convenient to work with an equivalent formulation obtained by shifting $\vec{k} \rightarrow \vec{k} - \vec{\delta}$ and successively $2 \vec{\delta} \rightarrow \vec{\delta}$ with $\vec{k}, \vec{\delta} \in \mathrm{BZ}$ such that
\begin{equation}
    Q\pdag_{\vec{\delta}}(\vec{r}) = \sum_{\vec{k} \in \mathrm{BZ}} \tilde{\Phi}_{\vec{k}}(\vec{r}) \tilde{\Phi}_{-\vec{k}+\vec{\delta}}(\vec{r}) \chi\pdag_{\vec{k},-\vec{k}+\vec{\delta}}.
\end{equation}

The restriction to the singlet sector is implemented by adding a strong $S^2$ term to $H$. Since the Hamiltonian is $\mathrm{SU}(2)$-symmetric, this effectively splits the energy levels by their total spin, with only the singlets remaining unchanged.

\end{appendices}


\begin{thebibliography}{91}%
	\makeatletter
	\providecommand \@ifxundefined [1]{%
		\@ifx{#1\undefined}
	}%
	\providecommand \@ifnum [1]{%
		\ifnum #1\expandafter \@firstoftwo
		\else \expandafter \@secondoftwo
		\fi
	}%
	\providecommand \@ifx [1]{%
		\ifx #1\expandafter \@firstoftwo
		\else \expandafter \@secondoftwo
		\fi
	}%
	\providecommand \natexlab [1]{#1}%
	\providecommand \enquote  [1]{``#1''}%
	\providecommand \bibnamefont  [1]{#1}%
	\providecommand \bibfnamefont [1]{#1}%
	\providecommand \citenamefont [1]{#1}%
	\providecommand \href@noop [0]{\@secondoftwo}%
	\providecommand \href [0]{\begingroup \@sanitize@url \@href}%
	\providecommand \@href[1]{\@@startlink{#1}\@@href}%
	\providecommand \@@href[1]{\endgroup#1\@@endlink}%
	\providecommand \@sanitize@url [0]{\catcode `\\12\catcode `\$12\catcode
		`\&12\catcode `\#12\catcode `\^12\catcode `\_12\catcode `\%12\relax}%
	\providecommand \@@startlink[1]{}%
	\providecommand \@@endlink[0]{}%
	\providecommand \url  [0]{\begingroup\@sanitize@url \@url }%
	\providecommand \@url [1]{\endgroup\@href {#1}{\urlprefix }}%
	\providecommand \urlprefix  [0]{URL }%
	\providecommand \Eprint [0]{\href }%
	\providecommand \doibase [0]{https://doi.org/}%
	\providecommand \selectlanguage [0]{\@gobble}%
	\providecommand \bibinfo  [0]{\@secondoftwo}%
	\providecommand \bibfield  [0]{\@secondoftwo}%
	\providecommand \translation [1]{[#1]}%
	\providecommand \BibitemOpen [0]{}%
	\providecommand \bibitemStop [0]{}%
	\providecommand \bibitemNoStop [0]{.\EOS\space}%
	\providecommand \EOS [0]{\spacefactor3000\relax}%
	\providecommand \BibitemShut  [1]{\csname bibitem#1\endcsname}%
	\let\auto@bib@innerbib\@empty
	\bibitem [{\citenamefont {Bergholtz}\ and\ \citenamefont
		{Liu}(2013)}]{BergholtzReview}%
	\BibitemOpen
	\bibfield  {author} {\bibinfo {author} {\bibfnamefont {E.~J.}\ \bibnamefont
			{Bergholtz}}\ and\ \bibinfo {author} {\bibfnamefont {Z.}~\bibnamefont
			{Liu}},\ }\bibfield  {title} {\bibinfo {title} {Topological flat band models
			and fractional chern insulators},\ }\href
	{https://doi.org/10.1142/S021797921330017X} {\bibfield  {journal} {\bibinfo
			{journal} {International Journal of Modern Physics B}\ }\textbf {\bibinfo
			{volume} {27}},\ \bibinfo {pages} {1330017} (\bibinfo {year}
		{2013})}\BibitemShut {NoStop}%
	\bibitem [{\citenamefont {Parameswaran}\ \emph {et~al.}(2013)\citenamefont
		{Parameswaran}, \citenamefont {Roy},\ and\ \citenamefont
		{Sondhi}}]{SidReview}%
	\BibitemOpen
	\bibfield  {author} {\bibinfo {author} {\bibfnamefont {S.~A.}\ \bibnamefont
			{Parameswaran}}, \bibinfo {author} {\bibfnamefont {R.}~\bibnamefont {Roy}},\
		and\ \bibinfo {author} {\bibfnamefont {S.~L.}\ \bibnamefont {Sondhi}},\
	}\bibfield  {title} {\bibinfo {title} {Fractional quantum {Hall} physics in
			topological flat bands},\ }\href {https://doi.org/10.1016/j.crhy.2013.04.003}
	{\bibfield  {journal} {\bibinfo  {journal} {Comptes Rendus. Physique}\
		}\textbf {\bibinfo {volume} {14}},\ \bibinfo {pages} {816} (\bibinfo {year}
		{2013})}\BibitemShut {NoStop}%
	\bibitem [{\citenamefont {T{\"o}rm{\"a}}\ \emph {et~al.}(2022)\citenamefont
		{T{\"o}rm{\"a}}, \citenamefont {Peotta},\ and\ \citenamefont
		{Bernevig}}]{BalentSCReview}%
	\BibitemOpen
	\bibfield  {author} {\bibinfo {author} {\bibfnamefont {P.}~\bibnamefont
			{T{\"o}rm{\"a}}}, \bibinfo {author} {\bibfnamefont {S.}~\bibnamefont
			{Peotta}},\ and\ \bibinfo {author} {\bibfnamefont {B.~A.}\ \bibnamefont
			{Bernevig}},\ }\bibfield  {title} {\bibinfo {title} {Superconductivity,
			superfluidity and quantum geometry in twisted multilayer systems},\ }\href
	{https://doi.org/10.1038/s42254-022-00466-y} {\bibfield  {journal} {\bibinfo
			{journal} {Nature Reviews Physics}\ }\textbf {\bibinfo {volume} {4}},\
		\bibinfo {pages} {528} (\bibinfo {year} {2022})}\BibitemShut {NoStop}%
	\bibitem [{\citenamefont {Kitaev}(2003)}]{Kitaev2003}%
	\BibitemOpen
	\bibfield  {author} {\bibinfo {author} {\bibfnamefont {A.}~\bibnamefont
			{Kitaev}},\ }\bibfield  {title} {\bibinfo {title} {Fault-tolerant quantum
			computation by anyons},\ }\href
	{https://doi.org/https://doi.org/10.1016/S0003-4916(02)00018-0} {\bibfield
		{journal} {\bibinfo  {journal} {Annals of Physics}\ }\textbf {\bibinfo
			{volume} {303}},\ \bibinfo {pages} {2} (\bibinfo {year} {2003})}\BibitemShut
	{NoStop}%
	\bibitem [{\citenamefont {Levin}\ and\ \citenamefont {Wen}(2005)}]{Levin2005}%
	\BibitemOpen
	\bibfield  {author} {\bibinfo {author} {\bibfnamefont {M.~A.}\ \bibnamefont
			{Levin}}\ and\ \bibinfo {author} {\bibfnamefont {X.-G.}\ \bibnamefont
			{Wen}},\ }\bibfield  {title} {\bibinfo {title} {String-net condensation: A
			physical mechanism for topological phases},\ }\href
	{https://doi.org/10.1103/PhysRevB.71.045110} {\bibfield  {journal} {\bibinfo
			{journal} {Phys. Rev. B}\ }\textbf {\bibinfo {volume} {71}},\ \bibinfo
		{pages} {045110} (\bibinfo {year} {2005})}\BibitemShut {NoStop}%
	\bibitem [{\citenamefont {Affleck}\ \emph {et~al.}(1987)\citenamefont
		{Affleck}, \citenamefont {Kennedy}, \citenamefont {Lieb},\ and\ \citenamefont
		{Tasaki}}]{Affleck1987}%
	\BibitemOpen
	\bibfield  {author} {\bibinfo {author} {\bibfnamefont {I.}~\bibnamefont
			{Affleck}}, \bibinfo {author} {\bibfnamefont {T.}~\bibnamefont {Kennedy}},
		\bibinfo {author} {\bibfnamefont {E.~H.}\ \bibnamefont {Lieb}},\ and\
		\bibinfo {author} {\bibfnamefont {H.}~\bibnamefont {Tasaki}},\ }\bibfield
	{title} {\bibinfo {title} {Rigorous results on valence-bond ground states in
			antiferromagnets},\ }\href {https://doi.org/10.1103/PhysRevLett.59.799}
	{\bibfield  {journal} {\bibinfo  {journal} {Phys. Rev. Lett.}\ }\textbf
		{\bibinfo {volume} {59}},\ \bibinfo {pages} {799} (\bibinfo {year}
		{1987})}\BibitemShut {NoStop}%
	\bibitem [{\citenamefont {Rokhsar}\ and\ \citenamefont
		{Kivelson}(1988)}]{Rokhsar1988}%
	\BibitemOpen
	\bibfield  {author} {\bibinfo {author} {\bibfnamefont {D.~S.}\ \bibnamefont
			{Rokhsar}}\ and\ \bibinfo {author} {\bibfnamefont {S.~A.}\ \bibnamefont
			{Kivelson}},\ }\bibfield  {title} {\bibinfo {title} {Superconductivity and
			the quantum hard-core dimer gas},\ }\href
	{https://doi.org/10.1103/PhysRevLett.61.2376} {\bibfield  {journal} {\bibinfo
			{journal} {Phys. Rev. Lett.}\ }\textbf {\bibinfo {volume} {61}},\ \bibinfo
		{pages} {2376} (\bibinfo {year} {1988})}\BibitemShut {NoStop}%
	\bibitem [{\citenamefont {Castelnovo}\ \emph {et~al.}(2005)\citenamefont
		{Castelnovo}, \citenamefont {Chamon}, \citenamefont {Mudry},\ and\
		\citenamefont {Pujol}}]{Castelnovo2005}%
	\BibitemOpen
	\bibfield  {author} {\bibinfo {author} {\bibfnamefont {C.}~\bibnamefont
			{Castelnovo}}, \bibinfo {author} {\bibfnamefont {C.}~\bibnamefont {Chamon}},
		\bibinfo {author} {\bibfnamefont {C.}~\bibnamefont {Mudry}},\ and\ \bibinfo
		{author} {\bibfnamefont {P.}~\bibnamefont {Pujol}},\ }\bibfield  {title}
	{\bibinfo {title} {{From quantum mechanics to classical statistical physics:
				Generalized Rokhsar{\textendash}Kivelson Hamiltonians and the
				{\textquotedblleft}Stochastic Matrix Form{\textquotedblright}
				decomposition}},\ }\href {https://doi.org/10.1016/j.aop.2005.01.006}
	{\bibfield  {journal} {\bibinfo  {journal} {Ann. Phys.}\ }\textbf {\bibinfo
			{volume} {318}},\ \bibinfo {pages} {316} (\bibinfo {year}
		{2005})}\BibitemShut {NoStop}%
	\bibitem [{\citenamefont {Andrei}\ and\ \citenamefont
		{MacDonald}(2020)}]{andreiGrapheneBilayersTwist2020}%
	\BibitemOpen
	\bibfield  {author} {\bibinfo {author} {\bibfnamefont {E.~Y.}\ \bibnamefont
			{Andrei}}\ and\ \bibinfo {author} {\bibfnamefont {A.~H.}\ \bibnamefont
			{MacDonald}},\ }\bibfield  {title} {\bibinfo {title} {Graphene bilayers with
			a twist},\ }\href {https://doi.org/10.1038/s41563-020-00840-0} {\bibfield
		{journal} {\bibinfo  {journal} {Nature Materials}\ }\textbf {\bibinfo
			{volume} {19}},\ \bibinfo {pages} {1265} (\bibinfo {year}
		{2020})}\BibitemShut {NoStop}%
	\bibitem [{\citenamefont {Balents}\ \emph {et~al.}(2020)\citenamefont
		{Balents}, \citenamefont {Dean}, \citenamefont {Efetov},\ and\ \citenamefont
		{Young}}]{balentsSuperconductivityStrongCorrelations2020}%
	\BibitemOpen
	\bibfield  {author} {\bibinfo {author} {\bibfnamefont {L.}~\bibnamefont
			{Balents}}, \bibinfo {author} {\bibfnamefont {C.~R.}\ \bibnamefont {Dean}},
		\bibinfo {author} {\bibfnamefont {D.~K.}\ \bibnamefont {Efetov}},\ and\
		\bibinfo {author} {\bibfnamefont {A.~F.}\ \bibnamefont {Young}},\ }\bibfield
	{title} {\bibinfo {title} {Superconductivity and strong correlations in
			moir\'e flat bands},\ }\href {https://doi.org/10.1038/s41567-020-0906-9}
	{\bibfield  {journal} {\bibinfo  {journal} {Nature Physics}\ }\textbf
		{\bibinfo {volume} {16}},\ \bibinfo {pages} {725} (\bibinfo {year}
		{2020})}\BibitemShut {NoStop}%
	\bibitem [{\citenamefont {Kennes}\ \emph {et~al.}(2021)\citenamefont {Kennes},
		\citenamefont {Claassen}, \citenamefont {Xian}, \citenamefont {Georges},
		\citenamefont {Millis}, \citenamefont {Hone}, \citenamefont {Dean},
		\citenamefont {Basov}, \citenamefont {Pasupathy},\ and\ \citenamefont
		{Rubio}}]{kennes2020moir}%
	\BibitemOpen
	\bibfield  {author} {\bibinfo {author} {\bibfnamefont {D.~M.}\ \bibnamefont
			{Kennes}}, \bibinfo {author} {\bibfnamefont {M.}~\bibnamefont {Claassen}},
		\bibinfo {author} {\bibfnamefont {L.}~\bibnamefont {Xian}}, \bibinfo {author}
		{\bibfnamefont {A.}~\bibnamefont {Georges}}, \bibinfo {author} {\bibfnamefont
			{A.~J.}\ \bibnamefont {Millis}}, \bibinfo {author} {\bibfnamefont
			{J.}~\bibnamefont {Hone}}, \bibinfo {author} {\bibfnamefont {C.~R.}\
			\bibnamefont {Dean}}, \bibinfo {author} {\bibfnamefont {D.~N.}\ \bibnamefont
			{Basov}}, \bibinfo {author} {\bibfnamefont {A.~N.}\ \bibnamefont
			{Pasupathy}},\ and\ \bibinfo {author} {\bibfnamefont {A.}~\bibnamefont
			{Rubio}},\ }\bibfield  {title} {\bibinfo {title} {Moiré heterostructures as
			a condensed-matter quantum simulator},\ }\href
	{https://doi.org/10.1038/s41567-020-01154-3} {\bibfield  {journal} {\bibinfo
			{journal} {Nat. Phys.}\ }\textbf {\bibinfo {volume} {17}},\ \bibinfo {pages}
		{155} (\bibinfo {year} {2021})}\BibitemShut {NoStop}%
	\bibitem [{\citenamefont {Adak}\ \emph {et~al.}(2024)\citenamefont {Adak},
		\citenamefont {Sinha}, \citenamefont {Agarwal},\ and\ \citenamefont
		{Deshmukh}}]{MoireTopology}%
	\BibitemOpen
	\bibfield  {author} {\bibinfo {author} {\bibfnamefont {P.~C.}\ \bibnamefont
			{Adak}}, \bibinfo {author} {\bibfnamefont {S.}~\bibnamefont {Sinha}},
		\bibinfo {author} {\bibfnamefont {A.}~\bibnamefont {Agarwal}},\ and\ \bibinfo
		{author} {\bibfnamefont {M.~M.}\ \bibnamefont {Deshmukh}},\ }\bibfield
	{title} {\bibinfo {title} {{Tunable moir{\ifmmode\acute{e}\else\'{e}\fi}
				materials for probing Berry physics and topology}},\ }\href
	{https://doi.org/10.1038/s41578-024-00671-4} {\bibfield  {journal} {\bibinfo
			{journal} {Nat. Rev. Mater.}\ ,\ \bibinfo {pages} {1}} (\bibinfo {year}
		{2024})}\BibitemShut {NoStop}%
	\bibitem [{\citenamefont {Ledwith}\ \emph {et~al.}(2023)\citenamefont
		{Ledwith}, \citenamefont {Vishwanath},\ and\ \citenamefont
		{Parker}}]{Ledwith2023}%
	\BibitemOpen
	\bibfield  {author} {\bibinfo {author} {\bibfnamefont {P.~J.}\ \bibnamefont
			{Ledwith}}, \bibinfo {author} {\bibfnamefont {A.}~\bibnamefont
			{Vishwanath}},\ and\ \bibinfo {author} {\bibfnamefont {D.~E.}\ \bibnamefont
			{Parker}},\ }\bibfield  {title} {\bibinfo {title} {Vortexability: A unifying
			criterion for ideal fractional chern insulators},\ }\href
	{https://doi.org/10.1103/PhysRevB.108.205144} {\bibfield  {journal} {\bibinfo
			{journal} {Phys. Rev. B}\ }\textbf {\bibinfo {volume} {108}},\ \bibinfo
		{pages} {205144} (\bibinfo {year} {2023})}\BibitemShut {NoStop}%
	\bibitem [{\citenamefont {Okuma}(2024)}]{Okuma2024}%
	\BibitemOpen
	\bibfield  {author} {\bibinfo {author} {\bibfnamefont {N.}~\bibnamefont
			{Okuma}},\ }\href@noop {} {\bibinfo {title} {Constructing vortex functions
			and basis states of chern insulators: ideal condition, inequality from index
			theorem, and coherent-like states on von neumann lattice}} (\bibinfo {year}
	{2024}),\ \Eprint {https://arxiv.org/abs/2405.11796} {arXiv:2405.11796}
	\BibitemShut {NoStop}%
	\bibitem [{\citenamefont {Tarnopolsky}\ \emph {et~al.}(2019)\citenamefont
		{Tarnopolsky}, \citenamefont {Kruchkov},\ and\ \citenamefont
		{Vishwanath}}]{Tarnopolsky2019}%
	\BibitemOpen
	\bibfield  {author} {\bibinfo {author} {\bibfnamefont {G.}~\bibnamefont
			{Tarnopolsky}}, \bibinfo {author} {\bibfnamefont {A.~J.}\ \bibnamefont
			{Kruchkov}},\ and\ \bibinfo {author} {\bibfnamefont {A.}~\bibnamefont
			{Vishwanath}},\ }\bibfield  {title} {\bibinfo {title} {Origin of magic angles
			in twisted bilayer graphene},\ }\href
	{https://doi.org/10.1103/PhysRevLett.122.106405} {\bibfield  {journal}
		{\bibinfo  {journal} {Phys. Rev. Lett.}\ }\textbf {\bibinfo {volume} {122}},\
		\bibinfo {pages} {106405} (\bibinfo {year} {2019})}\BibitemShut {NoStop}%
	\bibitem [{\citenamefont {Wang}\ \emph
		{et~al.}(2021{\natexlab{a}})\citenamefont {Wang}, \citenamefont {Zheng},
		\citenamefont {Millis},\ and\ \citenamefont {Cano}}]{Wang2021a}%
	\BibitemOpen
	\bibfield  {author} {\bibinfo {author} {\bibfnamefont {J.}~\bibnamefont
			{Wang}}, \bibinfo {author} {\bibfnamefont {Y.}~\bibnamefont {Zheng}},
		\bibinfo {author} {\bibfnamefont {A.~J.}\ \bibnamefont {Millis}},\ and\
		\bibinfo {author} {\bibfnamefont {J.}~\bibnamefont {Cano}},\ }\bibfield
	{title} {\bibinfo {title} {Chiral approximation to twisted bilayer graphene:
			Exact intravalley inversion symmetry, nodal structure, and implications for
			higher magic angles},\ }\href
	{https://doi.org/10.1103/PhysRevResearch.3.023155} {\bibfield  {journal}
		{\bibinfo  {journal} {Phys. Rev. Res.}\ }\textbf {\bibinfo {volume} {3}},\
		\bibinfo {pages} {023155} (\bibinfo {year} {2021}{\natexlab{a}})}\BibitemShut
	{NoStop}%
	\bibitem [{\citenamefont {Wang}\ \emph
		{et~al.}(2021{\natexlab{b}})\citenamefont {Wang}, \citenamefont {Cano},
		\citenamefont {Millis}, \citenamefont {Liu},\ and\ \citenamefont
		{Yang}}]{Wang2021}%
	\BibitemOpen
	\bibfield  {author} {\bibinfo {author} {\bibfnamefont {J.}~\bibnamefont
			{Wang}}, \bibinfo {author} {\bibfnamefont {J.}~\bibnamefont {Cano}}, \bibinfo
		{author} {\bibfnamefont {A.~J.}\ \bibnamefont {Millis}}, \bibinfo {author}
		{\bibfnamefont {Z.}~\bibnamefont {Liu}},\ and\ \bibinfo {author}
		{\bibfnamefont {B.}~\bibnamefont {Yang}},\ }\bibfield  {title} {\bibinfo
		{title} {Exact landau level description of geometry and interaction in a
			flatband},\ }\href {https://doi.org/10.1103/PhysRevLett.127.246403}
	{\bibfield  {journal} {\bibinfo  {journal} {Phys. Rev. Lett.}\ }\textbf
		{\bibinfo {volume} {127}},\ \bibinfo {pages} {246403} (\bibinfo {year}
		{2021}{\natexlab{b}})}\BibitemShut {NoStop}%
	\bibitem [{\citenamefont {Haldane}(2018)}]{Haldane2018}%
	\BibitemOpen
	\bibfield  {author} {\bibinfo {author} {\bibfnamefont {F.~D.~M.}\
			\bibnamefont {Haldane}},\ }\bibfield  {title} {\bibinfo {title} {{A
				modular-invariant modified Weierstrass sigma-function as a building block for
				lowest-Landau-level wavefunctions on the torus}},\ }\href
	{https://doi.org/10.1063/1.5042618} {\bibfield  {journal} {\bibinfo
			{journal} {Journal of Mathematical Physics}\ }\textbf {\bibinfo {volume}
			{59}},\ \bibinfo {pages} {071901} (\bibinfo {year} {2018})}\BibitemShut
	{NoStop}%
	\bibitem [{\citenamefont {Bultinck}\ \emph {et~al.}(2020)\citenamefont
		{Bultinck}, \citenamefont {Khalaf}, \citenamefont {Liu}, \citenamefont
		{Chatterjee}, \citenamefont {Vishwanath},\ and\ \citenamefont
		{Zaletel}}]{Bultinck2020}%
	\BibitemOpen
	\bibfield  {author} {\bibinfo {author} {\bibfnamefont {N.}~\bibnamefont
			{Bultinck}}, \bibinfo {author} {\bibfnamefont {E.}~\bibnamefont {Khalaf}},
		\bibinfo {author} {\bibfnamefont {S.}~\bibnamefont {Liu}}, \bibinfo {author}
		{\bibfnamefont {S.}~\bibnamefont {Chatterjee}}, \bibinfo {author}
		{\bibfnamefont {A.}~\bibnamefont {Vishwanath}},\ and\ \bibinfo {author}
		{\bibfnamefont {M.~P.}\ \bibnamefont {Zaletel}},\ }\bibfield  {title}
	{\bibinfo {title} {Ground state and hidden symmetry of magic-angle graphene
			at even integer filling},\ }\href
	{https://doi.org/10.1103/PhysRevX.10.031034} {\bibfield  {journal} {\bibinfo
			{journal} {Phys. Rev. X}\ }\textbf {\bibinfo {volume} {10}},\ \bibinfo
		{pages} {031034} (\bibinfo {year} {2020})}\BibitemShut {NoStop}%
	\bibitem [{\citenamefont {Polshyn}\ \emph {et~al.}(2022)\citenamefont
		{Polshyn}, \citenamefont {Zhang}, \citenamefont {Kumar}, \citenamefont
		{Soejima}, \citenamefont {Ledwith}, \citenamefont {Watanabe}, \citenamefont
		{Taniguchi}, \citenamefont {Vishwanath}, \citenamefont {Zaletel},\ and\
		\citenamefont {Young}}]{Polshyn2022}%
	\BibitemOpen
	\bibfield  {author} {\bibinfo {author} {\bibfnamefont {H.}~\bibnamefont
			{Polshyn}}, \bibinfo {author} {\bibfnamefont {Y.}~\bibnamefont {Zhang}},
		\bibinfo {author} {\bibfnamefont {M.~A.}\ \bibnamefont {Kumar}}, \bibinfo
		{author} {\bibfnamefont {T.}~\bibnamefont {Soejima}}, \bibinfo {author}
		{\bibfnamefont {P.}~\bibnamefont {Ledwith}}, \bibinfo {author} {\bibfnamefont
			{K.}~\bibnamefont {Watanabe}}, \bibinfo {author} {\bibfnamefont
			{T.}~\bibnamefont {Taniguchi}}, \bibinfo {author} {\bibfnamefont
			{A.}~\bibnamefont {Vishwanath}}, \bibinfo {author} {\bibfnamefont {M.~P.}\
			\bibnamefont {Zaletel}},\ and\ \bibinfo {author} {\bibfnamefont {A.~F.}\
			\bibnamefont {Young}},\ }\bibfield  {title} {\bibinfo {title} {{Topological
				charge density waves at half-integer filling of a
				moir{\ifmmode\acute{e}\else\'{e}\fi} superlattice}},\ }\href
	{https://doi.org/10.1038/s41567-021-01418-6} {\bibfield  {journal} {\bibinfo
			{journal} {Nat. Phys.}\ }\textbf {\bibinfo {volume} {18}},\ \bibinfo {pages}
		{42} (\bibinfo {year} {2022})}\BibitemShut {NoStop}%
	\bibitem [{\citenamefont {Wilhelm}\ \emph
		{et~al.}(2023{\natexlab{a}})\citenamefont {Wilhelm}, \citenamefont {Lang},
		\citenamefont {Scheurer},\ and\ \citenamefont {Läuchli}}]{Wilhelm2023}%
	\BibitemOpen
	\bibfield  {author} {\bibinfo {author} {\bibfnamefont {P.~H.}\ \bibnamefont
			{Wilhelm}}, \bibinfo {author} {\bibfnamefont {T.~C.}\ \bibnamefont {Lang}},
		\bibinfo {author} {\bibfnamefont {M.~S.}\ \bibnamefont {Scheurer}},\ and\
		\bibinfo {author} {\bibfnamefont {A.~M.}\ \bibnamefont {Läuchli}},\
	}\bibfield  {title} {\bibinfo {title} {{Non-coplanar magnetism, topological
				density wave order and emergent symmetry at half-integer filling of moir\'e
				Chern bands}},\ }\href {https://doi.org/10.21468/SciPostPhys.14.3.040}
	{\bibfield  {journal} {\bibinfo  {journal} {SciPost Phys.}\ }\textbf
		{\bibinfo {volume} {14}},\ \bibinfo {pages} {040} (\bibinfo {year}
		{2023}{\natexlab{a}})}\BibitemShut {NoStop}%
	\bibitem [{\citenamefont {Dong}\ \emph
		{et~al.}(2023{\natexlab{a}})\citenamefont {Dong}, \citenamefont {Ledwith},
		\citenamefont {Khalaf}, \citenamefont {Lee},\ and\ \citenamefont
		{Vishwanath}}]{Dong2023}%
	\BibitemOpen
	\bibfield  {author} {\bibinfo {author} {\bibfnamefont {J.}~\bibnamefont
			{Dong}}, \bibinfo {author} {\bibfnamefont {P.~J.}\ \bibnamefont {Ledwith}},
		\bibinfo {author} {\bibfnamefont {E.}~\bibnamefont {Khalaf}}, \bibinfo
		{author} {\bibfnamefont {J.~Y.}\ \bibnamefont {Lee}},\ and\ \bibinfo {author}
		{\bibfnamefont {A.}~\bibnamefont {Vishwanath}},\ }\bibfield  {title}
	{\bibinfo {title} {{Many-body ground states from decomposition of ideal
				higher Chern bands: Applications to chirally twisted graphene multilayers}},\
	}\href {https://doi.org/10.1103/PhysRevResearch.5.023166} {\bibfield
		{journal} {\bibinfo  {journal} {Phys. Rev. Res.}\ }\textbf {\bibinfo {volume}
			{5}},\ \bibinfo {pages} {023166} (\bibinfo {year}
		{2023}{\natexlab{a}})}\BibitemShut {NoStop}%
	\bibitem [{\citenamefont {Wang}\ \emph {et~al.}(2023)\citenamefont {Wang},
		\citenamefont {Klevtsov},\ and\ \citenamefont {Liu}}]{Wang2023}%
	\BibitemOpen
	\bibfield  {author} {\bibinfo {author} {\bibfnamefont {J.}~\bibnamefont
			{Wang}}, \bibinfo {author} {\bibfnamefont {S.}~\bibnamefont {Klevtsov}},\
		and\ \bibinfo {author} {\bibfnamefont {Z.}~\bibnamefont {Liu}},\ }\bibfield
	{title} {\bibinfo {title} {{Origin of model fractional Chern insulators in
				all topological ideal flatbands: Explicit color-entangled wave function and
				exact density algebra}},\ }\href
	{https://doi.org/10.1103/PhysRevResearch.5.023167} {\bibfield  {journal}
		{\bibinfo  {journal} {Phys. Rev. Res.}\ }\textbf {\bibinfo {volume} {5}},\
		\bibinfo {pages} {023167} (\bibinfo {year} {2023})}\BibitemShut {NoStop}%
	\bibitem [{\citenamefont {Wang}\ \emph
		{et~al.}(2024{\natexlab{a}})\citenamefont {Wang}, \citenamefont {Zhou},
		\citenamefont {Lin}, \citenamefont {Feng}, \citenamefont {Wang},
		\citenamefont {Liang}, \citenamefont {Zhang}, \citenamefont {Wu},
		\citenamefont {Liu}, \citenamefont {Watanabe}, \citenamefont {Taniguchi},
		\citenamefont {Yang}, \citenamefont {Zhang}, \citenamefont {Liu},
		\citenamefont {Gao}, \citenamefont {Liu}, \citenamefont {Xie}, \citenamefont
		{Song},\ and\ \citenamefont {Lu}}]{Wang2024}%
	\BibitemOpen
	\bibfield  {author} {\bibinfo {author} {\bibfnamefont {W.}~\bibnamefont
			{Wang}}, \bibinfo {author} {\bibfnamefont {G.}~\bibnamefont {Zhou}}, \bibinfo
		{author} {\bibfnamefont {W.}~\bibnamefont {Lin}}, \bibinfo {author}
		{\bibfnamefont {Z.}~\bibnamefont {Feng}}, \bibinfo {author} {\bibfnamefont
			{Y.}~\bibnamefont {Wang}}, \bibinfo {author} {\bibfnamefont {M.}~\bibnamefont
			{Liang}}, \bibinfo {author} {\bibfnamefont {Z.}~\bibnamefont {Zhang}},
		\bibinfo {author} {\bibfnamefont {M.}~\bibnamefont {Wu}}, \bibinfo {author}
		{\bibfnamefont {L.}~\bibnamefont {Liu}}, \bibinfo {author} {\bibfnamefont
			{K.}~\bibnamefont {Watanabe}}, \bibinfo {author} {\bibfnamefont
			{T.}~\bibnamefont {Taniguchi}}, \bibinfo {author} {\bibfnamefont
			{W.}~\bibnamefont {Yang}}, \bibinfo {author} {\bibfnamefont {G.}~\bibnamefont
			{Zhang}}, \bibinfo {author} {\bibfnamefont {K.}~\bibnamefont {Liu}}, \bibinfo
		{author} {\bibfnamefont {J.}~\bibnamefont {Gao}}, \bibinfo {author}
		{\bibfnamefont {Y.}~\bibnamefont {Liu}}, \bibinfo {author} {\bibfnamefont
			{X.~C.}\ \bibnamefont {Xie}}, \bibinfo {author} {\bibfnamefont
			{Z.}~\bibnamefont {Song}},\ and\ \bibinfo {author} {\bibfnamefont
			{X.}~\bibnamefont {Lu}},\ }\href@noop {} {\bibinfo {title} {Correlated charge
			density wave insulators in chirally twisted triple bilayer graphene}}
	(\bibinfo {year} {2024}{\natexlab{a}}),\ \Eprint
	{https://arxiv.org/abs/2405.15742} {arXiv:2405.15742} \BibitemShut {NoStop}%
	\bibitem [{\citenamefont {Ledwith}\ \emph {et~al.}(2020)\citenamefont
		{Ledwith}, \citenamefont {Tarnopolsky}, \citenamefont {Khalaf},\ and\
		\citenamefont {Vishwanath}}]{Ledwith2020}%
	\BibitemOpen
	\bibfield  {author} {\bibinfo {author} {\bibfnamefont {P.~J.}\ \bibnamefont
			{Ledwith}}, \bibinfo {author} {\bibfnamefont {G.}~\bibnamefont
			{Tarnopolsky}}, \bibinfo {author} {\bibfnamefont {E.}~\bibnamefont
			{Khalaf}},\ and\ \bibinfo {author} {\bibfnamefont {A.}~\bibnamefont
			{Vishwanath}},\ }\bibfield  {title} {\bibinfo {title} {{Fractional Chern
				insulator states in twisted bilayer graphene: An analytical approach}},\
	}\href {https://doi.org/10.1103/PhysRevResearch.2.023237} {\bibfield
		{journal} {\bibinfo  {journal} {Phys. Rev. Res.}\ }\textbf {\bibinfo {volume}
			{2}},\ \bibinfo {pages} {023237} (\bibinfo {year} {2020})}\BibitemShut
	{NoStop}%
	\bibitem [{\citenamefont {Abouelkomsan}\ \emph {et~al.}(2020)\citenamefont
		{Abouelkomsan}, \citenamefont {Liu},\ and\ \citenamefont
		{Bergholtz}}]{Abouelkomsan2020}%
	\BibitemOpen
	\bibfield  {author} {\bibinfo {author} {\bibfnamefont {A.}~\bibnamefont
			{Abouelkomsan}}, \bibinfo {author} {\bibfnamefont {Z.}~\bibnamefont {Liu}},\
		and\ \bibinfo {author} {\bibfnamefont {E.~J.}\ \bibnamefont {Bergholtz}},\
	}\bibfield  {title} {\bibinfo {title} {{Particle-Hole Duality, Emergent Fermi
				Liquids, and Fractional Chern Insulators in Moir\'e Flatbands}},\ }\href
	{https://doi.org/10.1103/PhysRevLett.124.106803} {\bibfield  {journal}
		{\bibinfo  {journal} {Phys. Rev. Lett.}\ }\textbf {\bibinfo {volume} {124}},\
		\bibinfo {pages} {106803} (\bibinfo {year} {2020})}\BibitemShut {NoStop}%
	\bibitem [{\citenamefont {Repellin}\ and\ \citenamefont
		{Senthil}(2020)}]{Repellin2020}%
	\BibitemOpen
	\bibfield  {author} {\bibinfo {author} {\bibfnamefont {C.}~\bibnamefont
			{Repellin}}\ and\ \bibinfo {author} {\bibfnamefont {T.}~\bibnamefont
			{Senthil}},\ }\bibfield  {title} {\bibinfo {title} {{Chern bands of twisted
				bilayer graphene: Fractional Chern insulators and spin phase transition}},\
	}\href {https://doi.org/10.1103/PhysRevResearch.2.023238} {\bibfield
		{journal} {\bibinfo  {journal} {Phys. Rev. Res.}\ }\textbf {\bibinfo {volume}
			{2}},\ \bibinfo {pages} {023238} (\bibinfo {year} {2020})}\BibitemShut
	{NoStop}%
	\bibitem [{\citenamefont {Wilhelm}\ \emph {et~al.}(2021)\citenamefont
		{Wilhelm}, \citenamefont {Lang},\ and\ \citenamefont
		{L\"auchli}}]{Wilhelm2021}%
	\BibitemOpen
	\bibfield  {author} {\bibinfo {author} {\bibfnamefont {P.}~\bibnamefont
			{Wilhelm}}, \bibinfo {author} {\bibfnamefont {T.~C.}\ \bibnamefont {Lang}},\
		and\ \bibinfo {author} {\bibfnamefont {A.~M.}\ \bibnamefont {L\"auchli}},\
	}\bibfield  {title} {\bibinfo {title} {Interplay of fractional chern
			insulator and charge density wave phases in twisted bilayer graphene},\
	}\href {https://doi.org/10.1103/PhysRevB.103.125406} {\bibfield  {journal}
		{\bibinfo  {journal} {Phys. Rev. B}\ }\textbf {\bibinfo {volume} {103}},\
		\bibinfo {pages} {125406} (\bibinfo {year} {2021})}\BibitemShut {NoStop}%
	\bibitem [{\citenamefont {Liu}\ \emph {et~al.}(2021)\citenamefont {Liu},
		\citenamefont {Abouelkomsan},\ and\ \citenamefont {Bergholtz}}]{Zhao2021}%
	\BibitemOpen
	\bibfield  {author} {\bibinfo {author} {\bibfnamefont {Z.}~\bibnamefont
			{Liu}}, \bibinfo {author} {\bibfnamefont {A.}~\bibnamefont {Abouelkomsan}},\
		and\ \bibinfo {author} {\bibfnamefont {E.~J.}\ \bibnamefont {Bergholtz}},\
	}\bibfield  {title} {\bibinfo {title} {Gate-tunable fractional chern
			insulators in twisted double bilayer graphene},\ }\href
	{https://doi.org/10.1103/PhysRevLett.126.026801} {\bibfield  {journal}
		{\bibinfo  {journal} {Phys. Rev. Lett.}\ }\textbf {\bibinfo {volume} {126}},\
		\bibinfo {pages} {026801} (\bibinfo {year} {2021})}\BibitemShut {NoStop}%
	\bibitem [{\citenamefont {Wang}\ and\ \citenamefont {Liu}(2022)}]{Wang2022}%
	\BibitemOpen
	\bibfield  {author} {\bibinfo {author} {\bibfnamefont {J.}~\bibnamefont
			{Wang}}\ and\ \bibinfo {author} {\bibfnamefont {Z.}~\bibnamefont {Liu}},\
	}\bibfield  {title} {\bibinfo {title} {Hierarchy of ideal flatbands in chiral
			twisted multilayer graphene models},\ }\href
	{https://doi.org/10.1103/PhysRevLett.128.176403} {\bibfield  {journal}
		{\bibinfo  {journal} {Phys. Rev. Lett.}\ }\textbf {\bibinfo {volume} {128}},\
		\bibinfo {pages} {176403} (\bibinfo {year} {2022})}\BibitemShut {NoStop}%
	\bibitem [{\citenamefont {Sharma}\ \emph {et~al.}(2024)\citenamefont {Sharma},
		\citenamefont {Peng},\ and\ \citenamefont {Sheng}}]{Sharma2024}%
	\BibitemOpen
	\bibfield  {author} {\bibinfo {author} {\bibfnamefont {P.}~\bibnamefont
			{Sharma}}, \bibinfo {author} {\bibfnamefont {Y.}~\bibnamefont {Peng}},\ and\
		\bibinfo {author} {\bibfnamefont {D.~N.}\ \bibnamefont {Sheng}},\ }\href@noop
	{} {\bibinfo {title} {{Topological quantum phase transitions driven by
				displacement fields in the twisted MoTe2 bilayers}}} (\bibinfo {year}
	{2024}),\ \Eprint {https://arxiv.org/abs/2405.08181} {arXiv:2405.08181}
	\BibitemShut {NoStop}%
	\bibitem [{\citenamefont {Liu}\ \emph {et~al.}(2024)\citenamefont {Liu},
		\citenamefont {Liu},\ and\ \citenamefont {Bergholtz}}]{Liu2024}%
	\BibitemOpen
	\bibfield  {author} {\bibinfo {author} {\bibfnamefont {H.}~\bibnamefont
			{Liu}}, \bibinfo {author} {\bibfnamefont {Z.}~\bibnamefont {Liu}},\ and\
		\bibinfo {author} {\bibfnamefont {E.~J.}\ \bibnamefont {Bergholtz}},\
	}\href@noop {} {\bibinfo {title} {{Non-Abelian Fractional Chern Insulators
				and Competing States in Flat Moir\'e Bands}}} (\bibinfo {year} {2024}),\
	\Eprint {https://arxiv.org/abs/2405.08887} {arXiv:2405.08887} \BibitemShut
	{NoStop}%
	\bibitem [{\citenamefont {Shen}\ \emph {et~al.}(2024)\citenamefont {Shen},
		\citenamefont {Wang}, \citenamefont {Guo}, \citenamefont {Xu}, \citenamefont
		{Duan},\ and\ \citenamefont {Xu}}]{Shen2024}%
	\BibitemOpen
	\bibfield  {author} {\bibinfo {author} {\bibfnamefont {X.}~\bibnamefont
			{Shen}}, \bibinfo {author} {\bibfnamefont {C.}~\bibnamefont {Wang}}, \bibinfo
		{author} {\bibfnamefont {R.}~\bibnamefont {Guo}}, \bibinfo {author}
		{\bibfnamefont {Z.}~\bibnamefont {Xu}}, \bibinfo {author} {\bibfnamefont
			{W.}~\bibnamefont {Duan}},\ and\ \bibinfo {author} {\bibfnamefont
			{Y.}~\bibnamefont {Xu}},\ }\href@noop {} {\bibinfo {title} {Stabilizing
			fractional chern insulators via exchange interaction in moir\'e systems}}
	(\bibinfo {year} {2024}),\ \Eprint {https://arxiv.org/abs/2405.12294}
	{arXiv:2405.12294} \BibitemShut {NoStop}%
	\bibitem [{\citenamefont {Reddy}\ and\ \citenamefont {Fu}(2023)}]{Reddy2023}%
	\BibitemOpen
	\bibfield  {author} {\bibinfo {author} {\bibfnamefont {A.~P.}\ \bibnamefont
			{Reddy}}\ and\ \bibinfo {author} {\bibfnamefont {L.}~\bibnamefont {Fu}},\
	}\bibfield  {title} {\bibinfo {title} {Toward a global phase diagram of the
			fractional quantum anomalous hall effect},\ }\href
	{https://doi.org/10.1103/PhysRevB.108.245159} {\bibfield  {journal} {\bibinfo
			{journal} {Phys. Rev. B}\ }\textbf {\bibinfo {volume} {108}},\ \bibinfo
		{pages} {245159} (\bibinfo {year} {2023})}\BibitemShut {NoStop}%
	\bibitem [{\citenamefont {Dong}\ \emph
		{et~al.}(2023{\natexlab{b}})\citenamefont {Dong}, \citenamefont {Wang},
		\citenamefont {Ledwith}, \citenamefont {Vishwanath},\ and\ \citenamefont
		{Parker}}]{Dong2023a}%
	\BibitemOpen
	\bibfield  {author} {\bibinfo {author} {\bibfnamefont {J.}~\bibnamefont
			{Dong}}, \bibinfo {author} {\bibfnamefont {J.}~\bibnamefont {Wang}}, \bibinfo
		{author} {\bibfnamefont {P.~J.}\ \bibnamefont {Ledwith}}, \bibinfo {author}
		{\bibfnamefont {A.}~\bibnamefont {Vishwanath}},\ and\ \bibinfo {author}
		{\bibfnamefont {D.~E.}\ \bibnamefont {Parker}},\ }\bibfield  {title}
	{\bibinfo {title} {Composite fermi liquid at zero magnetic field in twisted
			${\mathrm{mote}}_{2}$},\ }\href
	{https://doi.org/10.1103/PhysRevLett.131.136502} {\bibfield  {journal}
		{\bibinfo  {journal} {Phys. Rev. Lett.}\ }\textbf {\bibinfo {volume} {131}},\
		\bibinfo {pages} {136502} (\bibinfo {year} {2023}{\natexlab{b}})}\BibitemShut
	{NoStop}%
	\bibitem [{\citenamefont {Xie}\ \emph {et~al.}(2021)\citenamefont {Xie},
		\citenamefont {Pierce}, \citenamefont {Park}, \citenamefont {Parker},
		\citenamefont {Khalaf}, \citenamefont {Ledwith}, \citenamefont {Cao},
		\citenamefont {Lee}, \citenamefont {Chen}, \citenamefont {Forrester},
		\citenamefont {Watanabe}, \citenamefont {Taniguchi}, \citenamefont
		{Vishwanath}, \citenamefont {Jarillo-Herrero},\ and\ \citenamefont
		{Yacoby}}]{Xie2021}%
	\BibitemOpen
	\bibfield  {author} {\bibinfo {author} {\bibfnamefont {Y.}~\bibnamefont
			{Xie}}, \bibinfo {author} {\bibfnamefont {A.~T.}\ \bibnamefont {Pierce}},
		\bibinfo {author} {\bibfnamefont {J.~M.}\ \bibnamefont {Park}}, \bibinfo
		{author} {\bibfnamefont {D.~E.}\ \bibnamefont {Parker}}, \bibinfo {author}
		{\bibfnamefont {E.}~\bibnamefont {Khalaf}}, \bibinfo {author} {\bibfnamefont
			{P.}~\bibnamefont {Ledwith}}, \bibinfo {author} {\bibfnamefont
			{Y.}~\bibnamefont {Cao}}, \bibinfo {author} {\bibfnamefont {S.~H.}\
			\bibnamefont {Lee}}, \bibinfo {author} {\bibfnamefont {S.}~\bibnamefont
			{Chen}}, \bibinfo {author} {\bibfnamefont {P.~R.}\ \bibnamefont {Forrester}},
		\bibinfo {author} {\bibfnamefont {K.}~\bibnamefont {Watanabe}}, \bibinfo
		{author} {\bibfnamefont {T.}~\bibnamefont {Taniguchi}}, \bibinfo {author}
		{\bibfnamefont {A.}~\bibnamefont {Vishwanath}}, \bibinfo {author}
		{\bibfnamefont {P.}~\bibnamefont {Jarillo-Herrero}},\ and\ \bibinfo {author}
		{\bibfnamefont {A.}~\bibnamefont {Yacoby}},\ }\bibfield  {title} {\bibinfo
		{title} {{Fractional Chern insulators in magic-angle twisted bilayer
				graphene}},\ }\href {https://doi.org/10.1038/s41586-021-04002-3} {\bibfield
		{journal} {\bibinfo  {journal} {Nature}\ }\textbf {\bibinfo {volume} {600}},\
		\bibinfo {pages} {439} (\bibinfo {year} {2021})}\BibitemShut {NoStop}%
	\bibitem [{\citenamefont {Lu}\ \emph {et~al.}(2024)\citenamefont {Lu},
		\citenamefont {Han}, \citenamefont {Yao}, \citenamefont {Reddy},
		\citenamefont {Yang}, \citenamefont {Seo}, \citenamefont {Watanabe},
		\citenamefont {Taniguchi}, \citenamefont {Fu},\ and\ \citenamefont
		{Ju}}]{Lu2024}%
	\BibitemOpen
	\bibfield  {author} {\bibinfo {author} {\bibfnamefont {Z.}~\bibnamefont
			{Lu}}, \bibinfo {author} {\bibfnamefont {T.}~\bibnamefont {Han}}, \bibinfo
		{author} {\bibfnamefont {Y.}~\bibnamefont {Yao}}, \bibinfo {author}
		{\bibfnamefont {A.~P.}\ \bibnamefont {Reddy}}, \bibinfo {author}
		{\bibfnamefont {J.}~\bibnamefont {Yang}}, \bibinfo {author} {\bibfnamefont
			{J.}~\bibnamefont {Seo}}, \bibinfo {author} {\bibfnamefont {K.}~\bibnamefont
			{Watanabe}}, \bibinfo {author} {\bibfnamefont {T.}~\bibnamefont {Taniguchi}},
		\bibinfo {author} {\bibfnamefont {L.}~\bibnamefont {Fu}},\ and\ \bibinfo
		{author} {\bibfnamefont {L.}~\bibnamefont {Ju}},\ }\bibfield  {title}
	{\bibinfo {title} {{Fractional quantum anomalous Hall effect in multilayer
				graphene}},\ }\href {https://doi.org/10.1038/s41586-023-07010-7} {\bibfield
		{journal} {\bibinfo  {journal} {Nature}\ }\textbf {\bibinfo {volume} {626}},\
		\bibinfo {pages} {759} (\bibinfo {year} {2024})}\BibitemShut {NoStop}%
	\bibitem [{\citenamefont {Xie}\ \emph {et~al.}(2024)\citenamefont {Xie},
		\citenamefont {Huo}, \citenamefont {Lu}, \citenamefont {Feng}, \citenamefont
		{Zhang}, \citenamefont {Wang}, \citenamefont {Yang}, \citenamefont
		{Watanabe}, \citenamefont {Taniguchi}, \citenamefont {Liu}, \citenamefont
		{Song}, \citenamefont {Xie}, \citenamefont {Liu},\ and\ \citenamefont
		{Lu}}]{Xie2024}%
	\BibitemOpen
	\bibfield  {author} {\bibinfo {author} {\bibfnamefont {J.}~\bibnamefont
			{Xie}}, \bibinfo {author} {\bibfnamefont {Z.}~\bibnamefont {Huo}}, \bibinfo
		{author} {\bibfnamefont {X.}~\bibnamefont {Lu}}, \bibinfo {author}
		{\bibfnamefont {Z.}~\bibnamefont {Feng}}, \bibinfo {author} {\bibfnamefont
			{Z.}~\bibnamefont {Zhang}}, \bibinfo {author} {\bibfnamefont
			{W.}~\bibnamefont {Wang}}, \bibinfo {author} {\bibfnamefont {Q.}~\bibnamefont
			{Yang}}, \bibinfo {author} {\bibfnamefont {K.}~\bibnamefont {Watanabe}},
		\bibinfo {author} {\bibfnamefont {T.}~\bibnamefont {Taniguchi}}, \bibinfo
		{author} {\bibfnamefont {K.}~\bibnamefont {Liu}}, \bibinfo {author}
		{\bibfnamefont {Z.}~\bibnamefont {Song}}, \bibinfo {author} {\bibfnamefont
			{X.~C.}\ \bibnamefont {Xie}}, \bibinfo {author} {\bibfnamefont
			{J.}~\bibnamefont {Liu}},\ and\ \bibinfo {author} {\bibfnamefont
			{X.}~\bibnamefont {Lu}},\ }\href@noop {} {\bibinfo {title} {Even- and
			odd-denominator fractional quantum anomalous hall effect in graphene moire
			superlattices}} (\bibinfo {year} {2024}),\ \Eprint
	{https://arxiv.org/abs/2405.16944} {arXiv:2405.16944} \BibitemShut {NoStop}%
	\bibitem [{\citenamefont {Cai}\ \emph {et~al.}(2023)\citenamefont {Cai},
		\citenamefont {Anderson}, \citenamefont {Wang}, \citenamefont {Zhang},
		\citenamefont {Liu}, \citenamefont {Holtzmann}, \citenamefont {Zhang},
		\citenamefont {Fan}, \citenamefont {Taniguchi}, \citenamefont {Watanabe},
		\citenamefont {Ran}, \citenamefont {Cao}, \citenamefont {Fu}, \citenamefont
		{Xiao}, \citenamefont {Yao},\ and\ \citenamefont {Xu}}]{Cai2023}%
	\BibitemOpen
	\bibfield  {author} {\bibinfo {author} {\bibfnamefont {J.}~\bibnamefont
			{Cai}}, \bibinfo {author} {\bibfnamefont {E.}~\bibnamefont {Anderson}},
		\bibinfo {author} {\bibfnamefont {C.}~\bibnamefont {Wang}}, \bibinfo {author}
		{\bibfnamefont {X.}~\bibnamefont {Zhang}}, \bibinfo {author} {\bibfnamefont
			{X.}~\bibnamefont {Liu}}, \bibinfo {author} {\bibfnamefont {W.}~\bibnamefont
			{Holtzmann}}, \bibinfo {author} {\bibfnamefont {Y.}~\bibnamefont {Zhang}},
		\bibinfo {author} {\bibfnamefont {F.}~\bibnamefont {Fan}}, \bibinfo {author}
		{\bibfnamefont {T.}~\bibnamefont {Taniguchi}}, \bibinfo {author}
		{\bibfnamefont {K.}~\bibnamefont {Watanabe}}, \bibinfo {author}
		{\bibfnamefont {Y.}~\bibnamefont {Ran}}, \bibinfo {author} {\bibfnamefont
			{T.}~\bibnamefont {Cao}}, \bibinfo {author} {\bibfnamefont {L.}~\bibnamefont
			{Fu}}, \bibinfo {author} {\bibfnamefont {D.}~\bibnamefont {Xiao}}, \bibinfo
		{author} {\bibfnamefont {W.}~\bibnamefont {Yao}},\ and\ \bibinfo {author}
		{\bibfnamefont {X.}~\bibnamefont {Xu}},\ }\bibfield  {title} {\bibinfo
		{title} {{Signatures of fractional quantum anomalous Hall states in twisted
				MoTe2}},\ }\href {https://doi.org/10.1038/s41586-023-06289-w} {\bibfield
		{journal} {\bibinfo  {journal} {Nature}\ }\textbf {\bibinfo {volume} {622}},\
		\bibinfo {pages} {63} (\bibinfo {year} {2023})}\BibitemShut {NoStop}%
	\bibitem [{\citenamefont {Zeng}\ \emph {et~al.}(2023)\citenamefont {Zeng},
		\citenamefont {Xia}, \citenamefont {Kang}, \citenamefont {Zhu}, \citenamefont
		{Kn{\ifmmode\ddot{u}\else\"{u}\fi}ppel}, \citenamefont {Vaswani},
		\citenamefont {Watanabe}, \citenamefont {Taniguchi}, \citenamefont {Mak},\
		and\ \citenamefont {Shan}}]{Zeng2023}%
	\BibitemOpen
	\bibfield  {author} {\bibinfo {author} {\bibfnamefont {Y.}~\bibnamefont
			{Zeng}}, \bibinfo {author} {\bibfnamefont {Z.}~\bibnamefont {Xia}}, \bibinfo
		{author} {\bibfnamefont {K.}~\bibnamefont {Kang}}, \bibinfo {author}
		{\bibfnamefont {J.}~\bibnamefont {Zhu}}, \bibinfo {author} {\bibfnamefont
			{P.}~\bibnamefont {Kn{\ifmmode\ddot{u}\else\"{u}\fi}ppel}}, \bibinfo {author}
		{\bibfnamefont {C.}~\bibnamefont {Vaswani}}, \bibinfo {author} {\bibfnamefont
			{K.}~\bibnamefont {Watanabe}}, \bibinfo {author} {\bibfnamefont
			{T.}~\bibnamefont {Taniguchi}}, \bibinfo {author} {\bibfnamefont {K.~F.}\
			\bibnamefont {Mak}},\ and\ \bibinfo {author} {\bibfnamefont {J.}~\bibnamefont
			{Shan}},\ }\bibfield  {title} {\bibinfo {title} {{Thermodynamic evidence of
				fractional Chern insulator in moir{\ifmmode\acute{e}\else\'{e}\fi} MoTe2}},\
	}\href {https://doi.org/10.1038/s41586-023-06452-3} {\bibfield  {journal}
		{\bibinfo  {journal} {Nature}\ }\textbf {\bibinfo {volume} {622}},\ \bibinfo
		{pages} {69} (\bibinfo {year} {2023})}\BibitemShut {NoStop}%
	\bibitem [{\citenamefont {Park}\ \emph {et~al.}(2023)\citenamefont {Park},
		\citenamefont {Cai}, \citenamefont {Anderson}, \citenamefont {Zhang},
		\citenamefont {Zhu}, \citenamefont {Liu}, \citenamefont {Wang}, \citenamefont
		{Holtzmann}, \citenamefont {Hu}, \citenamefont {Liu}, \citenamefont
		{Taniguchi}, \citenamefont {Watanabe}, \citenamefont {Chu}, \citenamefont
		{Cao}, \citenamefont {Fu}, \citenamefont {Yao}, \citenamefont {Chang},
		\citenamefont {Cobden}, \citenamefont {Xiao},\ and\ \citenamefont
		{Xu}}]{Park2023}%
	\BibitemOpen
	\bibfield  {author} {\bibinfo {author} {\bibfnamefont {H.}~\bibnamefont
			{Park}}, \bibinfo {author} {\bibfnamefont {J.}~\bibnamefont {Cai}}, \bibinfo
		{author} {\bibfnamefont {E.}~\bibnamefont {Anderson}}, \bibinfo {author}
		{\bibfnamefont {Y.}~\bibnamefont {Zhang}}, \bibinfo {author} {\bibfnamefont
			{J.}~\bibnamefont {Zhu}}, \bibinfo {author} {\bibfnamefont {X.}~\bibnamefont
			{Liu}}, \bibinfo {author} {\bibfnamefont {C.}~\bibnamefont {Wang}}, \bibinfo
		{author} {\bibfnamefont {W.}~\bibnamefont {Holtzmann}}, \bibinfo {author}
		{\bibfnamefont {C.}~\bibnamefont {Hu}}, \bibinfo {author} {\bibfnamefont
			{Z.}~\bibnamefont {Liu}}, \bibinfo {author} {\bibfnamefont {T.}~\bibnamefont
			{Taniguchi}}, \bibinfo {author} {\bibfnamefont {K.}~\bibnamefont {Watanabe}},
		\bibinfo {author} {\bibfnamefont {J.-H.}\ \bibnamefont {Chu}}, \bibinfo
		{author} {\bibfnamefont {T.}~\bibnamefont {Cao}}, \bibinfo {author}
		{\bibfnamefont {L.}~\bibnamefont {Fu}}, \bibinfo {author} {\bibfnamefont
			{W.}~\bibnamefont {Yao}}, \bibinfo {author} {\bibfnamefont {C.-Z.}\
			\bibnamefont {Chang}}, \bibinfo {author} {\bibfnamefont {D.}~\bibnamefont
			{Cobden}}, \bibinfo {author} {\bibfnamefont {D.}~\bibnamefont {Xiao}},\ and\
		\bibinfo {author} {\bibfnamefont {X.}~\bibnamefont {Xu}},\ }\bibfield
	{title} {\bibinfo {title} {{Observation of fractionally quantized anomalous
				Hall effect}},\ }\href {https://doi.org/10.1038/s41586-023-06536-0}
	{\bibfield  {journal} {\bibinfo  {journal} {Nature}\ }\textbf {\bibinfo
			{volume} {622}},\ \bibinfo {pages} {74} (\bibinfo {year} {2023})}\BibitemShut
	{NoStop}%
	\bibitem [{\citenamefont {Xu}\ \emph {et~al.}(2023)\citenamefont {Xu},
		\citenamefont {Sun}, \citenamefont {Jia}, \citenamefont {Liu}, \citenamefont
		{Xu}, \citenamefont {Li}, \citenamefont {Gu}, \citenamefont {Watanabe},
		\citenamefont {Taniguchi}, \citenamefont {Tong}, \citenamefont {Jia},
		\citenamefont {Shi}, \citenamefont {Jiang}, \citenamefont {Zhang},
		\citenamefont {Liu},\ and\ \citenamefont {Li}}]{Xu2023}%
	\BibitemOpen
	\bibfield  {author} {\bibinfo {author} {\bibfnamefont {F.}~\bibnamefont
			{Xu}}, \bibinfo {author} {\bibfnamefont {Z.}~\bibnamefont {Sun}}, \bibinfo
		{author} {\bibfnamefont {T.}~\bibnamefont {Jia}}, \bibinfo {author}
		{\bibfnamefont {C.}~\bibnamefont {Liu}}, \bibinfo {author} {\bibfnamefont
			{C.}~\bibnamefont {Xu}}, \bibinfo {author} {\bibfnamefont {C.}~\bibnamefont
			{Li}}, \bibinfo {author} {\bibfnamefont {Y.}~\bibnamefont {Gu}}, \bibinfo
		{author} {\bibfnamefont {K.}~\bibnamefont {Watanabe}}, \bibinfo {author}
		{\bibfnamefont {T.}~\bibnamefont {Taniguchi}}, \bibinfo {author}
		{\bibfnamefont {B.}~\bibnamefont {Tong}}, \bibinfo {author} {\bibfnamefont
			{J.}~\bibnamefont {Jia}}, \bibinfo {author} {\bibfnamefont {Z.}~\bibnamefont
			{Shi}}, \bibinfo {author} {\bibfnamefont {S.}~\bibnamefont {Jiang}}, \bibinfo
		{author} {\bibfnamefont {Y.}~\bibnamefont {Zhang}}, \bibinfo {author}
		{\bibfnamefont {X.}~\bibnamefont {Liu}},\ and\ \bibinfo {author}
		{\bibfnamefont {T.}~\bibnamefont {Li}},\ }\bibfield  {title} {\bibinfo
		{title} {{Observation of Integer and Fractional Quantum Anomalous Hall
				Effects in Twisted Bilayer ${\mathrm{MoTe}}_{2}$}},\ }\href
	{https://doi.org/10.1103/PhysRevX.13.031037} {\bibfield  {journal} {\bibinfo
			{journal} {Phys. Rev. X}\ }\textbf {\bibinfo {volume} {13}},\ \bibinfo
		{pages} {031037} (\bibinfo {year} {2023})}\BibitemShut {NoStop}%
	\bibitem [{\citenamefont {Li}\ and\ \citenamefont {Wu}(2024)}]{Li2024}%
	\BibitemOpen
	\bibfield  {author} {\bibinfo {author} {\bibfnamefont {B.}~\bibnamefont
			{Li}}\ and\ \bibinfo {author} {\bibfnamefont {F.}~\bibnamefont {Wu}},\
	}\href@noop {} {\bibinfo {title} {Variational mapping of chern bands to
			landau levels: Application to fractional chern insulators in twisted
			mote$_2$}} (\bibinfo {year} {2024}),\ \Eprint
	{https://arxiv.org/abs/2405.20307} {arXiv:2405.20307} \BibitemShut {NoStop}%
	\bibitem [{\citenamefont {Yu}\ \emph {et~al.}(2024)\citenamefont {Yu},
		\citenamefont {Herzog-Arbeitman}, \citenamefont {Wang}, \citenamefont
		{Vafek}, \citenamefont {Bernevig},\ and\ \citenamefont {Regnault}}]{Yu2024}%
	\BibitemOpen
	\bibfield  {author} {\bibinfo {author} {\bibfnamefont {J.}~\bibnamefont
			{Yu}}, \bibinfo {author} {\bibfnamefont {J.}~\bibnamefont
			{Herzog-Arbeitman}}, \bibinfo {author} {\bibfnamefont {M.}~\bibnamefont
			{Wang}}, \bibinfo {author} {\bibfnamefont {O.}~\bibnamefont {Vafek}},
		\bibinfo {author} {\bibfnamefont {B.~A.}\ \bibnamefont {Bernevig}},\ and\
		\bibinfo {author} {\bibfnamefont {N.}~\bibnamefont {Regnault}},\ }\bibfield
	{title} {\bibinfo {title} {Fractional chern insulators versus nonmagnetic
			states in twisted bilayer ${\mathrm{mote}}_{2}$},\ }\href
	{https://doi.org/10.1103/PhysRevB.109.045147} {\bibfield  {journal} {\bibinfo
			{journal} {Phys. Rev. B}\ }\textbf {\bibinfo {volume} {109}},\ \bibinfo
		{pages} {045147} (\bibinfo {year} {2024})}\BibitemShut {NoStop}%
	\bibitem [{\citenamefont {Jia}\ \emph {et~al.}(2024)\citenamefont {Jia},
		\citenamefont {Yu}, \citenamefont {Liu}, \citenamefont {Herzog-Arbeitman},
		\citenamefont {Qi}, \citenamefont {Pi}, \citenamefont {Regnault},
		\citenamefont {Weng}, \citenamefont {Bernevig},\ and\ \citenamefont
		{Wu}}]{Jia2024}%
	\BibitemOpen
	\bibfield  {author} {\bibinfo {author} {\bibfnamefont {Y.}~\bibnamefont
			{Jia}}, \bibinfo {author} {\bibfnamefont {J.}~\bibnamefont {Yu}}, \bibinfo
		{author} {\bibfnamefont {J.}~\bibnamefont {Liu}}, \bibinfo {author}
		{\bibfnamefont {J.}~\bibnamefont {Herzog-Arbeitman}}, \bibinfo {author}
		{\bibfnamefont {Z.}~\bibnamefont {Qi}}, \bibinfo {author} {\bibfnamefont
			{H.}~\bibnamefont {Pi}}, \bibinfo {author} {\bibfnamefont {N.}~\bibnamefont
			{Regnault}}, \bibinfo {author} {\bibfnamefont {H.}~\bibnamefont {Weng}},
		\bibinfo {author} {\bibfnamefont {B.~A.}\ \bibnamefont {Bernevig}},\ and\
		\bibinfo {author} {\bibfnamefont {Q.}~\bibnamefont {Wu}},\ }\bibfield
	{title} {\bibinfo {title} {Moir\'e fractional chern insulators. i.
			first-principles calculations and continuum models of twisted bilayer
			${\mathrm{mote}}_{2}$},\ }\href {https://doi.org/10.1103/PhysRevB.109.205121}
	{\bibfield  {journal} {\bibinfo  {journal} {Phys. Rev. B}\ }\textbf {\bibinfo
			{volume} {109}},\ \bibinfo {pages} {205121} (\bibinfo {year}
		{2024})}\BibitemShut {NoStop}%
	\bibitem [{\citenamefont {Wang}\ \emph
		{et~al.}(2024{\natexlab{b}})\citenamefont {Wang}, \citenamefont {Zhang},
		\citenamefont {Liu}, \citenamefont {He}, \citenamefont {Xu}, \citenamefont
		{Ran}, \citenamefont {Cao},\ and\ \citenamefont {Xiao}}]{Wang2024a}%
	\BibitemOpen
	\bibfield  {author} {\bibinfo {author} {\bibfnamefont {C.}~\bibnamefont
			{Wang}}, \bibinfo {author} {\bibfnamefont {X.-W.}\ \bibnamefont {Zhang}},
		\bibinfo {author} {\bibfnamefont {X.}~\bibnamefont {Liu}}, \bibinfo {author}
		{\bibfnamefont {Y.}~\bibnamefont {He}}, \bibinfo {author} {\bibfnamefont
			{X.}~\bibnamefont {Xu}}, \bibinfo {author} {\bibfnamefont {Y.}~\bibnamefont
			{Ran}}, \bibinfo {author} {\bibfnamefont {T.}~\bibnamefont {Cao}},\ and\
		\bibinfo {author} {\bibfnamefont {D.}~\bibnamefont {Xiao}},\ }\bibfield
	{title} {\bibinfo {title} {Fractional chern insulator in twisted bilayer
			${\mathrm{mote}}_{2}$},\ }\href
	{https://doi.org/10.1103/PhysRevLett.132.036501} {\bibfield  {journal}
		{\bibinfo  {journal} {Phys. Rev. Lett.}\ }\textbf {\bibinfo {volume} {132}},\
		\bibinfo {pages} {036501} (\bibinfo {year} {2024}{\natexlab{b}})}\BibitemShut
	{NoStop}%
	\bibitem [{\citenamefont {Lu}\ and\ \citenamefont {Santos}(2024)}]{Lu2024a}%
	\BibitemOpen
	\bibfield  {author} {\bibinfo {author} {\bibfnamefont {T.}~\bibnamefont
			{Lu}}\ and\ \bibinfo {author} {\bibfnamefont {L.~H.}\ \bibnamefont
			{Santos}},\ }\href@noop {} {\bibinfo {title} {Fractional chern insulators in
			twisted bilayer mote$_2$: A composite fermion perspective}} (\bibinfo {year}
	{2024}),\ \Eprint {https://arxiv.org/abs/2406.03530} {arXiv:2406.03530}
	\BibitemShut {NoStop}%
	\bibitem [{\citenamefont {Cao}\ \emph {et~al.}(2018)\citenamefont {Cao},
		\citenamefont {Fatemi}, \citenamefont {Fang}, \citenamefont {Watanabe},
		\citenamefont {Taniguchi}, \citenamefont {Kaxiras},\ and\ \citenamefont
		{Jarillo-Herrero}}]{Cao2018}%
	\BibitemOpen
	\bibfield  {author} {\bibinfo {author} {\bibfnamefont {Y.}~\bibnamefont
			{Cao}}, \bibinfo {author} {\bibfnamefont {V.}~\bibnamefont {Fatemi}},
		\bibinfo {author} {\bibfnamefont {S.}~\bibnamefont {Fang}}, \bibinfo {author}
		{\bibfnamefont {K.}~\bibnamefont {Watanabe}}, \bibinfo {author}
		{\bibfnamefont {T.}~\bibnamefont {Taniguchi}}, \bibinfo {author}
		{\bibfnamefont {E.}~\bibnamefont {Kaxiras}},\ and\ \bibinfo {author}
		{\bibfnamefont {P.}~\bibnamefont {Jarillo-Herrero}},\ }\bibfield  {title}
	{\bibinfo {title} {{Unconventional superconductivity in magic-angle graphene
				superlattices}},\ }\href {https://doi.org/10.1038/nature26160} {\bibfield
		{journal} {\bibinfo  {journal} {Nature}\ }\textbf {\bibinfo {volume} {556}},\
		\bibinfo {pages} {43} (\bibinfo {year} {2018})}\BibitemShut {NoStop}%
	\bibitem [{\citenamefont {Lu}\ \emph {et~al.}(2019)\citenamefont {Lu},
		\citenamefont {Stepanov}, \citenamefont {Yang}, \citenamefont {Xie},
		\citenamefont {Aamir}, \citenamefont {Das}, \citenamefont {Urgell},
		\citenamefont {Watanabe}, \citenamefont {Taniguchi}, \citenamefont {Zhang},
		\citenamefont {Bachtold}, \citenamefont {MacDonald},\ and\ \citenamefont
		{Efetov}}]{Lu2019}%
	\BibitemOpen
	\bibfield  {author} {\bibinfo {author} {\bibfnamefont {X.}~\bibnamefont
			{Lu}}, \bibinfo {author} {\bibfnamefont {P.}~\bibnamefont {Stepanov}},
		\bibinfo {author} {\bibfnamefont {W.}~\bibnamefont {Yang}}, \bibinfo {author}
		{\bibfnamefont {M.}~\bibnamefont {Xie}}, \bibinfo {author} {\bibfnamefont
			{M.~A.}\ \bibnamefont {Aamir}}, \bibinfo {author} {\bibfnamefont
			{I.}~\bibnamefont {Das}}, \bibinfo {author} {\bibfnamefont {C.}~\bibnamefont
			{Urgell}}, \bibinfo {author} {\bibfnamefont {K.}~\bibnamefont {Watanabe}},
		\bibinfo {author} {\bibfnamefont {T.}~\bibnamefont {Taniguchi}}, \bibinfo
		{author} {\bibfnamefont {G.}~\bibnamefont {Zhang}}, \bibinfo {author}
		{\bibfnamefont {A.}~\bibnamefont {Bachtold}}, \bibinfo {author}
		{\bibfnamefont {A.~H.}\ \bibnamefont {MacDonald}},\ and\ \bibinfo {author}
		{\bibfnamefont {D.~K.}\ \bibnamefont {Efetov}},\ }\bibfield  {title}
	{\bibinfo {title} {{Superconductors, orbital magnets and correlated states in
				magic-angle bilayer graphene}},\ }\href
	{https://doi.org/10.1038/s41586-019-1695-0} {\bibfield  {journal} {\bibinfo
			{journal} {Nature}\ }\textbf {\bibinfo {volume} {574}},\ \bibinfo {pages}
		{653} (\bibinfo {year} {2019})}\BibitemShut {NoStop}%
	\bibitem [{\citenamefont {Yankowitz}\ \emph {et~al.}(2019)\citenamefont
		{Yankowitz}, \citenamefont {Chen}, \citenamefont {Polshyn}, \citenamefont
		{Zhang}, \citenamefont {Watanabe}, \citenamefont {Taniguchi}, \citenamefont
		{Graf}, \citenamefont {Young},\ and\ \citenamefont {Dean}}]{Yankowitz2019}%
	\BibitemOpen
	\bibfield  {author} {\bibinfo {author} {\bibfnamefont {M.}~\bibnamefont
			{Yankowitz}}, \bibinfo {author} {\bibfnamefont {S.}~\bibnamefont {Chen}},
		\bibinfo {author} {\bibfnamefont {H.}~\bibnamefont {Polshyn}}, \bibinfo
		{author} {\bibfnamefont {Y.}~\bibnamefont {Zhang}}, \bibinfo {author}
		{\bibfnamefont {K.}~\bibnamefont {Watanabe}}, \bibinfo {author}
		{\bibfnamefont {T.}~\bibnamefont {Taniguchi}}, \bibinfo {author}
		{\bibfnamefont {D.}~\bibnamefont {Graf}}, \bibinfo {author} {\bibfnamefont
			{A.~F.}\ \bibnamefont {Young}},\ and\ \bibinfo {author} {\bibfnamefont
			{C.~R.}\ \bibnamefont {Dean}},\ }\bibfield  {title} {\bibinfo {title} {Tuning
			superconductivity in twisted bilayer graphene},\ }\href
	{https://doi.org/10.1126/science.aav1910} {\bibfield  {journal} {\bibinfo
			{journal} {Science}\ }\textbf {\bibinfo {volume} {363}},\ \bibinfo {pages}
		{1059} (\bibinfo {year} {2019})}\BibitemShut {NoStop}%
	\bibitem [{\citenamefont {Arora}\ \emph {et~al.}(2020)\citenamefont {Arora},
		\citenamefont {Polski}, \citenamefont {Zhang}, \citenamefont {Thomson},
		\citenamefont {Choi}, \citenamefont {Kim}, \citenamefont {Lin}, \citenamefont
		{Wilson}, \citenamefont {Xu}, \citenamefont {Chu}, \citenamefont {Watanabe},
		\citenamefont {Taniguchi}, \citenamefont {Alicea},\ and\ \citenamefont
		{Nadj-Perge}}]{Arora2020}%
	\BibitemOpen
	\bibfield  {author} {\bibinfo {author} {\bibfnamefont {H.~S.}\ \bibnamefont
			{Arora}}, \bibinfo {author} {\bibfnamefont {R.}~\bibnamefont {Polski}},
		\bibinfo {author} {\bibfnamefont {Y.}~\bibnamefont {Zhang}}, \bibinfo
		{author} {\bibfnamefont {A.}~\bibnamefont {Thomson}}, \bibinfo {author}
		{\bibfnamefont {Y.}~\bibnamefont {Choi}}, \bibinfo {author} {\bibfnamefont
			{H.}~\bibnamefont {Kim}}, \bibinfo {author} {\bibfnamefont {Z.}~\bibnamefont
			{Lin}}, \bibinfo {author} {\bibfnamefont {I.~Z.}\ \bibnamefont {Wilson}},
		\bibinfo {author} {\bibfnamefont {X.}~\bibnamefont {Xu}}, \bibinfo {author}
		{\bibfnamefont {J.-H.}\ \bibnamefont {Chu}}, \bibinfo {author} {\bibfnamefont
			{K.}~\bibnamefont {Watanabe}}, \bibinfo {author} {\bibfnamefont
			{T.}~\bibnamefont {Taniguchi}}, \bibinfo {author} {\bibfnamefont
			{J.}~\bibnamefont {Alicea}},\ and\ \bibinfo {author} {\bibfnamefont
			{S.}~\bibnamefont {Nadj-Perge}},\ }\bibfield  {title} {\bibinfo {title}
		{{Superconductivity in metallic twisted bilayer graphene stabilized by
				WSe2}},\ }\href {https://doi.org/10.1038/s41586-020-2473-8} {\bibfield
		{journal} {\bibinfo  {journal} {Nature}\ }\textbf {\bibinfo {volume} {583}},\
		\bibinfo {pages} {379} (\bibinfo {year} {2020})}\BibitemShut {NoStop}%
	\bibitem [{\citenamefont {Park}\ \emph {et~al.}(2021)\citenamefont {Park},
		\citenamefont {Cao}, \citenamefont {Watanabe}, \citenamefont {Taniguchi},\
		and\ \citenamefont {Jarillo-Herrero}}]{Park2021}%
	\BibitemOpen
	\bibfield  {author} {\bibinfo {author} {\bibfnamefont {J.~M.}\ \bibnamefont
			{Park}}, \bibinfo {author} {\bibfnamefont {Y.}~\bibnamefont {Cao}}, \bibinfo
		{author} {\bibfnamefont {K.}~\bibnamefont {Watanabe}}, \bibinfo {author}
		{\bibfnamefont {T.}~\bibnamefont {Taniguchi}},\ and\ \bibinfo {author}
		{\bibfnamefont {P.}~\bibnamefont {Jarillo-Herrero}},\ }\bibfield  {title}
	{\bibinfo {title} {{Tunable strongly coupled superconductivity in magic-angle
				twisted trilayer graphene}},\ }\href
	{https://doi.org/10.1038/s41586-021-03192-0} {\bibfield  {journal} {\bibinfo
			{journal} {Nature}\ }\textbf {\bibinfo {volume} {590}},\ \bibinfo {pages}
		{249} (\bibinfo {year} {2021})}\BibitemShut {NoStop}%
	\bibitem [{\citenamefont {Hao}\ \emph {et~al.}(2021)\citenamefont {Hao},
		\citenamefont {Zimmerman}, \citenamefont {Ledwith}, \citenamefont {Khalaf},
		\citenamefont {Najafabadi}, \citenamefont {Watanabe}, \citenamefont
		{Taniguchi}, \citenamefont {Vishwanath},\ and\ \citenamefont
		{Kim}}]{Hao2021}%
	\BibitemOpen
	\bibfield  {author} {\bibinfo {author} {\bibfnamefont {Z.}~\bibnamefont
			{Hao}}, \bibinfo {author} {\bibfnamefont {A.~M.}\ \bibnamefont {Zimmerman}},
		\bibinfo {author} {\bibfnamefont {P.}~\bibnamefont {Ledwith}}, \bibinfo
		{author} {\bibfnamefont {E.}~\bibnamefont {Khalaf}}, \bibinfo {author}
		{\bibfnamefont {D.~H.}\ \bibnamefont {Najafabadi}}, \bibinfo {author}
		{\bibfnamefont {K.}~\bibnamefont {Watanabe}}, \bibinfo {author}
		{\bibfnamefont {T.}~\bibnamefont {Taniguchi}}, \bibinfo {author}
		{\bibfnamefont {A.}~\bibnamefont {Vishwanath}},\ and\ \bibinfo {author}
		{\bibfnamefont {P.}~\bibnamefont {Kim}},\ }\bibfield  {title} {\bibinfo
		{title} {Electric field–tunable superconductivity in alternating-twist
			magic-angle trilayer graphene},\ }\href
	{https://doi.org/10.1126/science.abg0399} {\bibfield  {journal} {\bibinfo
			{journal} {Science}\ }\textbf {\bibinfo {volume} {371}},\ \bibinfo {pages}
		{1133} (\bibinfo {year} {2021})}\BibitemShut {NoStop}%
	\bibitem [{\citenamefont {Oh}\ \emph {et~al.}(2021)\citenamefont {Oh},
		\citenamefont {Nuckolls}, \citenamefont {Wong}, \citenamefont {Lee},
		\citenamefont {Liu}, \citenamefont {Watanabe}, \citenamefont {Taniguchi},\
		and\ \citenamefont {Yazdani}}]{Oh2021}%
	\BibitemOpen
	\bibfield  {author} {\bibinfo {author} {\bibfnamefont {M.}~\bibnamefont
			{Oh}}, \bibinfo {author} {\bibfnamefont {K.~P.}\ \bibnamefont {Nuckolls}},
		\bibinfo {author} {\bibfnamefont {D.}~\bibnamefont {Wong}}, \bibinfo {author}
		{\bibfnamefont {R.~L.}\ \bibnamefont {Lee}}, \bibinfo {author} {\bibfnamefont
			{X.}~\bibnamefont {Liu}}, \bibinfo {author} {\bibfnamefont {K.}~\bibnamefont
			{Watanabe}}, \bibinfo {author} {\bibfnamefont {T.}~\bibnamefont
			{Taniguchi}},\ and\ \bibinfo {author} {\bibfnamefont {A.}~\bibnamefont
			{Yazdani}},\ }\bibfield  {title} {\bibinfo {title} {{Evidence for
				unconventional superconductivity in twisted bilayer graphene}},\ }\href
	{https://doi.org/10.1038/s41586-021-04121-x} {\bibfield  {journal} {\bibinfo
			{journal} {Nature}\ }\textbf {\bibinfo {volume} {600}},\ \bibinfo {pages}
		{240} (\bibinfo {year} {2021})}\BibitemShut {NoStop}%
	\bibitem [{\citenamefont {Liu}\ \emph {et~al.}(2022)\citenamefont {Liu},
		\citenamefont {Zhang}, \citenamefont {Watanabe}, \citenamefont {Taniguchi},\
		and\ \citenamefont {Li}}]{JIAsTrilayerScreening}%
	\BibitemOpen
	\bibfield  {author} {\bibinfo {author} {\bibfnamefont {X.}~\bibnamefont
			{Liu}}, \bibinfo {author} {\bibfnamefont {N.~J.}\ \bibnamefont {Zhang}},
		\bibinfo {author} {\bibfnamefont {K.}~\bibnamefont {Watanabe}}, \bibinfo
		{author} {\bibfnamefont {T.}~\bibnamefont {Taniguchi}},\ and\ \bibinfo
		{author} {\bibfnamefont {J.~I.~A.}\ \bibnamefont {Li}},\ }\bibfield  {title}
	{\bibinfo {title} {Isospin order in superconducting magic-angle twisted
			trilayer graphene},\ }\href {https://doi.org/10.1038/s41567-022-01515-0}
	{\bibfield  {journal} {\bibinfo  {journal} {Nature Physics}\ }\textbf
		{\bibinfo {volume} {18}},\ \bibinfo {pages} {522} (\bibinfo {year}
		{2022})}\BibitemShut {NoStop}%
	\bibitem [{\citenamefont {Kim}\ \emph {et~al.}(2022)\citenamefont {Kim},
		\citenamefont {Choi}, \citenamefont {Lewandowski}, \citenamefont {Thomson},
		\citenamefont {Zhang}, \citenamefont {Polski}, \citenamefont {Watanabe},
		\citenamefont {Taniguchi}, \citenamefont {Alicea},\ and\ \citenamefont
		{Nadj-Perge}}]{Kim2022}%
	\BibitemOpen
	\bibfield  {author} {\bibinfo {author} {\bibfnamefont {H.}~\bibnamefont
			{Kim}}, \bibinfo {author} {\bibfnamefont {Y.}~\bibnamefont {Choi}}, \bibinfo
		{author} {\bibfnamefont {C.}~\bibnamefont {Lewandowski}}, \bibinfo {author}
		{\bibfnamefont {A.}~\bibnamefont {Thomson}}, \bibinfo {author} {\bibfnamefont
			{Y.}~\bibnamefont {Zhang}}, \bibinfo {author} {\bibfnamefont
			{R.}~\bibnamefont {Polski}}, \bibinfo {author} {\bibfnamefont
			{K.}~\bibnamefont {Watanabe}}, \bibinfo {author} {\bibfnamefont
			{T.}~\bibnamefont {Taniguchi}}, \bibinfo {author} {\bibfnamefont
			{J.}~\bibnamefont {Alicea}},\ and\ \bibinfo {author} {\bibfnamefont
			{S.}~\bibnamefont {Nadj-Perge}},\ }\bibfield  {title} {\bibinfo {title}
		{{Evidence for unconventional superconductivity in twisted trilayer
				graphene}},\ }\href {https://doi.org/10.1038/s41586-022-04715-z} {\bibfield
		{journal} {\bibinfo  {journal} {Nature}\ }\textbf {\bibinfo {volume} {606}},\
		\bibinfo {pages} {494} (\bibinfo {year} {2022})}\BibitemShut {NoStop}%
	\bibitem [{\citenamefont {Lin}\ \emph {et~al.}(2022)\citenamefont {Lin},
		\citenamefont {Siriviboon}, \citenamefont {Scammell}, \citenamefont {Liu},
		\citenamefont {Rhodes}, \citenamefont {Watanabe}, \citenamefont {Taniguchi},
		\citenamefont {Hone}, \citenamefont {Scheurer},\ and\ \citenamefont
		{Li}}]{SCDiodesMoireExp}%
	\BibitemOpen
	\bibfield  {author} {\bibinfo {author} {\bibfnamefont {J.-X.}\ \bibnamefont
			{Lin}}, \bibinfo {author} {\bibfnamefont {P.}~\bibnamefont {Siriviboon}},
		\bibinfo {author} {\bibfnamefont {H.~D.}\ \bibnamefont {Scammell}}, \bibinfo
		{author} {\bibfnamefont {S.}~\bibnamefont {Liu}}, \bibinfo {author}
		{\bibfnamefont {D.}~\bibnamefont {Rhodes}}, \bibinfo {author} {\bibfnamefont
			{K.}~\bibnamefont {Watanabe}}, \bibinfo {author} {\bibfnamefont
			{T.}~\bibnamefont {Taniguchi}}, \bibinfo {author} {\bibfnamefont
			{J.}~\bibnamefont {Hone}}, \bibinfo {author} {\bibfnamefont {M.~S.}\
			\bibnamefont {Scheurer}},\ and\ \bibinfo {author} {\bibfnamefont {J.~I.~A.}\
			\bibnamefont {Li}},\ }\bibfield  {title} {\bibinfo {title} {Zero-field
			superconducting diode effect in small-twist-angle trilayer graphene},\ }\href
	{https://doi.org/10.1038/s41567-022-01700-1} {\bibfield  {journal} {\bibinfo
			{journal} {Nature Physics}\ }\textbf {\bibinfo {volume} {18}},\ \bibinfo
		{pages} {1221} (\bibinfo {year} {2022})}\BibitemShut {NoStop}%
	\bibitem [{\citenamefont {Xia}\ \emph {et~al.}(2024)\citenamefont {Xia},
		\citenamefont {Han}, \citenamefont {Watanabe}, \citenamefont {Taniguchi},
		\citenamefont {Shan},\ and\ \citenamefont {Mak}}]{Xia2024}%
	\BibitemOpen
	\bibfield  {author} {\bibinfo {author} {\bibfnamefont {Y.}~\bibnamefont
			{Xia}}, \bibinfo {author} {\bibfnamefont {Z.}~\bibnamefont {Han}}, \bibinfo
		{author} {\bibfnamefont {K.}~\bibnamefont {Watanabe}}, \bibinfo {author}
		{\bibfnamefont {T.}~\bibnamefont {Taniguchi}}, \bibinfo {author}
		{\bibfnamefont {J.}~\bibnamefont {Shan}},\ and\ \bibinfo {author}
		{\bibfnamefont {K.~F.}\ \bibnamefont {Mak}},\ }\href@noop {} {\bibinfo
		{title} {Unconventional superconductivity in twisted bilayer wse2}} (\bibinfo
	{year} {2024}),\ \Eprint {https://arxiv.org/abs/2405.14784}
	{arXiv:2405.14784} \BibitemShut {NoStop}%
	\bibitem [{\citenamefont {{Guo}}\ \emph {et~al.}(2024)\citenamefont {{Guo}},
		\citenamefont {{Pack}}, \citenamefont {{Swann}}, \citenamefont {{Holtzman}},
		\citenamefont {{Cothrine}}, \citenamefont {{Watanabe}}, \citenamefont
		{{Taniguchi}}, \citenamefont {{Mandrus}}, \citenamefont {{Barmak}},
		\citenamefont {{Hone}}, \citenamefont {{Millis}}, \citenamefont
		{{Pasupathy}},\ and\ \citenamefont {{Dean}}}]{WSe2Experiment2}%
	\BibitemOpen
	\bibfield  {author} {\bibinfo {author} {\bibfnamefont {Y.}~\bibnamefont
			{{Guo}}}, \bibinfo {author} {\bibfnamefont {J.}~\bibnamefont {{Pack}}},
		\bibinfo {author} {\bibfnamefont {J.}~\bibnamefont {{Swann}}}, \bibinfo
		{author} {\bibfnamefont {L.}~\bibnamefont {{Holtzman}}}, \bibinfo {author}
		{\bibfnamefont {M.}~\bibnamefont {{Cothrine}}}, \bibinfo {author}
		{\bibfnamefont {K.}~\bibnamefont {{Watanabe}}}, \bibinfo {author}
		{\bibfnamefont {T.}~\bibnamefont {{Taniguchi}}}, \bibinfo {author}
		{\bibfnamefont {D.}~\bibnamefont {{Mandrus}}}, \bibinfo {author}
		{\bibfnamefont {K.}~\bibnamefont {{Barmak}}}, \bibinfo {author}
		{\bibfnamefont {J.}~\bibnamefont {{Hone}}}, \bibinfo {author} {\bibfnamefont
			{A.~J.}\ \bibnamefont {{Millis}}}, \bibinfo {author} {\bibfnamefont {A.~N.}\
			\bibnamefont {{Pasupathy}}},\ and\ \bibinfo {author} {\bibfnamefont {C.~R.}\
			\bibnamefont {{Dean}}},\ }\bibfield  {title} {\bibinfo {title}
		{{Superconductivity in twisted bilayer WSe$_2$}},\ }\href@noop {} {\bibfield
		{journal} {\bibinfo  {journal} {arXiv e-prints}\ } (\bibinfo {year}
		{2024})},\ \Eprint {https://arxiv.org/abs/2406.03418} {arXiv:2406.03418}
	\BibitemShut {NoStop}%
	\bibitem [{\citenamefont {Zhou}\ \emph {et~al.}(2021)\citenamefont {Zhou},
		\citenamefont {Xie}, \citenamefont {Taniguchi}, \citenamefont {Watanabe},\
		and\ \citenamefont {Young}}]{rhombohedralgrapheneSC}%
	\BibitemOpen
	\bibfield  {author} {\bibinfo {author} {\bibfnamefont {H.}~\bibnamefont
			{Zhou}}, \bibinfo {author} {\bibfnamefont {T.}~\bibnamefont {Xie}}, \bibinfo
		{author} {\bibfnamefont {T.}~\bibnamefont {Taniguchi}}, \bibinfo {author}
		{\bibfnamefont {K.}~\bibnamefont {Watanabe}},\ and\ \bibinfo {author}
		{\bibfnamefont {A.~F.}\ \bibnamefont {Young}},\ }\bibfield  {title} {\bibinfo
		{title} {Superconductivity in rhombohedral trilayer graphene},\ }\href
	{https://doi.org/10.1038/s41586-021-03926-0} {\bibfield  {journal} {\bibinfo
			{journal} {Nature}\ }\textbf {\bibinfo {volume} {598}},\ \bibinfo {pages}
		{434} (\bibinfo {year} {2021})}\BibitemShut {NoStop}%
	\bibitem [{\citenamefont {Han}\ \emph {et~al.}(2024{\natexlab{a}})\citenamefont
		{Han}, \citenamefont {Lu}, \citenamefont {Yao}, \citenamefont {Shi},
		\citenamefont {Yang}, \citenamefont {Seo}, \citenamefont {Ye}, \citenamefont
		{Wu}, \citenamefont {Zhou}, \citenamefont {Liu}, \citenamefont {Shi},
		\citenamefont {Hua}, \citenamefont {Watanabe}, \citenamefont {Taniguchi},
		\citenamefont {Xiong}, \citenamefont {Fu},\ and\ \citenamefont
		{Ju}}]{Han2024a}%
	\BibitemOpen
	\bibfield  {author} {\bibinfo {author} {\bibfnamefont {T.}~\bibnamefont
			{Han}}, \bibinfo {author} {\bibfnamefont {Z.}~\bibnamefont {Lu}}, \bibinfo
		{author} {\bibfnamefont {Y.}~\bibnamefont {Yao}}, \bibinfo {author}
		{\bibfnamefont {L.}~\bibnamefont {Shi}}, \bibinfo {author} {\bibfnamefont
			{J.}~\bibnamefont {Yang}}, \bibinfo {author} {\bibfnamefont {J.}~\bibnamefont
			{Seo}}, \bibinfo {author} {\bibfnamefont {S.}~\bibnamefont {Ye}}, \bibinfo
		{author} {\bibfnamefont {Z.}~\bibnamefont {Wu}}, \bibinfo {author}
		{\bibfnamefont {M.}~\bibnamefont {Zhou}}, \bibinfo {author} {\bibfnamefont
			{H.}~\bibnamefont {Liu}}, \bibinfo {author} {\bibfnamefont {G.}~\bibnamefont
			{Shi}}, \bibinfo {author} {\bibfnamefont {Z.}~\bibnamefont {Hua}}, \bibinfo
		{author} {\bibfnamefont {K.}~\bibnamefont {Watanabe}}, \bibinfo {author}
		{\bibfnamefont {T.}~\bibnamefont {Taniguchi}}, \bibinfo {author}
		{\bibfnamefont {P.}~\bibnamefont {Xiong}}, \bibinfo {author} {\bibfnamefont
			{L.}~\bibnamefont {Fu}},\ and\ \bibinfo {author} {\bibfnamefont
			{L.}~\bibnamefont {Ju}},\ }\href@noop {} {\bibinfo {title} {Signatures of
			chiral superconductivity in rhombohedral graphene}} (\bibinfo {year}
	{2024}{\natexlab{a}}),\ \Eprint {https://arxiv.org/abs/2408.15233}
	{arXiv:2408.15233} \BibitemShut {NoStop}%
	\bibitem [{\citenamefont {Han}\ \emph {et~al.}(2024{\natexlab{b}})\citenamefont
		{Han}, \citenamefont {Herzog-Arbeitman}, \citenamefont {Bernevig},\ and\
		\citenamefont {Kivelson}}]{Han2024}%
	\BibitemOpen
	\bibfield  {author} {\bibinfo {author} {\bibfnamefont {Z.}~\bibnamefont
			{Han}}, \bibinfo {author} {\bibfnamefont {J.}~\bibnamefont
			{Herzog-Arbeitman}}, \bibinfo {author} {\bibfnamefont {B.~A.}\ \bibnamefont
			{Bernevig}},\ and\ \bibinfo {author} {\bibfnamefont {S.~A.}\ \bibnamefont
			{Kivelson}},\ }\href@noop {} {\bibinfo {title} {"quantum geometric nesting''
			and solvable model flat-band systems}} (\bibinfo {year}
	{2024}{\natexlab{b}}),\ \Eprint {https://arxiv.org/abs/2401.04163}
	{arXiv:2401.04163} \BibitemShut {NoStop}%
	\bibitem [{\citenamefont {Scammell}\ \emph {et~al.}(2022)\citenamefont
		{Scammell}, \citenamefont {Li},\ and\ \citenamefont
		{Scheurer}}]{Scammell_2022}%
	\BibitemOpen
	\bibfield  {author} {\bibinfo {author} {\bibfnamefont {H.~D.}\ \bibnamefont
			{Scammell}}, \bibinfo {author} {\bibfnamefont {J.~I.~A.}\ \bibnamefont
			{Li}},\ and\ \bibinfo {author} {\bibfnamefont {M.~S.}\ \bibnamefont
			{Scheurer}},\ }\bibfield  {title} {\bibinfo {title} {Theory of zero-field
			superconducting diode effect in twisted trilayer graphene},\ }\href
	{https://doi.org/10.1088/2053-1583/ac5b16} {\bibfield  {journal} {\bibinfo
			{journal} {2D Materials}\ }\textbf {\bibinfo {volume} {9}},\ \bibinfo {pages}
		{025027} (\bibinfo {year} {2022})}\BibitemShut {NoStop}%
	\bibitem [{\citenamefont {Bravyi}(2006)}]{Bravyi2006}%
	\BibitemOpen
	\bibfield  {author} {\bibinfo {author} {\bibfnamefont {S.}~\bibnamefont
			{Bravyi}},\ }\href@noop {} {\bibinfo {title} {{Efficient algorithm for a
				quantum analogue of 2-SAT}}} (\bibinfo {year} {2006}),\ \Eprint
	{https://arxiv.org/abs/quant-ph/0602108} {arXiv:quant-ph/0602108}
	\BibitemShut {NoStop}%
	\bibitem [{\citenamefont {Bravyi}\ \emph {et~al.}(2010)\citenamefont {Bravyi},
		\citenamefont {Moore},\ and\ \citenamefont {Russell}}]{Bravyi2010}%
	\BibitemOpen
	\bibfield  {author} {\bibinfo {author} {\bibfnamefont {S.}~\bibnamefont
			{Bravyi}}, \bibinfo {author} {\bibfnamefont {C.}~\bibnamefont {Moore}},\ and\
		\bibinfo {author} {\bibfnamefont {A.}~\bibnamefont {Russell}},\ }\href@noop
	{} {\emph {\bibinfo {title} {{Bounds on the Quantum Satisfiability
					Threshold}}}}\ (\bibinfo  {publisher} {Tsinghua University Press},\ \bibinfo
	{year} {2010})\BibitemShut {NoStop}%
	\bibitem [{\citenamefont {Laumann}\ \emph {et~al.}(2010)\citenamefont
		{Laumann}, \citenamefont {L\"auchli}, \citenamefont {Moessner}, \citenamefont
		{Scardicchio},\ and\ \citenamefont {Sondhi}}]{Laumann2010}%
	\BibitemOpen
	\bibfield  {author} {\bibinfo {author} {\bibfnamefont {C.~R.}\ \bibnamefont
			{Laumann}}, \bibinfo {author} {\bibfnamefont {A.~M.}\ \bibnamefont
			{L\"auchli}}, \bibinfo {author} {\bibfnamefont {R.}~\bibnamefont {Moessner}},
		\bibinfo {author} {\bibfnamefont {A.}~\bibnamefont {Scardicchio}},\ and\
		\bibinfo {author} {\bibfnamefont {S.~L.}\ \bibnamefont {Sondhi}},\ }\bibfield
	{title} {\bibinfo {title} {Product, generic, and random generic quantum
			satisfiability},\ }\href {https://doi.org/10.1103/PhysRevA.81.062345}
	{\bibfield  {journal} {\bibinfo  {journal} {Phys. Rev. A}\ }\textbf {\bibinfo
			{volume} {81}},\ \bibinfo {pages} {062345} (\bibinfo {year}
		{2010})}\BibitemShut {NoStop}%
	\bibitem [{\citenamefont {Sattath}\ \emph {et~al.}(2016)\citenamefont
		{Sattath}, \citenamefont {Morampudi}, \citenamefont {Laumann},\ and\
		\citenamefont {Moessner}}]{Sattath2016}%
	\BibitemOpen
	\bibfield  {author} {\bibinfo {author} {\bibfnamefont {O.}~\bibnamefont
			{Sattath}}, \bibinfo {author} {\bibfnamefont {S.~C.}\ \bibnamefont
			{Morampudi}}, \bibinfo {author} {\bibfnamefont {C.~R.}\ \bibnamefont
			{Laumann}},\ and\ \bibinfo {author} {\bibfnamefont {R.}~\bibnamefont
			{Moessner}},\ }\bibfield  {title} {\bibinfo {title} {{When a local
				Hamiltonian must be frustration-free}},\ }\href
	{https://doi.org/10.1073/pnas.1519833113} {\bibfield  {journal} {\bibinfo
			{journal} {Proc. Natl. Acad. Sci. U.S.A.}\ }\textbf {\bibinfo {volume}
			{113}},\ \bibinfo {pages} {6433} (\bibinfo {year} {2016})}\BibitemShut
	{NoStop}%
	\bibitem [{Note1()}]{Note1}%
	\BibitemOpen
	\bibinfo {note} {Although the excitation gap above $E=0$ in Fig.~\ref
		{fig:spectrum_filling} decreases until $\nu \simeq 0.8$, it does not appear
		to vanish for $\nu \rightarrow 1$. Still, our finite-size data is
		insufficient for definite statements about the regime just below $\nu
		=1$.}\BibitemShut {Stop}%
	\bibitem [{\citenamefont {Fendley}\ \emph {et~al.}(2003)\citenamefont
		{Fendley}, \citenamefont {Schoutens},\ and\ \citenamefont
		{de~Boer}}]{Fendley2003}%
	\BibitemOpen
	\bibfield  {author} {\bibinfo {author} {\bibfnamefont {P.}~\bibnamefont
			{Fendley}}, \bibinfo {author} {\bibfnamefont {K.}~\bibnamefont {Schoutens}},\
		and\ \bibinfo {author} {\bibfnamefont {J.}~\bibnamefont {de~Boer}},\
	}\bibfield  {title} {\bibinfo {title} {{Lattice Models with $\mathcal{N}=2$
				Supersymmetry}},\ }\href {https://doi.org/10.1103/PhysRevLett.90.120402}
	{\bibfield  {journal} {\bibinfo  {journal} {Phys. Rev. Lett.}\ }\textbf
		{\bibinfo {volume} {90}},\ \bibinfo {pages} {120402} (\bibinfo {year}
		{2003})}\BibitemShut {NoStop}%
	\bibitem [{\citenamefont {Huijse}\ \emph {et~al.}(2008)\citenamefont {Huijse},
		\citenamefont {Halverson}, \citenamefont {Fendley},\ and\ \citenamefont
		{Schoutens}}]{Huijse2008}%
	\BibitemOpen
	\bibfield  {author} {\bibinfo {author} {\bibfnamefont {L.}~\bibnamefont
			{Huijse}}, \bibinfo {author} {\bibfnamefont {J.}~\bibnamefont {Halverson}},
		\bibinfo {author} {\bibfnamefont {P.}~\bibnamefont {Fendley}},\ and\ \bibinfo
		{author} {\bibfnamefont {K.}~\bibnamefont {Schoutens}},\ }\bibfield  {title}
	{\bibinfo {title} {Charge frustration and quantum criticality for strongly
			correlated fermions},\ }\href
	{https://doi.org/10.1103/PhysRevLett.101.146406} {\bibfield  {journal}
		{\bibinfo  {journal} {Phys. Rev. Lett.}\ }\textbf {\bibinfo {volume} {101}},\
		\bibinfo {pages} {146406} (\bibinfo {year} {2008})}\BibitemShut {NoStop}%
	\bibitem [{\citenamefont {Huijse}\ \emph {et~al.}(2012)\citenamefont {Huijse},
		\citenamefont {Mehta}, \citenamefont {Moran}, \citenamefont {Schoutens},\
		and\ \citenamefont {Vala}}]{Huijse2012}%
	\BibitemOpen
	\bibfield  {author} {\bibinfo {author} {\bibfnamefont {L.}~\bibnamefont
			{Huijse}}, \bibinfo {author} {\bibfnamefont {D.}~\bibnamefont {Mehta}},
		\bibinfo {author} {\bibfnamefont {N.}~\bibnamefont {Moran}}, \bibinfo
		{author} {\bibfnamefont {K.}~\bibnamefont {Schoutens}},\ and\ \bibinfo
		{author} {\bibfnamefont {J.}~\bibnamefont {Vala}},\ }\bibfield  {title}
	{\bibinfo {title} {Supersymmetric lattice fermions on the triangular lattice:
			superfrustration and criticality},\ }\href
	{https://doi.org/10.1088/1367-2630/14/7/073002} {\bibfield  {journal}
		{\bibinfo  {journal} {New Journal of Physics}\ }\textbf {\bibinfo {volume}
			{14}},\ \bibinfo {pages} {073002} (\bibinfo {year} {2012})}\BibitemShut
	{NoStop}%
	\bibitem [{\citenamefont {Galanakis}\ \emph {et~al.}(2012)\citenamefont
		{Galanakis}, \citenamefont {Henley},\ and\ \citenamefont
		{Papanikolaou}}]{Galanakis2012}%
	\BibitemOpen
	\bibfield  {author} {\bibinfo {author} {\bibfnamefont {D.}~\bibnamefont
			{Galanakis}}, \bibinfo {author} {\bibfnamefont {C.~L.}\ \bibnamefont
			{Henley}},\ and\ \bibinfo {author} {\bibfnamefont {S.}~\bibnamefont
			{Papanikolaou}},\ }\bibfield  {title} {\bibinfo {title} {Order and
			supersymmetry at high filling zero-energy states on the triangular lattice},\
	}\href {https://doi.org/10.1103/PhysRevB.86.195105} {\bibfield  {journal}
		{\bibinfo  {journal} {Phys. Rev. B}\ }\textbf {\bibinfo {volume} {86}},\
		\bibinfo {pages} {195105} (\bibinfo {year} {2012})}\BibitemShut {NoStop}%
	\bibitem [{\citenamefont {Wilhelm}\ \emph
		{et~al.}(2023{\natexlab{b}})\citenamefont {Wilhelm}, \citenamefont {Kwan},
		\citenamefont {Läuchli},\ and\ \citenamefont {Parameswaran}}]{Wilhelm2023a}%
	\BibitemOpen
	\bibfield  {author} {\bibinfo {author} {\bibfnamefont {P.~H.}\ \bibnamefont
			{Wilhelm}}, \bibinfo {author} {\bibfnamefont {Y.~H.}\ \bibnamefont {Kwan}},
		\bibinfo {author} {\bibfnamefont {A.~M.}\ \bibnamefont {Läuchli}},\ and\
		\bibinfo {author} {\bibfnamefont {S.~A.}\ \bibnamefont {Parameswaran}},\
	}\href@noop {} {\bibinfo {title} {Supersymmetry on the honeycomb lattice:
			resonating charge stripes, superfrustration, and domain walls}} (\bibinfo
	{year} {2023}{\natexlab{b}}),\ \Eprint {https://arxiv.org/abs/2307.13031}
	{arXiv:2307.13031} \BibitemShut {NoStop}%
	\bibitem [{Note2()}]{Note2}%
	\BibitemOpen
	\bibinfo {note} {For simplicity of notation, we assume $S_j$ in $ \gamma
		^{\dagger }_{S_j,\protect \boldsymbol {\delta }}$ does not contain
		inversion-invariant momenta, as this would change the prefactor in the norm
		$1/\protect \sqrt {2} \rightarrow 1/2$.}\BibitemShut {Stop}%
	\bibitem [{\citenamefont {Anderson}(1967)}]{Anderson1967}%
	\BibitemOpen
	\bibfield  {author} {\bibinfo {author} {\bibfnamefont {P.~W.}\ \bibnamefont
			{Anderson}},\ }\bibfield  {title} {\bibinfo {title} {Infrared catastrophe in
			fermi gases with local scattering potentials},\ }\href
	{https://doi.org/10.1103/PhysRevLett.18.1049} {\bibfield  {journal} {\bibinfo
			{journal} {Phys. Rev. Lett.}\ }\textbf {\bibinfo {volume} {18}},\ \bibinfo
		{pages} {1049} (\bibinfo {year} {1967})}\BibitemShut {NoStop}%
	\bibitem [{Note3()}]{Note3}%
	\BibitemOpen
	\bibinfo {note} {Restricted to $N\leq 12$ due to the rapid increase in
		complexity of constructing $\protect \bar {\protect \mathcal
			{P}}$.}\BibitemShut {Stop}%
	\bibitem [{Note4()}]{Note4}%
	\BibitemOpen
	\bibinfo {note} {Due to the anticommutation of pairs of fermionic creation
		operators.}\BibitemShut {Stop}%
	\bibitem [{\citenamefont {Penrose}\ and\ \citenamefont
		{Onsager}(1956)}]{penroseBoseEinsteinCondensationLiquid1956}%
	\BibitemOpen
	\bibfield  {author} {\bibinfo {author} {\bibfnamefont {O.}~\bibnamefont
			{Penrose}}\ and\ \bibinfo {author} {\bibfnamefont {L.}~\bibnamefont
			{Onsager}},\ }\bibfield  {title} {\bibinfo {title} {Bose-{{Einstein
					Condensation}} and {{Liquid Helium}}},\ }\href
	{https://doi.org/10.1103/PhysRev.104.576} {\bibfield  {journal} {\bibinfo
			{journal} {Physical Review}\ }\textbf {\bibinfo {volume} {104}},\ \bibinfo
		{pages} {576} (\bibinfo {year} {1956})}\BibitemShut {NoStop}%
	\bibitem [{\citenamefont {Penrose}(1951)}]{penroseCXXXVIQuantumMechanics1951}%
	\BibitemOpen
	\bibfield  {author} {\bibinfo {author} {\bibfnamefont {O.}~\bibnamefont
			{Penrose}},\ }\bibfield  {title} {\bibinfo {title} {{{CXXXVI}}. {{On}} the
			quantum mechanics of helium {{II}}},\ }\href
	{https://doi.org/10.1080/14786445108560954} {\bibfield  {journal} {\bibinfo
			{journal} {The London, Edinburgh, and Dublin Philosophical Magazine and
				Journal of Science}\ }\textbf {\bibinfo {volume} {42}},\ \bibinfo {pages}
		{1373} (\bibinfo {year} {1951})}\BibitemShut {NoStop}%
	\bibitem [{\citenamefont {Yang}(1962)}]{yangConceptOffDiagonalLongRange1962}%
	\BibitemOpen
	\bibfield  {author} {\bibinfo {author} {\bibfnamefont {C.~N.}\ \bibnamefont
			{Yang}},\ }\bibfield  {title} {\bibinfo {title} {Concept of {{Off-Diagonal
					Long-Range Order}} and the {{Quantum Phases}} of {{Liquid He}} and of
			{{Superconductors}}},\ }\href {https://doi.org/10.1103/RevModPhys.34.694}
	{\bibfield  {journal} {\bibinfo  {journal} {Reviews of Modern Physics}\
		}\textbf {\bibinfo {volume} {34}},\ \bibinfo {pages} {694} (\bibinfo {year}
		{1962})}\BibitemShut {NoStop}%
	\bibitem [{\citenamefont {Sewell}(1990)}]{sewellOffdiagonalLongrangeOrder1990}%
	\BibitemOpen
	\bibfield  {author} {\bibinfo {author} {\bibfnamefont {G.~L.}\ \bibnamefont
			{Sewell}},\ }\bibfield  {title} {\bibinfo {title} {Off-diagonal long-range
			order and the {{Meissner}} effect},\ }\href
	{https://doi.org/10.1007/BF01013973} {\bibfield  {journal} {\bibinfo
			{journal} {Journal of Statistical Physics}\ }\textbf {\bibinfo {volume}
			{61}},\ \bibinfo {pages} {415} (\bibinfo {year} {1990})}\BibitemShut
	{NoStop}%
	\bibitem [{\citenamefont {Nieh}\ \emph {et~al.}(1995)\citenamefont {Nieh},
		\citenamefont {Su},\ and\ \citenamefont
		{Zhao}}]{niehOffdiagonalLongrangeOrder1995}%
	\BibitemOpen
	\bibfield  {author} {\bibinfo {author} {\bibfnamefont {H.~T.}\ \bibnamefont
			{Nieh}}, \bibinfo {author} {\bibfnamefont {G.}~\bibnamefont {Su}},\ and\
		\bibinfo {author} {\bibfnamefont {B.-H.}\ \bibnamefont {Zhao}},\ }\bibfield
	{title} {\bibinfo {title} {Off-diagonal long-range order: {{Meissner}} effect
			and flux quantization},\ }\href {https://doi.org/10.1103/PhysRevB.51.3760}
	{\bibfield  {journal} {\bibinfo  {journal} {Physical Review B}\ }\textbf
		{\bibinfo {volume} {51}},\ \bibinfo {pages} {3760} (\bibinfo {year}
		{1995})}\BibitemShut {NoStop}%
	\bibitem [{\citenamefont {Sewell}(1997)}]{sewellOffdiagonalLongRange1997}%
	\BibitemOpen
	\bibfield  {author} {\bibinfo {author} {\bibfnamefont {G.~L.}\ \bibnamefont
			{Sewell}},\ }\bibfield  {title} {\bibinfo {title} {Off-diagonal long range
			order and superconductive electrodynamics},\ }\href
	{https://doi.org/10.1063/1.532193} {\bibfield  {journal} {\bibinfo  {journal}
			{Journal of Mathematical Physics}\ }\textbf {\bibinfo {volume} {38}},\
		\bibinfo {pages} {2053} (\bibinfo {year} {1997})}\BibitemShut {NoStop}%
	\bibitem [{Note5()}]{Note5}%
	\BibitemOpen
	\bibinfo {note} {Note that we work in a single $\protect \mathcal {C}=1$
		band, such that time-reversal symmetry is already explicitly
		broken.}\BibitemShut {Stop}%
	\bibitem [{\citenamefont {Halperin}(1983)}]{Halperin1983}%
	\BibitemOpen
	\bibfield  {author} {\bibinfo {author} {\bibfnamefont {B.~I.}\ \bibnamefont
			{Halperin}},\ }\bibfield  {title} {\bibinfo {title} {{Theory of the quantized
				Hall conductance}},\ }\href@noop {} {\bibfield  {journal} {\bibinfo
			{journal} {Helv. Phys. Acta}\ }\textbf {\bibinfo {volume} {56}},\ \bibinfo
		{pages} {75} (\bibinfo {year} {1983})}\BibitemShut {NoStop}%
	\bibitem [{\citenamefont {Rubio-Verd{\ifmmode\acute{u}\else\'{u}\fi}}\ \emph
		{et~al.}(2022)\citenamefont {Rubio-Verd{\ifmmode\acute{u}\else\'{u}\fi}},
		\citenamefont {Turkel}, \citenamefont {Song}, \citenamefont {Klebl},
		\citenamefont {Samajdar}, \citenamefont {Scheurer}, \citenamefont
		{Venderbos}, \citenamefont {Watanabe}, \citenamefont {Taniguchi},
		\citenamefont {Ochoa}, \citenamefont {Xian}, \citenamefont {Kennes},
		\citenamefont {Fernandes}, \citenamefont {Rubio},\ and\ \citenamefont
		{Pasupathy}}]{Rubio2022}%
	\BibitemOpen
	\bibfield  {author} {\bibinfo {author} {\bibfnamefont {C.}~\bibnamefont
			{Rubio-Verd{\ifmmode\acute{u}\else\'{u}\fi}}}, \bibinfo {author}
		{\bibfnamefont {S.}~\bibnamefont {Turkel}}, \bibinfo {author} {\bibfnamefont
			{Y.}~\bibnamefont {Song}}, \bibinfo {author} {\bibfnamefont {L.}~\bibnamefont
			{Klebl}}, \bibinfo {author} {\bibfnamefont {R.}~\bibnamefont {Samajdar}},
		\bibinfo {author} {\bibfnamefont {M.~S.}\ \bibnamefont {Scheurer}}, \bibinfo
		{author} {\bibfnamefont {J.~W.~F.}\ \bibnamefont {Venderbos}}, \bibinfo
		{author} {\bibfnamefont {K.}~\bibnamefont {Watanabe}}, \bibinfo {author}
		{\bibfnamefont {T.}~\bibnamefont {Taniguchi}}, \bibinfo {author}
		{\bibfnamefont {H.}~\bibnamefont {Ochoa}}, \bibinfo {author} {\bibfnamefont
			{L.}~\bibnamefont {Xian}}, \bibinfo {author} {\bibfnamefont {D.~M.}\
			\bibnamefont {Kennes}}, \bibinfo {author} {\bibfnamefont {R.~M.}\
			\bibnamefont {Fernandes}}, \bibinfo {author} {\bibfnamefont
			{{\ifmmode\acute{A}\else\'{A}\fi}.}~\bibnamefont {Rubio}},\ and\ \bibinfo
		{author} {\bibfnamefont {A.~N.}\ \bibnamefont {Pasupathy}},\ }\bibfield
	{title} {\bibinfo {title} {{Moir{\ifmmode\acute{e}\else\'{e}\fi} nematic
				phase in twisted double bilayer graphene}},\ }\href
	{https://doi.org/10.1038/s41567-021-01438-2} {\bibfield  {journal} {\bibinfo
			{journal} {Nat. Phys.}\ }\textbf {\bibinfo {volume} {18}},\ \bibinfo {pages}
		{196} (\bibinfo {year} {2022})}\BibitemShut {NoStop}%
	\bibitem [{\citenamefont {Lawden}(1989)}]{Lawden1989}%
	\BibitemOpen
	\bibfield  {author} {\bibinfo {author} {\bibfnamefont {D.~F.}\ \bibnamefont
			{Lawden}},\ }\href {https://link.springer.com/book/10.1007/978-1-4757-3980-0}
	{\emph {\bibinfo {title} {{Elliptic Functions and Applications}}}}\ (\bibinfo
	{publisher} {Springer},\ \bibinfo {address} {New York, NY, USA},\ \bibinfo
	{year} {1989})\BibitemShut {NoStop}%
	\bibitem [{\citenamefont {Prange}\ and\ \citenamefont
		{Girvin}(1990)}]{Prange1990}%
	\BibitemOpen
	\bibfield  {author} {\bibinfo {author} {\bibfnamefont {R.~E.}\ \bibnamefont
			{Prange}}\ and\ \bibinfo {author} {\bibfnamefont {S.~M.}\ \bibnamefont
			{Girvin}},\ }\href {https://doi.org/10.1007/978-1-4612-3350-3} {\emph
		{\bibinfo {title} {{The Quantum Hall Effect}}}}\ (\bibinfo  {publisher}
	{Springer New York},\ \bibinfo {year} {1990})\BibitemShut {NoStop}%
	\bibitem [{\citenamefont {Yang}(1989)}]{Yang1989}%
	\BibitemOpen
	\bibfield  {author} {\bibinfo {author} {\bibfnamefont {C.~N.}\ \bibnamefont
			{Yang}},\ }\bibfield  {title} {\bibinfo {title} {\ensuremath{\eta} pairing
			and off-diagonal long-range order in a hubbard model},\ }\href
	{https://doi.org/10.1103/PhysRevLett.63.2144} {\bibfield  {journal} {\bibinfo
			{journal} {Phys. Rev. Lett.}\ }\textbf {\bibinfo {volume} {63}},\ \bibinfo
		{pages} {2144} (\bibinfo {year} {1989})}\BibitemShut {NoStop}%
	\bibitem [{\citenamefont {Meyer}(2023)}]{Meyer2023}%
	\BibitemOpen
	\bibfield  {author} {\bibinfo {author} {\bibfnamefont {C.~D.}\ \bibnamefont
			{Meyer}},\ }\href@noop {} {\emph {\bibinfo {title} {Matrix analysis and
				applied linear algebra}}}\ (\bibinfo  {publisher} {SIAM},\ \bibinfo {year}
	{2023})\BibitemShut {NoStop}%
	\bibitem [{\citenamefont {Koshino}\ \emph {et~al.}(2018)\citenamefont
		{Koshino}, \citenamefont {Yuan}, \citenamefont {Koretsune}, \citenamefont
		{Ochi}, \citenamefont {Kuroki},\ and\ \citenamefont {Fu}}]{Koshino2018}%
	\BibitemOpen
	\bibfield  {author} {\bibinfo {author} {\bibfnamefont {M.}~\bibnamefont
			{Koshino}}, \bibinfo {author} {\bibfnamefont {N.~F.~Q.}\ \bibnamefont
			{Yuan}}, \bibinfo {author} {\bibfnamefont {T.}~\bibnamefont {Koretsune}},
		\bibinfo {author} {\bibfnamefont {M.}~\bibnamefont {Ochi}}, \bibinfo {author}
		{\bibfnamefont {K.}~\bibnamefont {Kuroki}},\ and\ \bibinfo {author}
		{\bibfnamefont {L.}~\bibnamefont {Fu}},\ }\bibfield  {title} {\bibinfo
		{title} {Maximally localized wannier orbitals and the extended hubbard model
			for twisted bilayer graphene},\ }\href
	{https://doi.org/10.1103/physrevx.8.031087} {\bibfield  {journal} {\bibinfo
			{journal} {Phys. Rev. X}\ }\textbf {\bibinfo {volume} {8}},\ \bibinfo {pages}
		{031087} (\bibinfo {year} {2018})}\BibitemShut {NoStop}%
\end{thebibliography}
\end{document}